# Roadmap on multimode light shaping

**Marco Piccardo[1,2,26], Vincent Ginis[2,3,26], Andrew Forbes[4], Simon Mahler[5], Asher A. Friesem[5], Nir Davidson[5], Haoran Ren[6], Ahmed H. Dorrah[2], Federico Capasso[2], Firehun T. Dullo[7], Balpreet S. Ahluwalia[7], Antonio Ambrosio[1], Sylvain Gigan[8], Nicolas Treps[8], Markus Hiekkamäki[9], Robert Fickler[9], Michael Kues[10], David Moss[11], Roberto Morandotti[12], Johann Riemensberger[13], Tobias J. Kippenberg[13], Jérôme Faist[14], Giacomo Scalari[14], Nathalie Picqué[15], Theodor W. Hänsch[15], Giulio Cerullo[16], Cristian Manzoni[16], Luigi A. Lugiato[17], Massimo Brambilla[18], Lorenzo Columbo[19], Alessandra Gatti[17,20], Franco Prati[17], Abbas Shiri[21,22], Ayman F. Abouraddy[21,22], Andrea Alù[23], Emanuele Galiffi[24], J.B. Pendry[24] and Paloma A. Huidobro[25]**

1 Center for Nano Science and Technology, Fondazione Istituto Italiano di Tecnologia, Milano, Italy
2 Harvard John A. Paulson School of Engineering and Applied Sciences, Harvard University, Cambridge, MA, USA
3 Data Lab/Applied Physics, Vrije Universiteit Brussel, Belgium
4 School of Physics, University of the Witwatersrand, Johannesburg, South Africa
5 Department of Physics of Complex Systems, Weizmann Institute of Science, Rehovot, Israel
6 MQ Photonics Research Centre, Department of Physics and Astronomy, Macquarie University, Macquarie Park NSW, Australia
7 Department of Physics and Technology, UiT-The Arctic University of Norway, Tromsø, Norway
8 Laboratoire Kastler Brossel, ENS-PSL Research University, CNRS, Sorbonne Université, Collège de France, Paris, France
9 Tampere University, Photonics Laboratory, Physics Unit, Tampere, Finland
10 Institute of Photonics, Leibniz University Hannover, Germany
11 Optical Sciences Centre, Swinburne University of Technology, Australia
12 Énergie Matériaux Télécommunications, Institut National de la Recherche Scientifique, Canada
13 Laboratory of Photonics and Quantum Measurements, Swiss Federal Institute of Technology Lausanne (EPFL), Lausanne, Switzerland
14 Institute for Quantum Electronics, ETH Zurich, Zürich, Switzerland
15 Max-Planck Institute of Quantum Optics, Garching, Germany
16 Istituto di Fotonica e Nanotecnologia-CNR, Dipartimento di Fisica, Politecnico di Milano, Milano, Italy
17 Dipartimento di Scienza e Alta Tecnologia, Università dell'Insubria, Como, Italy
18 Dipartimento di Fisica Interateneo and CNR-IFN, Università e Politecnico di Bari, Bari, Italy
19 Dipartimento di Elettronica e Telecomunicazioni, Politecnico di Torino, Torino, Italy
20 Istituto di Fotonica e Nanotecnologie IFN-CNR, Milano, Italy
21 CREOL, The College of Optics & Photonics, University of Central Florida, Orlando, FL, USA
22 Department of Electrical and Computer Engineering, University of Central Florida, Orlando, FL, USA
23 Advanced Science Research Center, City University of New York
24 The Blackett Laboratory, Imperial College London, London, UK
25 Instituto de Telecomunicações, IST-University of Lisbon
26 Guest editors of the Roadmap

Email: piccardo@g.harvard.edu, ginis@seas.harvard.edu

## Abstract

Our ability to generate new distributions of light has been remarkably enhanced in recent years. At the most fundamental level, these light patterns are obtained by ingeniously combining different electromagnetic modes. Interestingly, the modal superposition occurs in the spatial, temporal as well as spatio-temporal domain. This generalized concept of structured light is being applied across the entire spectrum of optics: generating classical and quantum states of light, harnessing linear and nonlinear light-matter interactions, and advancing applications in microscopy, spectroscopy, holography, communication, and synchronization. This Roadmap highlights the common roots of these different techniques and thus establishes links between research areas that complement each other seamlessly. We provide an overview of all these areas, their backgrounds, current research, and future developments. We highlight the power of multimodal light manipulation and want to inspire new eclectic approaches in this vibrant research community.

## Contents

## 1. Introduction

M. Piccardo, Istituto Italiano di Tecnologia, Harvard University
V. Ginis, Vrije Universiteit Brussel, Harvard University

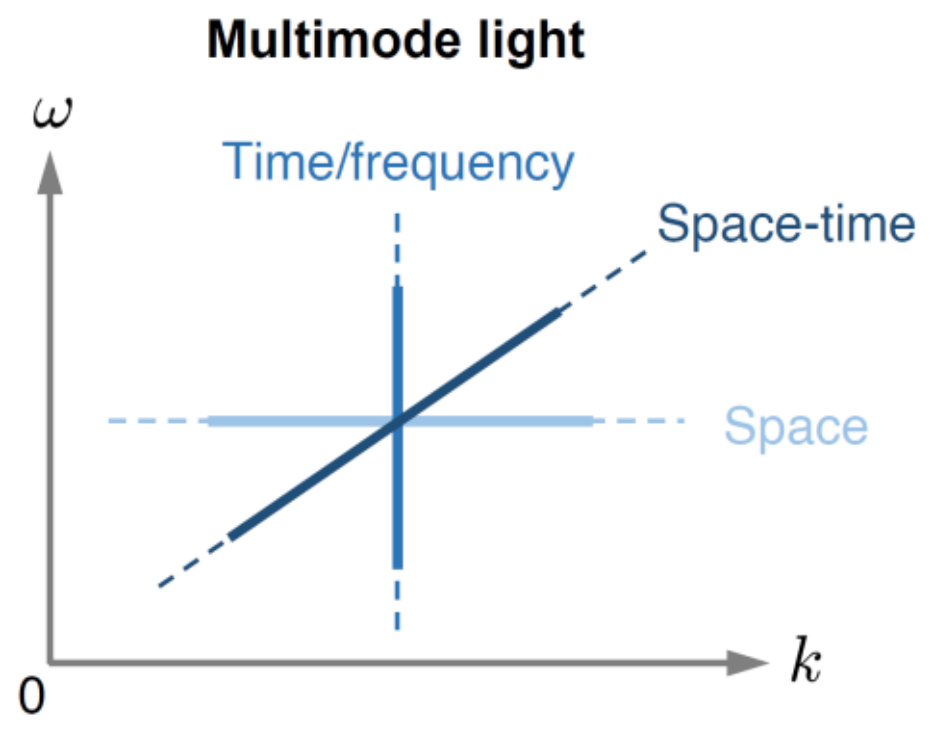

**Figure 1** A schematic showing the different cross-sections of the spectral (k,ω)-plane along which light distributions can be shaped. A superposition of modes with different wavevectors (k) at fixed frequency leads to structuration in space, while superpositions of modes with different frequency (ω) mold temporal waveforms. A combination of the two generates space-time beams.

The ability to engineer the distribution of light with high precision is the key to many applications in optics, from microscopy and material manipulation to spectroscopy and telecommunication. Whichever practical approach is chosen, the design always boils down to selecting and generating a superposition of electromagnetic modes. Traditionally, the concept of structured light [1] begins with the generation of intensity patterns at a given plane. This concept has gradually evolved over the past decades to include the other attributes of light, namely phase and polarization, at several propagation planes. By controlling different wavevectors (k), light can be shaped in space, while the selection of different frequency modes (ω) allows for the shaping of temporal waveforms [2] (Fig. 1). Even more recently, the simultaneous control of light's spatial and temporal properties has emerged, e.g., resulting in the notion of space-time beams.

This Roadmap consolidates this new paradigm of structured light and presents it as a more general concept of multimode light, integrating all degrees of freedom: amplitude, phase, polarization, and time/frequency.

The conceptual division of multimode light presented in Fig. 1 is reflected in the order of the different sections of the Roadmap: Sections 2 to 10 concern modes in space, Sections 11 to 15 concern modes in time and frequency, and Sections 16 to 19 focus on combinations of the two. The first series of contributions (spatial manipulation) begins with the study of light structuration at the source using one or many lasers (Secs. 2,3); then it presents the concept of orbital angular momentum as a new holographic method and with a three-dimensional shaping (Secs. 4,5); it concentrates on the applications of near-field and microscopy (Secs. 6-8); and finally, it passes from classical to quantum states of light (Secs. 9,10). The collection of works on time/frequency starts with quantum states of light but in the frequency domain (Sec. 11); subsequently it examines passive and active frequency comb sources, and in particular their spectroscopic applications (Secs. 12-14); and it concludes on temporal shaping using ultrafast lasers (Sec. 15). The last series of contributions is devoted to space-time optical patterns in cavities, waveguides and free space (Secs. 16,17), and to their specific realization in metamaterials (Secs. 18,19).

We believe that the pooling of the various existing approaches from different subcommunities in optics and photonics [1-5] can be inspiring and suggest new methods to increase our control of light. We also anticipate that many free-space optics approaches will be translated into integrated photonic schemes and that techniques developed in the classical and in the quantum realm may prove useful for each other. In this way, we wish to help establish connections to advance the research in optics and photonics, and to move towards more complex, useful, and holistic patterns of light.

## 2. Structured light at the source

Andrew Forbes, University of the Witwatersrand (South Africa)

### Status

Tailoring light at the source is gaining pace in sophistication, with exciting future prospects. All laser light has some spatial structure, sometime desired, as in the case of low divergence Gaussian beams, and sometimes not, such as thermal aberrations in high power laser beams. How can the desired structure be ensured, and the undesired structure avoided? What resources do we have in our structured light toolbox to achieve this? These are the questions we will address in this roadmap section.

In the early days of laser development it was thought that plane waves (also structured light by virtue of a linear phase gradient) would be the natural modes of Fabry-Perot resonators, but the role of diffraction quickly revealed the menagerie of beams that would be possible from even simple resonators. Since then, a comprehensive toolkit has steadily emerged for structuring light directly at the source: structured light from lasers [1]. Along the way, venerable topics have been reinvented and given a modern twist and once nascent tools have gain wider acceptance and applicability, taking us beyond the textbook Fabry-Perot lasers (see Figure 1). In the context of spatially structured light, this progress has been fuelled by concomitant advances in devices for amplitude and phase control of light, for example, from early diffractive optical elements (DOEs) to today’s liquid crystal (LC) and metasurface-based elements, from adaptive optics to today’s rewritable solutions that include digital micro-mirror devices (DMDs) and LC spatial light modulators. Further, the very geometry of the cavity has altered, from the interference of travelling waves in a resonant cavity, to more exotic solutions based on topological photonics for compact, robust and controllable structured light solutions. A significant driver of these advances in the past decade has been the interest in scalar and vectorial orbital angular momentum (OAM) [2]. With the exponential interest in structuring light for fundamental science and applications alike, the need to do so at the source has become essential.

### Advances, Challenges and Opportunities

Structured light from lasers has advanced tremendously over the past 60 years, accelerating in the 1990s by developments in DOEs, and gaining in pace ever since. Yet there exist many open challenges and exciting opportunities. These include new design principles for versatility and robustness, reconfigurable solutions to alter the structure on demand, and compact as well as high-power solutions to address real-world applications.

**Towards robust designs.** Lasing happens when gain and loss are balanced, a non-linear process of modal competition that determines the nature of the structured light that emerges from the laser. Early designs were based on amplitude (loss) control, typically with wires and stops, introducing high loss for the unwanted modes, while the flip side of the coin is gain control, ensuring higher gain for the desired mode, e.g., through structuring the pump (excitation) light to overlap with the desired lasing mode. Colloquially speaking, these design procedures give the desired mode the edge to win the competition, but this is not guaranteed and often more than just the desire structured mode will lase. Beyond Fabry-Perot cavities, gain-loss control at exceptional points has been exploited to make unidirectional lasers and chirality controlled lasers, while bringing in concepts from topological insulators has seen the design of structured light lasers where the lasing edge states are robust to unwanted perturbations and can be used to enhance the modal discrimination for single mode lasing [3].

Phase-only design approaches often use reciprocity to ensure a repeating structure at the output coupler end of the laser by specifying a phase conjugating element at the rear end. Using this approach, a variety of structured modes have been demonstrated from lasers, including Airy, flat-top and Bessel beams. Unfortunately, the modal discrimination is not guaranteed, the design solution is sensitive to cavity perturbations, only works well if one is near the Rayleigh length of the desired mode, while the procedure fails altogether for modes with spherical wavefronts, e.g., Gaussian as well as HG and LG beams (thus OAM lasers cannot be designed in this way). Recent advances have seen the design generalised to create arbitrarily defined structures at two planes within the cavity [4], useful for specifying both the desired laser output as well as the optimal structure in the gain region to minimise unwanted perturbations.

The challenge is to merge the design principles to ensure the desired mode with low loss and good discrimination, even in the presence of disorder, and to generalise the designs for all of light's degrees of freedom. For example, the exploding field of vectorially structured light has yet to be fully integrated into general laser design principles with only specific solutions existing, primarily for cylindrically symmetric vector vortex beams, e.g., radially polarised light. Designs can rarely be decoupled from implementation, and here the opportunity is to fully harness the promise of nano-featured optics (e.g., meta-surfaces) for high-purity spatial control (see Figure 2) and advanced 2D materials for more sophisticated temporal control [5], where conventional solid-state lasers have benefited from advances in topological insulators.

**Towards reconfigurable solutions.** Although adaptive mirrors have long been used in lasers, their efficacy has been limited to *removing* unwanted structure due to thermal effects. More recently it has been possible to exercise fast and reconfigurable control with a modern toolkit. Externally modulated gain control using DMDs has seen dynamic mode selection from bulk solid-state lasers [6], while chirality controlled pumps have been used to control the OAM twist [7] and value [8] from the output of micro-lasers. Dynamic intra-cavity solutions have gain traction since the early digital-laser demonstration, with both wavelength and OAM tuneability demonstrated in a hybrid free-space-to-fibre system, as well as dynamic intra-cavity vectorial control (see [1] for a review). However, existing solutions are low power and have slow switching speeds, while the link to chirality limits versatility. Adaptive optics may yet be revisited to overcome these restrictions, if they can be made fast, cheap and with a larger stroke. Soft-matter solutions are very much in their infancy, where optical manipulation provides some form of reconfiguration of the system, but could hold future promise.

**Towards compactness and high power.** Presently the challenge is to produce fast and compact on-chip solutions, primarily for communications, while on the other extreme the need is for high-power solutions for laser materials processing, where size is of little concern. On-chip solutions are becoming more versatile in the control of structured light [9], but on-chip integration remains challenging because the structured light toolkit is comprised mostly of bulk solutions for control and detection, an open challenge that needs to be addressed. Most structured light lasers are low power, with the exception the gas, disk and fibre laser systems that have produced cylindrical vector modes (radially and azimuthally polarised), reaching several kilowatts of average power and around 85 GW in peak power. A promising prospect is to produce the desired structured light at low power from a laser and then amplify it using a master-oscillator-power-amplifier (MOPA) approach. Both thin disk, fibres and bulk crystals have been explored in this regard, but with very little attention to the quality of the structure of the light. Compact fibre amplifiers are now explored for multimode amplification, but the challenge remains uniform gain across all modes without coupling or perturbation. A new development is the adaptation of chirped pulsed amplification to demonstrate vectorial light parametric amplification in a polarisation insensitive manner, reaching >1000 fold amplification factors for vector beams [10], bringing structured light a step closer to the extreme light regime.

**Concluding Remarks**

The reader would have noted that the subject matter here has concentrated on spatially structured light. Temporally structured light directly at the source still lags far behind its external control. Frequency modulation, q-switching and mode locking allow only limited tailoring of the temporal amplitude and phase, and while topological insulators have shown to be versatile in some temporal control, few of the existing design principles have been adapted for intra-cavity time-shaped light. While spatially structured light lasers are far more advanced, they exist primarily in research laboratories with few leaps into the commercial world. Thus, the scene is set for exciting prospects in advancing structured light from lasers, both in science and application.

**Acknowledgements**

I would like to thank Ms. Bereneice Sephton and Dr. Yijie Shen for assistance with the graphics.

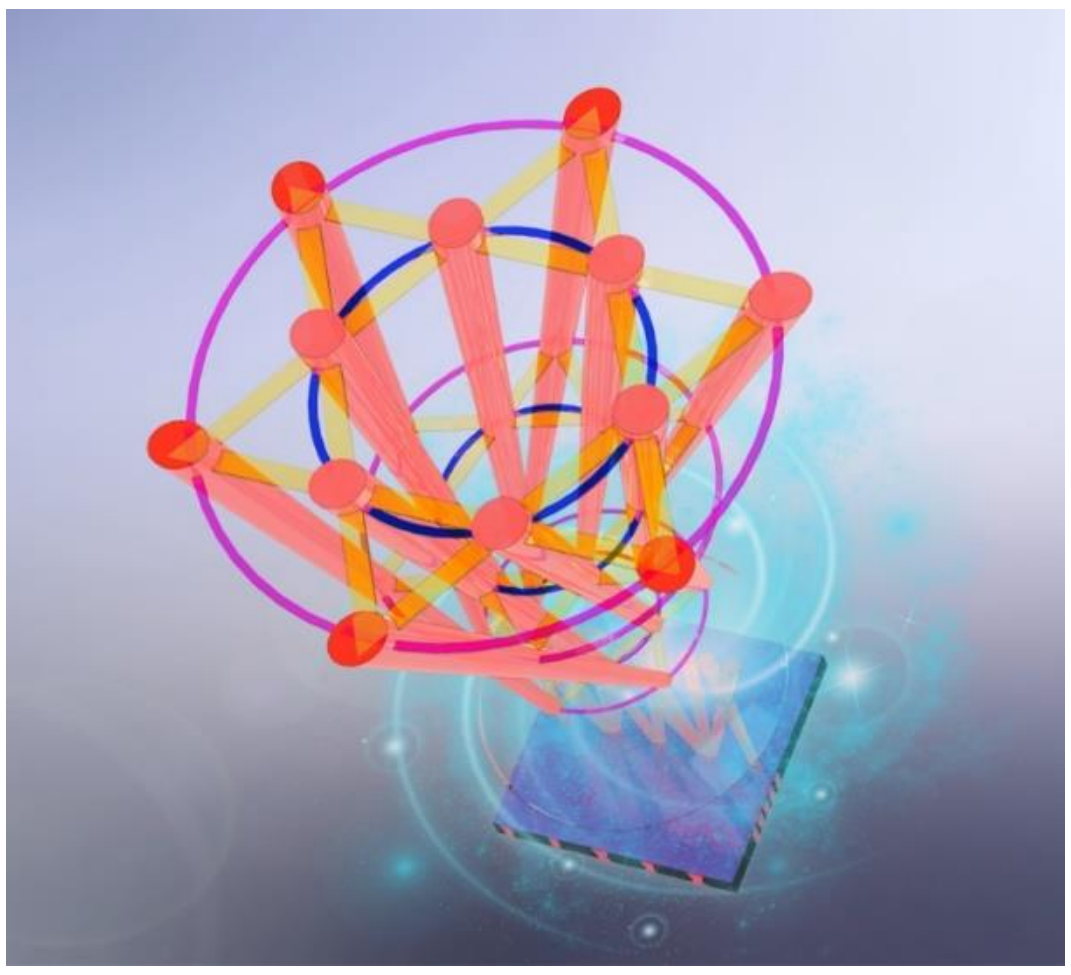

**Figure 1** The ubiquitous Fabry-Perot laser resonator has a history spanning more than half a century, but has been reinvented in the context of ray-wave duality structured light. Shown here is an artist's impression of a high-dimensional structured light mode from a simple Fabry-Perot micro-chip laser. Image reproduced from Y. Shen *et al.*, Optica 7, 820-831 (2020).

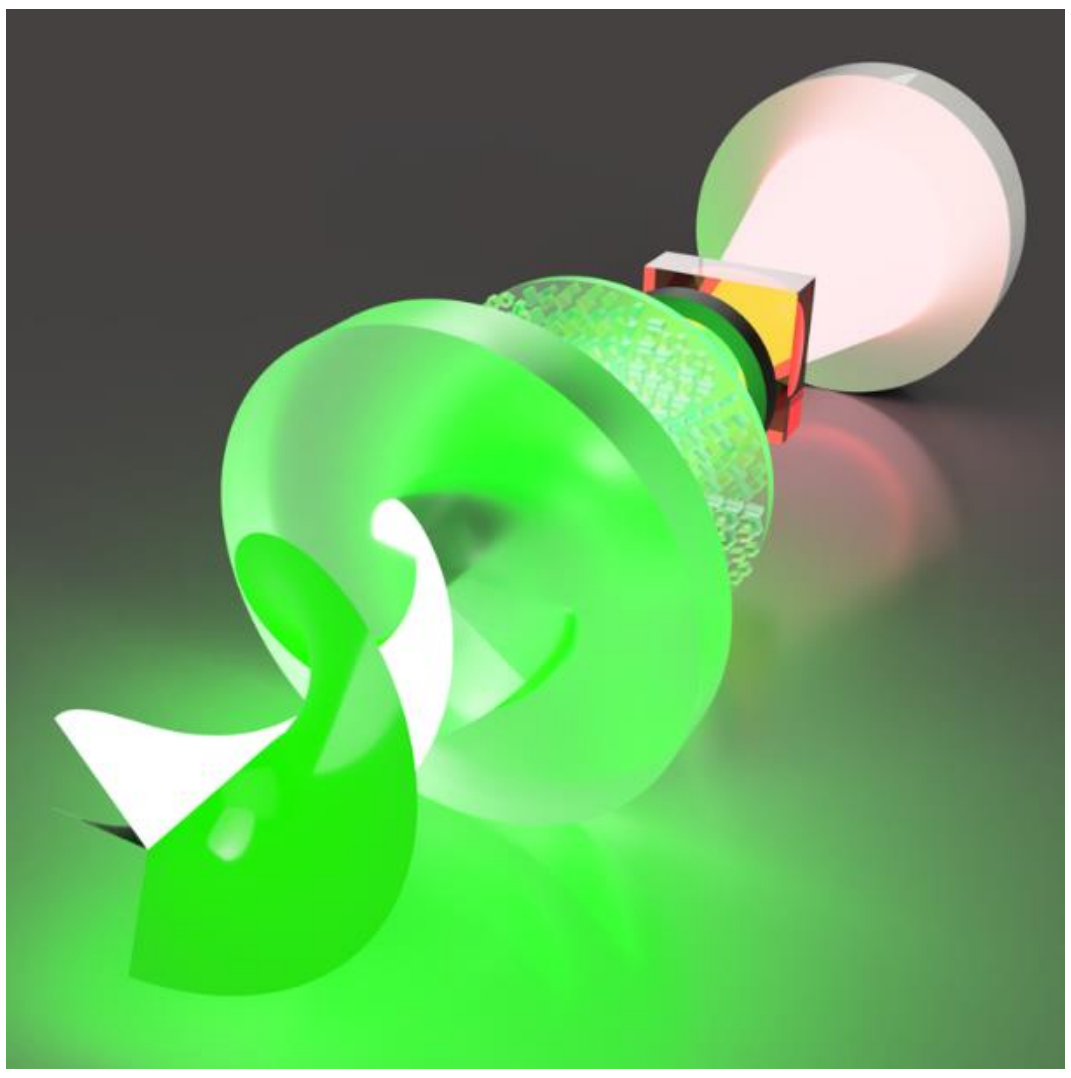

**Figure 2** An illustration of a metasurface laser based on J-plates for producing asymmetric vectorial states of light, as well as high OAM states (demonstrated up to 100) due to a highly resolved vortex centre by virtue of the nano-optical resolution.

## 3. Beam shaping with coupled lasers

Simon Mahler, Asher A. Friesem and Nir Davidson, Weizmann Institute of Science, Rehovot, Israel

**Status**

As noted in the introduction, shaping of laser beams is a continuously active and developing field. Many strides have been made with diffractive elements, metasurfaces and spatial light modulators on manipulating the amplitude, phase and polarization of fully coherent (single mode) lasers in order to successfully obtain a variety of coherent intensity patterns. Unfortunately, such manipulations are suitable for obtaining only certain intensity patterns, and difficult, if at all possible, to apply to high power multimode lasers. Over the past years, we developed a different approach to shape multimode laser beams in space.

It is based on phase locking an array of independent lasers of differing spatial and temporal (longitudinal) modes in order to obtain an overall laser output with controlled amplitude, phase and coherence distributions. As a result, it is possible to obtain a greater variety of optical patterns and deal with high power multimode lasers. In the following, we review some of our developments and results.

Our approach is reminiscent to that used in phased array antennas in the microwave regime, where the far-field beam shape and direction are tailored by controlling the phase of each-emitter in the array.

In our optical approach, the stable setting and control of the relative phases between the lasers in the array is done by phase locking them via dissipative coupling [1-3]. Dissipative coupling involves mode competition where modes with different losses compete for the same gain and only those with the lowest loss survive and lase. With such coupling, phase locking of the lasers can be achieved when all the lasers have a constant relative phase between them. Since the far-field distribution of the lasers is equivalent to the Fourier transform of the near-field distribution, the phase locked patterns of the far-field can be controlled indirectly by the near-field lasers that converge to the minimal loss solution. This process yields an increase of versatile phase locked intensity patterns.

**Current and Future Challenges**

Representative optical patterns with limited number of degrees of freedom, which have been obtained by dissipative coupling in an array of few tens of lasers [1, 2], are presented in figure 1. Figure 1(a) shows patterns that are comprised of sharp spots of high brightness, which correspond to strong coupling between the lasers so they are phase locked with a well-defined relative phase. When the coupling is positive, there is one bright spot at the centre of the far-field pattern that is separated by a distance $D = \lambda f/a$ from the higher order spots (where $\lambda$ the wavelength, $f$ the focal length of the Fourier lens, and $a$ the period of the array). When the coupling is negative, there are four bright spots that are separated by $D$ around a dark centre. When the coupling is one dimensional [e.g. horizontal], the far-field pattern is composed of [vertical] narrow lines separated by $D$. Recently, with nonlinear spatial-temporal coupling, it was shown that the peak intensities at the output could be increased and the stability of the system improved [3].

In general, the coupling can be varied to generate more complex phase locked far-field intensity patterns, such as shown in figure 1(b) [2]. Alternatively, other laser array geometries can also lead to interesting and complex phase locked intensity patterns. For example, the Kagome array in figure 1(c) left where frustration occurs [1], or in figure 1(c) middle with the less organized ring array or in figure 1(c) right with so called "opposite" coupling.

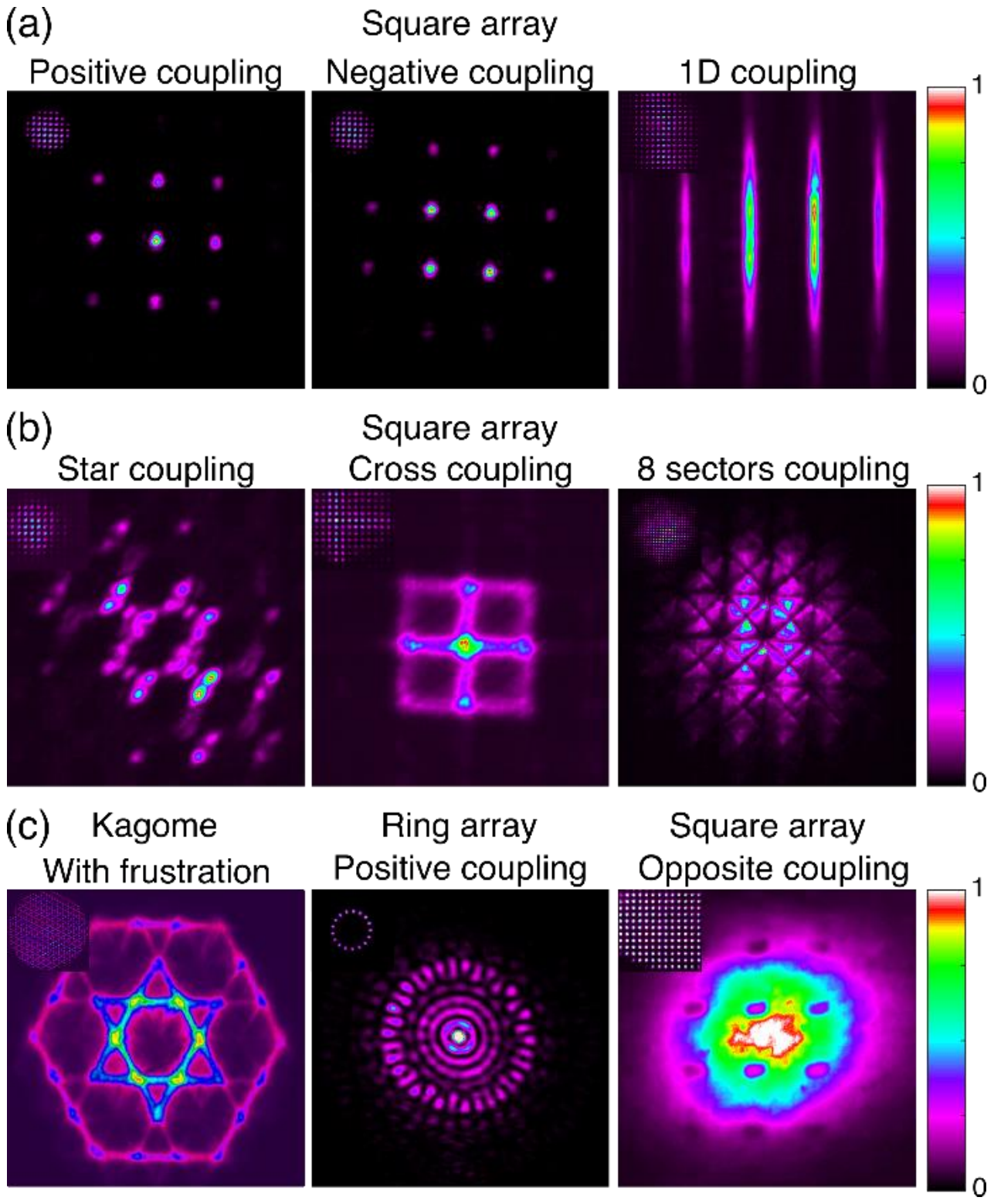

**Figure 1.** Phase locked far-field intensity patterns for arrays of coupled lasers. Insets – near-field intensity distributions of the lasers. (a) Square arrays with positive and negative coupling [1] and with one dimensional coupling [2]. (b) Square arrays with star, cross and 8 sectors couplings [2]. (c) Kagome array with frustration [1], ring array with positive coupling and square array with opposite coupling.

While arrays of coupled discrete lasers can indeed provide variety of beam shapes and far-field intensity patterns, they have limited independent degrees of freedom. For example, the number of lasers in the array that can be stably phase locked is limited to few tens and they often suffer from unwanted high orders of diffraction due to their inherently discrete near-field distribution.

**Advances in Science and Technology to Meet Challenges**

It is highly desirable to develop laser sources with arbitrary on-demand output-intensity patterns, without unwanted high diffraction orders and with many independent degrees of freedom. On the road to this goal, we recently introduced dissipative coupling between the different spatial modes of a highly multi-mode laser source. For a specific degenerate cavity laser (DCL), the number of lasing modes can exceed 100,000 [4], over 1000 times more than the typical number of coupled lasers. Such astronomical number of independent degrees of freedom provides a resource for controlled generation of complex patterns, as shown in figure 2.

Figure 2a shows high-resolution patterns generated with a digital DCL where an intra-cavity spatial light modulator (SLM) is incorporated [5, 6]. The resolution here is limited only by the resolution of the SLM.

We can then control the spatial coherence of these patterns with an intra-cavity aperture while maintaining near optimal image [5].

Figure 2b illustrates the ability to generate patterns that can be imaged through scattering media with minimal distortion [7]. A pattern of the digit "5" generated in the near field of a DCL is still identifiable after propagating through a random intra-cavity phase plate (thin diffuser) located in the far-field (figure 2(b), left). For this undistorted propagation, the laser modes were phase locked with a relative phase that ensured lasing only through selected areas of identical phase with low phase gradients on the phase diffuser (figure 2b, middle and right). Consequently, the loss and thereby the scattering of the far-field diffuser are minimized.

Finally, it is also possible to control the propagation of the laser beam pattern by means of dissipative coupling between the modes [5]. Figure 2c shows two possible patterns at two propagation distances separated by 10 cm inside a DCL. As evident, both patterns of an apple and a star are distinctly observed, where the laser beam gradually changes its shape from the near-field apple pattern (z = 0 cm) to the mid-field star pattern (z = 10 cm) [5].

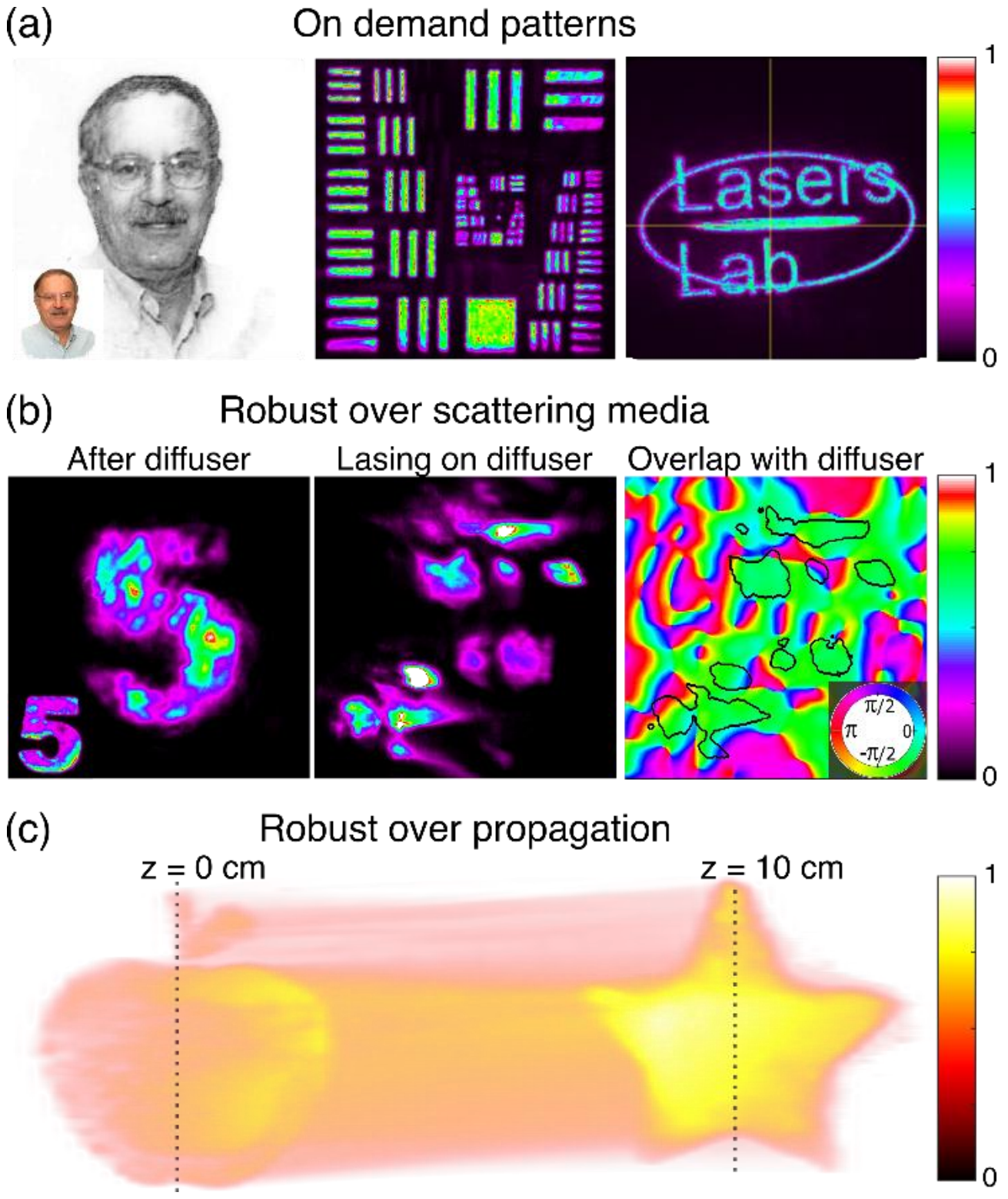

Figure 2. Intensity patterns obtained using dissipative coupling between modes in a digital DCL. (a) On demand high-resolution patterns [5]. (b) Imaging of a digit "5" (left inset) after passing through an intra-cavity scattering diffuser. In order to minimize the scattering effect of the diffuser, the modes phase lock [7]. (c) Simultaneous generation of two distinct intra-cavity apple and star patterns separated by 10 cm [5].

## Concluding Remarks

To conclude, we presented an approach to shape laser beams that is based on phase locking an array of many independent coupled lasers of differing spatial and longitudinal modes. The phase locking processes can follow temporal variations that occur on timescales as short as several cavity round trips, i.e. several nanoseconds for our digital DCL [4, 7]. We demonstrated that it possible to obtain an overall laser output with controlled amplitude, phase and coherence distributions. As a result, it is possible to obtain a great variety of uniquely shaped laser beams with high and low spatial coherence, and even control the shape of the beams at different propagation distances. Our approach and results clearly indicate that it is possible to obtain arbitrary shaped laser distributions, which could lead to new applications. In addition to the spatial control, it should be possible to obtain some temporal control by resorting to rapid spatial light modulation devices (e.g. digital micromirror devices).

**Acknowledgements**

The authors thank their students and colleagues from the Weizmann Institute of Science who participated in various parts of this research and Hui Cao and her team from Yale University for many helpful discussions. The authors acknowledge the Israel Science Foundation (ISF) (Grant No. 1881/17) and Israel Science Ministry for their support.

## 4. Orbital angular momentum holography

Haoran Ren, MQ Photonics Research Centre, Macquarie University, Australia

### Status

Holography offers a vital platform for 3D light-field displays, LiDAR, optical encryption, data storage, and artificial intelligence, allowing parallel read-out of image information at the speed of light. Recent advances in metasurface technologies have allowed the use of an ultrathin device to manipulate the amplitude, phase, and polarization of light, leading to a versatile platform for digitalising an optical hologram with nanoscale resolution. To increase the hologram bandwidth, essential for high-capacity holographic memory devices with potentially ultrafast switching of image frames, such degrees of freedom of light as polarization, wavelength, and incident angles have been exploited to reconstruct different holographic images. However, the bandwidth of a hologram has still remained too low for practical use. Multimode light shaping gives new hope to boost the hologram bandwidth via a new degree of freedom–twisted light carrying orbital angular momentum modes. Recently, the concept of orbital angular momentum (OAM) holography was introduced and demonstrated with a spatial light modulator [1], a phase-only metasurface hologram [2], and a complex-amplitude metasurface hologram [3].

Since the OAM of light has a theoretically unbounded set of orthogonal helical modes, OAM holography using incident OAM beams as independent information carriers holds great promise for largely improving the capacity of a single hologram. To achieve OAM holography, target image is sampled with a discrete sampling array (grid of points) in the Fourier plane of a hologram, creating an OAM-preserving hologram (Fig. 1a). When the image is sampled with a period larger than the Fourier transform of an incident OAM beam, the convolution at the image plane between the holographic image and the Fourier transform of the OAM beam results in an OAM-pixelated image. Adding a helical phase function to an OAM-preserving hologram leads to the OAM selectivity: only an OAM beam with an opposite helical mode index can convert OAM pixels into a Gaussian mode with stronger intensity and decode the OAM-dependent holographic image. Twenty OAM-dependent holographic images have been experimentally reconstructed from a complex-amplitude metasurface hologram (Fig. 1b). Future integrating OAM metasurfaces with on-chip OAM sources and extending OAM holography to 3D holography could offer a truly compact and ultrahigh-capacity holographic memory system for emerging holographic applications, such as portable holographic projectors, head-up displays, and wearable devices for augmented reality.

### Current and Future Challenges

Although OAM holography holds strong promise for increasing the bandwidth of a single hologram, certain challenges lie ahead for its mass deployment. Firstly, spatial light modulators have commonly been used to generate OAM beams for optically addressing holographic images, which makes current OAM holographic systems bulky, expensive, and relatively slow (up to 10 kHz refresh rate). More complications of using spatial light modulators in OAM holography include strong modal coupling and intermodal crosstalk, due to free-space turbulence and turbidity, large beam divergence, and misalignment between the OAM transmitter and receiver. These key challenges set a practical limit on the deployment of OAM in future holographic systems.

Secondly, even though metasurface OAM holography has recently been realized through either a planar metasurface [2] or a complex-amplitude metasurface [3], both approaches inevitably involve a costly and time-consuming nanofabrication process. More specifically, planar metasurface fabrication most usually relies on the use of electron-beam lithography and a reactive-ion dry etching process. Even though 3D direct laser writing technology offers a cheaper solution to the fabrication of a large-scale 3D metasurface by unlocking 3D design degrees of freedom of a meta-atom, a relatively low spatial resolution due to the diffraction limit (pitch distance ~1 µm) and a long processing time hinder its application for practical metasurface manufacturing.

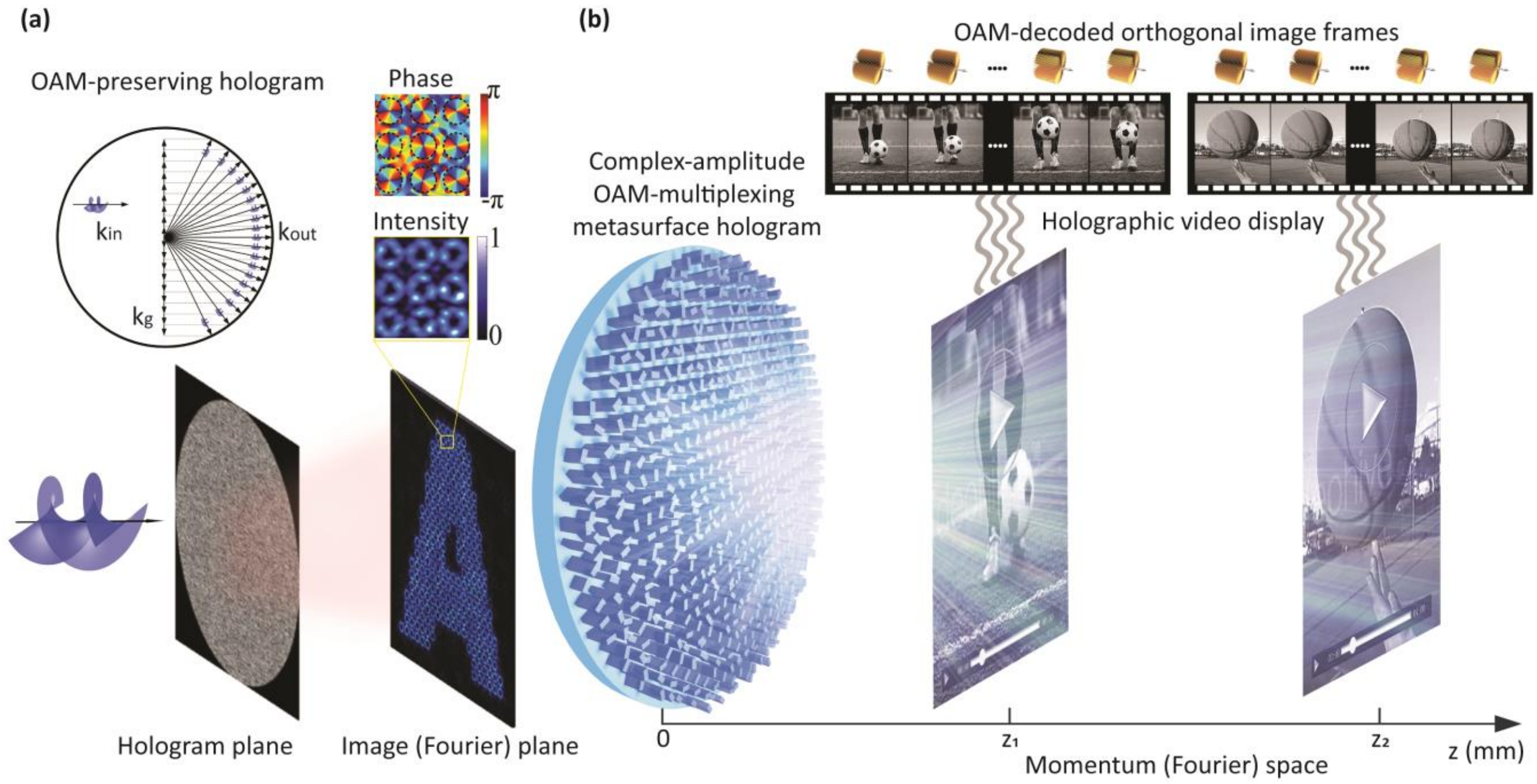

**Figure 1. Principle of OAM holography.** (**a**) Design of an OAM-preserving hologram through OAM-dependent spatial-frequency sampling in the image (Fourier) plane, leading to an OAM-pixelated holographic image. (**b**) An all-optical holographic video display based on a complex-amplitude OAM-multiplexing metasurface hologram, in which OAM-encoded image frames are selectively reconstructed through an incident vortex beam carrying different OAM modes [3]. (B) Reprinted by permission from Macmillan Publishers Ltd: Nature Nanotechnology [3], Copyright (2013).

Thirdly, the image quality of an OAM-decoded holographic image is fundamentally limited by the Nyquist-Shannon sampling theorem and the diffraction limit. To achieve high-resolution OAM-pixelated holographic image, OAM-dependent sampling constants derived from the diffraction of OAM pixels in the Fourier domain should always be smaller than the Nyquist-Shannon sampling constants, which sets an upper limit on the OAM-multiplexing channel number. This immediately makes a trade-off between the image quality and the bandwidth of an OAM-multiplexing hologram. Moreover, OAM-multiplexing holography is not immune to multiplexing crosstalk during the image reconstruction, and in particular OAM beams with a small helical mode index difference suffer from a strong multiplexing crosstalk. In addition, extending the OAM selectivity from a 2D Fourier plane to a 3D Fourier space for 3D OAM holography still remains a challenge, though it is highly demanded.

## Advances in Science and Technology to Meet Challenges

To leverage OAM holography from a lab demonstration to practical applications, photonic integration of OAM holographic systems is of great importance. Twisted light manipulation in nanophotonics has made many breakthroughs in rapid succession and led to a range of chip-scale OAM devices, including twisted light metasurfaces [4, 5], vortex micro-resonators [6, 7], vortex fibre and waveguides [8, 9]. In particular, a complex-amplitude OAM-multiplexing metasurface has the potential to largely increase the metasurface bandwidth, but current metasurface fabrication involves a costly and time-consuming process. Nanoimprinting lithography provides a viable approach to producing large-scale, low-cost, and high-yield metasurfaces. However, it faces two key challenges before it can be adopted for practical metasurface manufacturing: overlay alignment and template fabrication. Alternatively, 3D laser nanoprinting has allowed the printing of a millimetre-scale 3D metasurface with unlocked 3D design degrees of freedom. But time cost for laser nanoprinting was quite high, because a small hatching distance was required to provide sufficient mechanical strength to polymer nanopillars with high aspect ratios. Developing new photoresists with stronger mechanical property and implementing ultrahigh-throughput laser writing technique [10] could facilitate fast laser manufacturing of large-scale 3D metasurfaces.

Current metasurfaces still require external light excitation, making them difficult to be integrated on-chip. Integration of OAM metasurfaces with on-chip OAM sources, waveguides, and detectors will provide an exciting opportunity for future OAM holographic systems-on-a-chip. To tackle with the multiplexing crosstalk in OAM holography, it has been shown that a Fourier-based complex-amplitude hologram [3], with full control of amplitude and phase information, allows the linear superposition principle to be hold for a reduced multiplexing crosstalk. However, it is nontrivial to digitalise a Fourier-based complex-amplitude hologram, in which nearly 90% area of the Fourier hologram has normalised amplitudes smaller than 0.05. Even though an OAM diffuser array that consists of an OAM-dependent sampling array and a random phase function can be used to flatten the large amplitude variation in a complex-valued Fourier hologram, it results in a trade-off between the OAM multiplexing bandwidth and the image contrast. Notably, the random phase in the OAM diffuser array could also be employed for 3D OAM holography.

### Concluding Remarks

Recent demonstration of metasurface OAM holography provides unprecedented opportunities for the development of compact, low-cost, and high-capacity holographic memory devices and systems. While grand challenges still exist on physics/technology/material aspects, these challenges are rather likely the driving forces to push the field forward, eventually making metasurface OAM holography a promising platform for various application areas ranging from optical and quantum communications, holographic displays and encryption, all-optical machine learning, biological imaging, and astronomical observations.

### Acknowledgements

H. R. acknowledges the funding support from the Macquarie University Research Fellowship (MQRF) from Macquarie University.

## 5. Structured Light in 3D

Ahmed H. Dorrah and Federico Capasso, Harvard University

### Status

One need not look beyond the colours of a human eye or the wing of a butterfly to witness how nature has deployed diffraction, film interference, scattering, and photonic crystals to structure the optical spectrum. Engineering the optical properties of matter dates back to the 4$^{th}$-century Lycurgus cup which was made of silver and gold-doped glass and exhibited a mysterious dichroism. Similarly, the Grande Rose of the Reims Cathedral in France, a UNESCO cultural heritage monument from the 13th century, features stained glass with a remarkable range of colours, thanks to the presence of Co, Cu, Cr, Fe and Mn ions. Another instance of structuring the reflection properties of light is Lord Rayleigh's 1887 experiment with multi-layer stacks of dielectrics, known today as 1D photonic crystals. More recently metamaterials have been synthesized to mimic known optical responses and create artificial ones. However, the challenges associated with their mass-production at optical frequencies with subwavelength precision in 3D, has paved the path to metasurfaces — i.e., planar, or flat optics [1]. Moreover, the animate nature of digital holography, enabled by spatial light modulators (SLMs), has enriched wavefront shaping by allowing dynamic control of light, albeit at much lower spatial resolution. With the abundance of these tools, versatile control of light's frequency, wavevector, phase, amplitude, and polarization became possible and the field of structured light subsequently emerged. While structured light, or at least its first generation, has been primarily moulded in a transverse (i.e., 2D) plane, recent efforts have been devoted to structuring light over multiple planes along its optical path, i.e., in 3D. This might seem straight forward in confined media (such as photonic crystals or waveguides); however, the situation under unguided propagation is different.

The notion of structuring light in 3D is very appealing as it provides direct access to engineering the scattering forces of electromagnetic radiation which is at the heart of light-matter interaction. It also becomes particularly useful in biomedical imaging where compensating for propagation losses and scattering inside of living tissues cannot be afforded otherwise. Figures 1(a-b) exhibit two examples of structured light following arbitrary trajectories in 3D. Beams of this nature can either be inverse designed by backpropagating the target pattern to an initial plane that defines the desired hologram [2] or by sculpting the incident waveform into an ensemble of co-propagating modes with different longitudinal wavevectors which beat along the direction of propagation, thereby modulating the resulting envelope at-will via multimode interference. The latter approach is known in the literature as *Frozen Waves* and has recently been exploited in optical trapping [3]. Figure 1(c) depicts another popular class of structured light, vortex beams, created by a J-plate [4]. Such a device can couple any pair of incoming orthogonal polarizations into two different vortex beams with propagation-invariant orbital angular momentum (OAM); this constraint has been further relaxed with a new generation of metasurfaces, called TAM plates (Fig. 1(d)), that can generate vortex beams with longitudinally varying OAM. In this case the OAM varies only locally while being globally conserved across the entire transverse plane of the beam [5]. Other manifestations of 3D structured light include the Talbot effect, abruptly focusing Airy beams, and angular accelerating OAM modes, to name but a few.

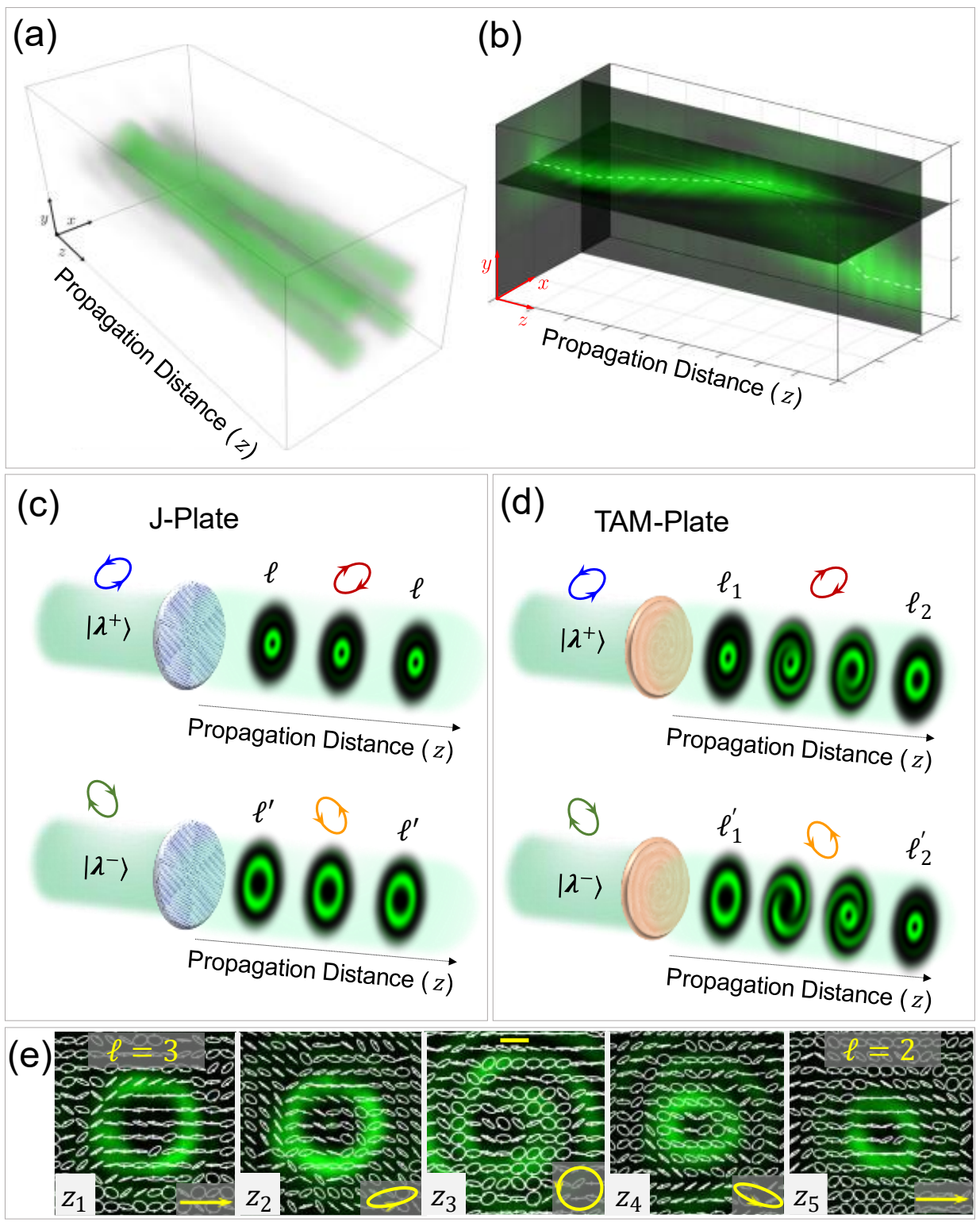

**Figure 1.** Structured light in 3D. a) Diffraction-resistant beam undergoes 4-way splitting under free space propagation. b) Structured beam tailored to traverse an arbitrarily chosen 3D trajectory in space. c) J-plate can couple the eigen polarizations $|\lambda^+\rangle$ and $|\lambda^-\rangle$ to two distinct propagation-invariant OAM beams of different $\ell$ values. d) TAM plates enabling 3D structured light with longitudinally varying OAM. e) Measured transverse intensity profiles of vector beams that change both their polarization and OAM with propagation. The arrows denote the local polarization state. The scale bar is 10 μm. Panels (a-b) are adapted with permission from Ref. [2] (©The Optical Society 2020).

## Current and Future Challenges

Structuring light in 3D is intrinsically constrained by the capabilities of the available wavefront shaping tools. Metasurfaces have enabled new functions that could not be afforded by conventional bulk optics. A prime example of this is dispersion engineering which widely enabled achromatic diffraction-limited focusing and novel devices with wavelength-dependent and multifunctional response — a feature which has recently been harnessed in 3D virtual and augmented reality. Polarization optics is another influential domain which has been revolutionized by metasurfaces by exploiting the shape birefringence of its “meta-atoms” to control the state of incident polarization, locally. This has enabled new capabilities beyond what bulk polarization optics and SLMs can achieve. For example, single multifunctional meta-optics can now project light from a scene into four diffraction orders behaving as analysers thus enabling a compact single shot, full-Stokes polarization camera without standard polarization optics [6] and can also structure light’s polarization at multiple planes along its direction of propagation [7]. Additionally, multiple degrees of freedom, for e.g., polarization and OAM, can be controlled simultaneously along the optical path with a single metasurface, as depicted in Fig. 1(e). Note that replicating these behaviours with bulk polarization optics or SLMs is inherently cumbersome (if not impossible) as the former can only modify incident polarization globally whereas the latter operates on one incident polarization at a time. Beyond these tools, multidimensional spatially varying vector vortex beams have been recently created by exploiting the anisotropy of microchip lasers (see section 2 by A. Forbes). Nevertheless, as the area of wavefront shaping started to mature, the bar for generating complex structured light has also raised, which created a new search for tools to mould the flow of light. The ultimate challenge is to shape all degrees of freedom of light, simultaneously or independently, with high resolution, efficiency, and broadband response, while achieving dynamic (time-varying) control. From the standpoint of structured light, many challenges still persist in terms of controlling highly non-paraxial beams and ultrashort pulses in space and time, and in overcoming the propagation loss, turbulence, and scattering in unguided communications, while enabling high-dimensional, secure, and *intense* structured light.

**Advances in Science and Technology to Meet Challenges**

The evolution of structured light from 2D to 3D has inspired several applications. The forerunners are optical communications, biomedical imaging, and micromanipulation. For example, Fig. 2(a) shows a configuration in which several microbeads are simultaneously trapped at multiple locations along the optical path [3] — a capability that could not be afforded with Gaussian beam traps but effectively realized here via a judicious interference of co-propagating Bessel beams. Additionally, optical trapping has itself been exploited in free-space volumetric displays in which a dynamic photophoretic optical trap drives an illuminated particle along a predefined trajectory in space thus enabling the scattered light to create arbitrary shapes in 3D (Fig, 2(b)); in this case, the displayed volumes can be viewed from any angle in space thus superseding the capabilities of conventional holography [8]. Longitudinal OAM multiplexing is another promising communications scheme which has outperformed conventional OAM division multiplexing by allowing multiple fixed receivers to detect spatially evolving information. Additionally, attenuation-compensated non-diffracting beams are now key enabling tools in light sheet microscopy [9] (Fig. 2(c)) — an area that can also benefit from the polarization diversity of metasurfaces to tailor the fluorescence response and its selectivity. Furthermore, multimode light shaping in 3D has inspired novel sensing schemes in which light changes its topology upon interacting with a fluid of an unknown refractive index [10] (Fig. 2(d)).

Structuring of light in 3D is sometimes also synonymous with controlling all three components of the electric field. In this pursuit, it became possible to shape the longitudinal component of the electric field ($E_z$) under tight focusing, giving rise to polarization Mobius strips, knotted structures, and photonic wheels with transverse angular momentum. Another rapidly growing field of research is nonlinear optics with structured light, owing to the role of spin-orbit coupling in defining the selection rules of non-linear processes. In this regard, several questions remain on how 3D structured light can redefine these nonlinear interactions. Choreographing 3D structured light in time is also a fertile area of research which will be covered later in this article. Relatedly, synthesizing space-time waveforms by precisely correlating the spatial and spectral degrees of freedom has led to new phenomena such as abnormal group velocity in free space, anomalous refraction, and others (see Section 17 by A. Shiri and A. Abouraddy) and is poised to reveal new behaviours. We envision that this area will benefit significantly from metasurfaces over the next few years thanks to dispersion engineering which can enable independent and parallel processing of the incident spectral lines, pixel-by-pixel, by imparting arbitrarily chosen 2D spatial distributions, as opposed to the limited 1D control of current SLM configurations. These are only a few examples, but the list goes on with numerous venues that will be unlocked within the realm of 3D structured light, especially with the projected evolution of inverse design, topology optimization, and time-modulated metasurfaces over the next few years.

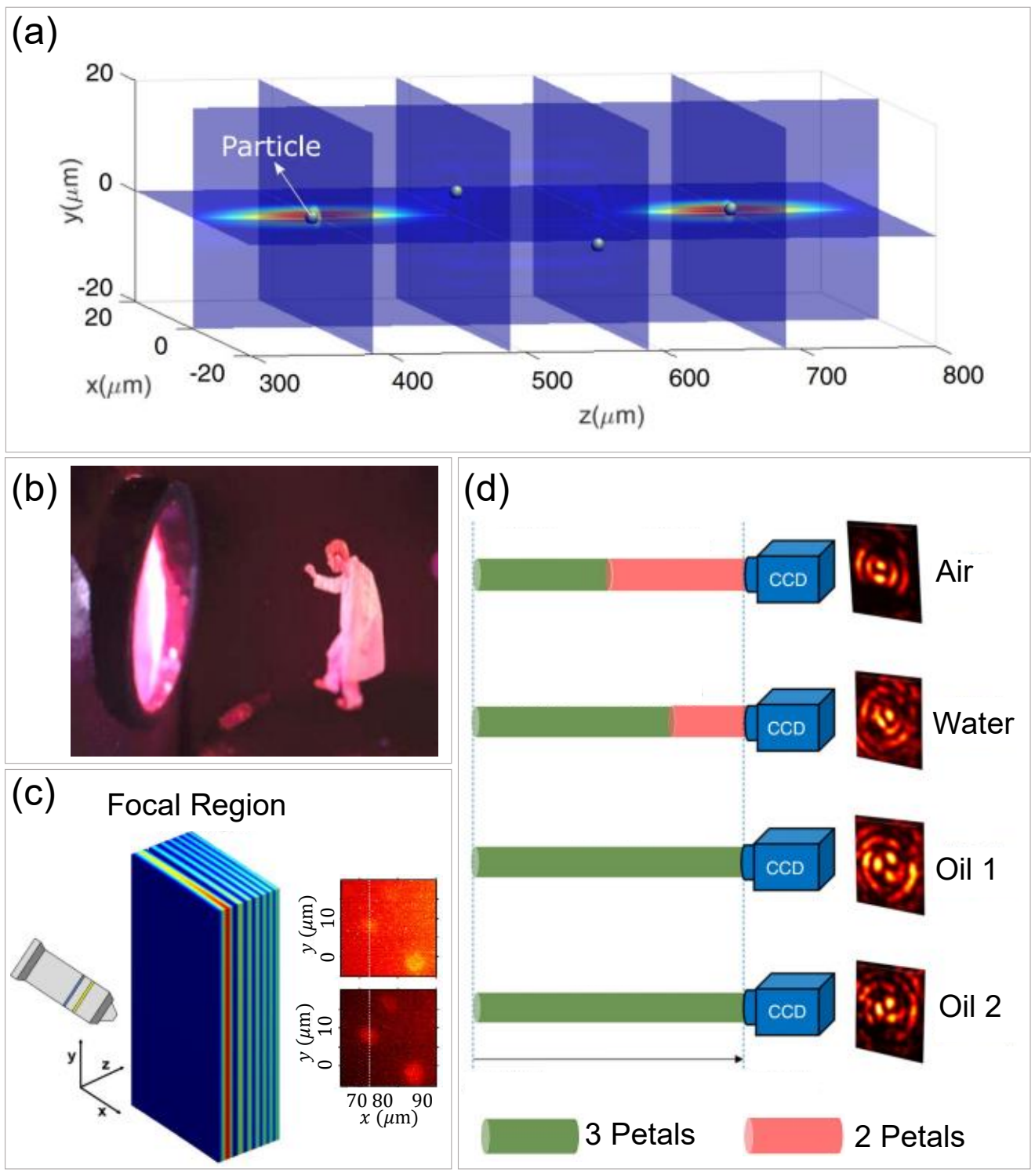

**Figure 2.** Applications of 3D structured light. a) Simultaneous trapping of microbeads at multiple planes along the direction of propagation. b) Photophoretic-trap volumetric display in which the trap and illumination beams are emitted from the circular aperture (left) to form a projected image at a distance. c) Light sheet microscopy relies on illuminating a biological sample laterally with a non-diffracting beam that in turn excites fluorescence plane-by-plane to render high resolution 3D images of the sample (left). Attenuation-compensated non-diffracting beams can be engineered to improve the signal-to-noise ratio when imaging, for e.g., the nuclei in the operculum of tubeworms (right). d) OAM-based refractometry by shining a propagation-dependent vortex beam through a fluid with unknown index of refraction. Panels (a-d) are adapted with permission from Refs. [3,8-10].

## Concluding Remarks

With recent advances in wavefront shaping, new classes of multidimensional structured light will keep emerging. As this rapidly growing area of optics begin to mature, new applications will also develop, and the quest for more sophisticated light shaping platforms will continue. The holy grail is to dynamically control all degrees of freedom of non-paraxial light with high spatial and temporal resolution in space and time using integrated and mass producible tools. With this versatile level of control, pushing structured light from 2D to 3D and from frozen to animate will help unlock a rich palette of optical phenomena throughout the next decade, ranging from the atomic to the astrophysical scale and will lead to the development of many new photonic components and optical techniques.

## Acknowledgements

F.C. and A.H.D. acknowledge financial support from NSF (grant no. 1541959), ONR (grant no. N00014-20-1-2450), AFOSR (grant no. FA95550-19-1-0135), and NSERC (grant no. PDF-533013-2019).

## 6. Sculpting near-field landscapes with multimode waveguides

Vincent Ginis, Vrije Universiteit Brussel, Harvard University
Marco Piccardo, Istituto Italiano di Tecnologia, Harvard University

### Status

The scientific and technological accomplishments of refractive and diffractive optics result from the ability to obtain complex field distributions with high precision. Usually, this is done utilizing a superposition of electromagnetic fields that correspond to different modes [1]. In a homogeneous medium with no material that can act as a scatterer, there are only propagating modes, characterized by purely real wavevectors. However, when one adds an inhomogeneity in a particular region, solutions of the Maxwell equations exist for which at least one component of the wave vector becomes imaginary [2,3]. These are known as evanescent modes. These modes play no role in the far-field regions but are all the more critical for the near field [2]. They determine the variations of the electromagnetic field in subwavelength regions surrounding the inhomogeneity. The electromagnetic near field is essential in applications such as microscopy, mode-switching, and particle detection and actuation. In recent years, the need to engineer specific variations of the near field, therefore, has become more and more compelling.

However, the techniques most commonly used to manipulate the far field are not suitable to control the near field. State-of-the-art techniques, therefore, use nano-antennas that locally convert a propagating incident field into a particular near-field distribution [3,4].

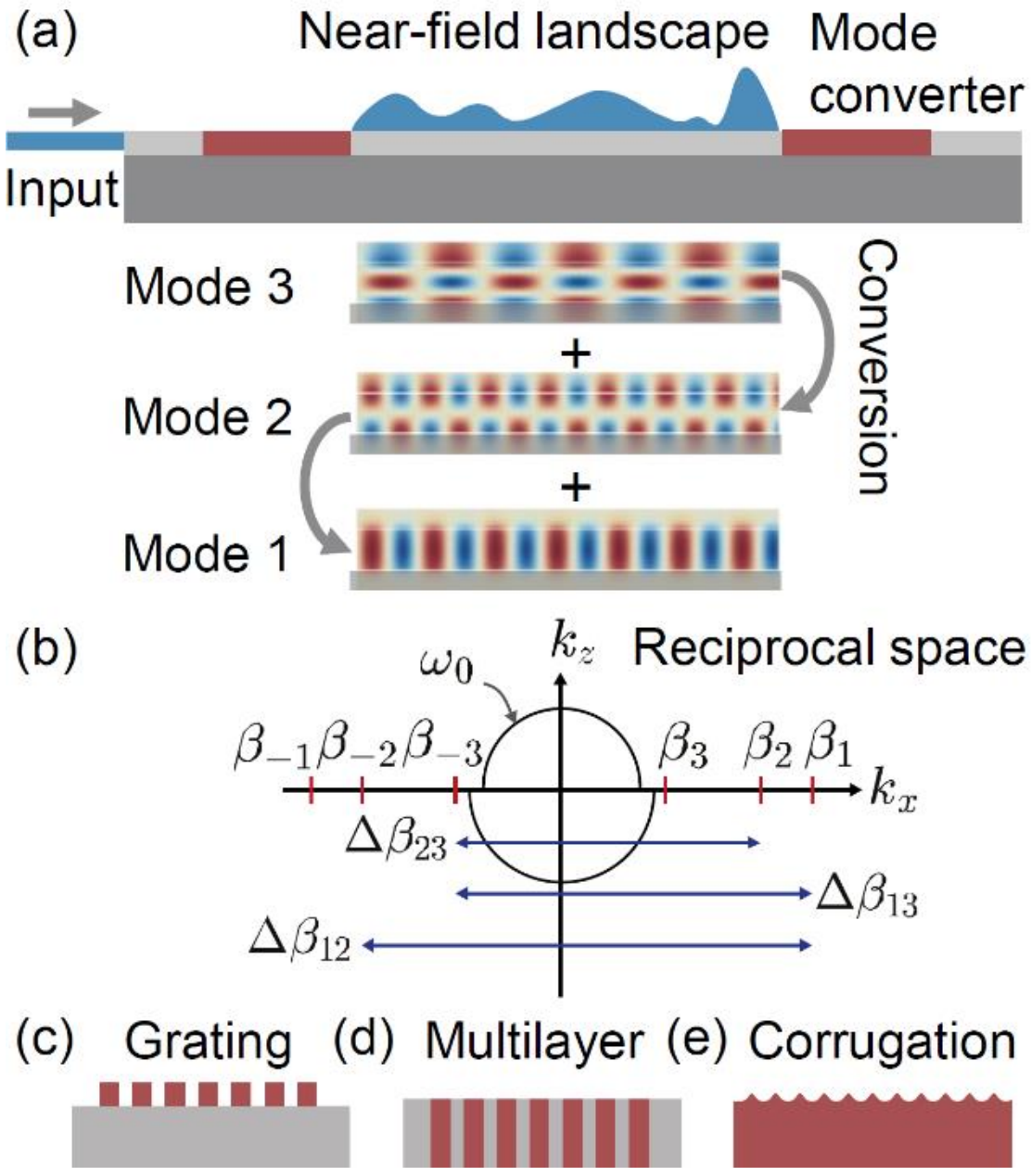

**Figure 1.** (a) The core mechanism underlying cascaded mode conversion for near-field shaping. A beam with a well-defined transverse mode profile is injected into a multimode waveguide and undergoes multiple reflections between two mode converters, in a region of the waveguide called arena. At every reflection, the transverse mode is converted, resulting in a superposition of different modes in the arena with defined amplitude and phase, generating an evanescent near-field landscape above the waveguide. (b) The visualization of the mechanism in the reciprocal $(k_x,k_z)$-plane. The different β's correspond to the longitudinal components of different guided-mode wavevectors. The arrows represent different possible conversions that can be implemented using mode converters to excite various transverse modes in the arena. (c)-(e) Various implementations of mode converters in a waveguide.

Recently, we proposed a different approach to manipulate the near field that does not involve the use of local scatterers [5]. In our work, we shape the near field remotely, analogous to the far-field diffractive techniques, using a multimode optical waveguide, a platform that naturally carries multiple evanescent waves. We define an "arena" delimited by two components, as the target region where the near field needs to appear (Fig. 1). The two components on both sides of the arena act as mode-converting reflectors. By engineering the mode conversion to occur in a cascaded manner, it becomes possible to excite all available modes—and their corresponding evanescent waves—in the arena. The properties of these modes (amplitude and phase) are entirely determined by the converters' properties, making it possible to generate a near-arbitrary linear combination of near-field waves above the arena (Fig. 2).

**Current and Future Challenges**

In the previous section, we introduced the mechanism of cascaded mode conversion for near-field engineering. The mode converters play a fundamental role. Most of the current challenges to achieving arbitrary near-field profiles reside in these elements. The mode converters can be implemented in a variety of ways, such as gratings, multilayer structures, and corrugated waveguide walls (Fig. 1c-e)—the latter being the approach used in the first demonstration of cascaded mode conversion [5].

However, a typical limitation of these approaches is the static nature of the mode converters, allowing only a specific near-field landscape to be realized for a given input. Indeed, in many applications it would be advantageous to have dynamically reconfigurable near-field profiles, for example by independently changing the amplitude and phase of each of the reflected and converted modes. Even more attractive would be the ability to choose different possible paths in reciprocal space between a pool of modes supported by the waveguide, e.g., in a 4-mode waveguide by choosing between the cascading mode conversion 1→3→4 and 1→2→3, etc.

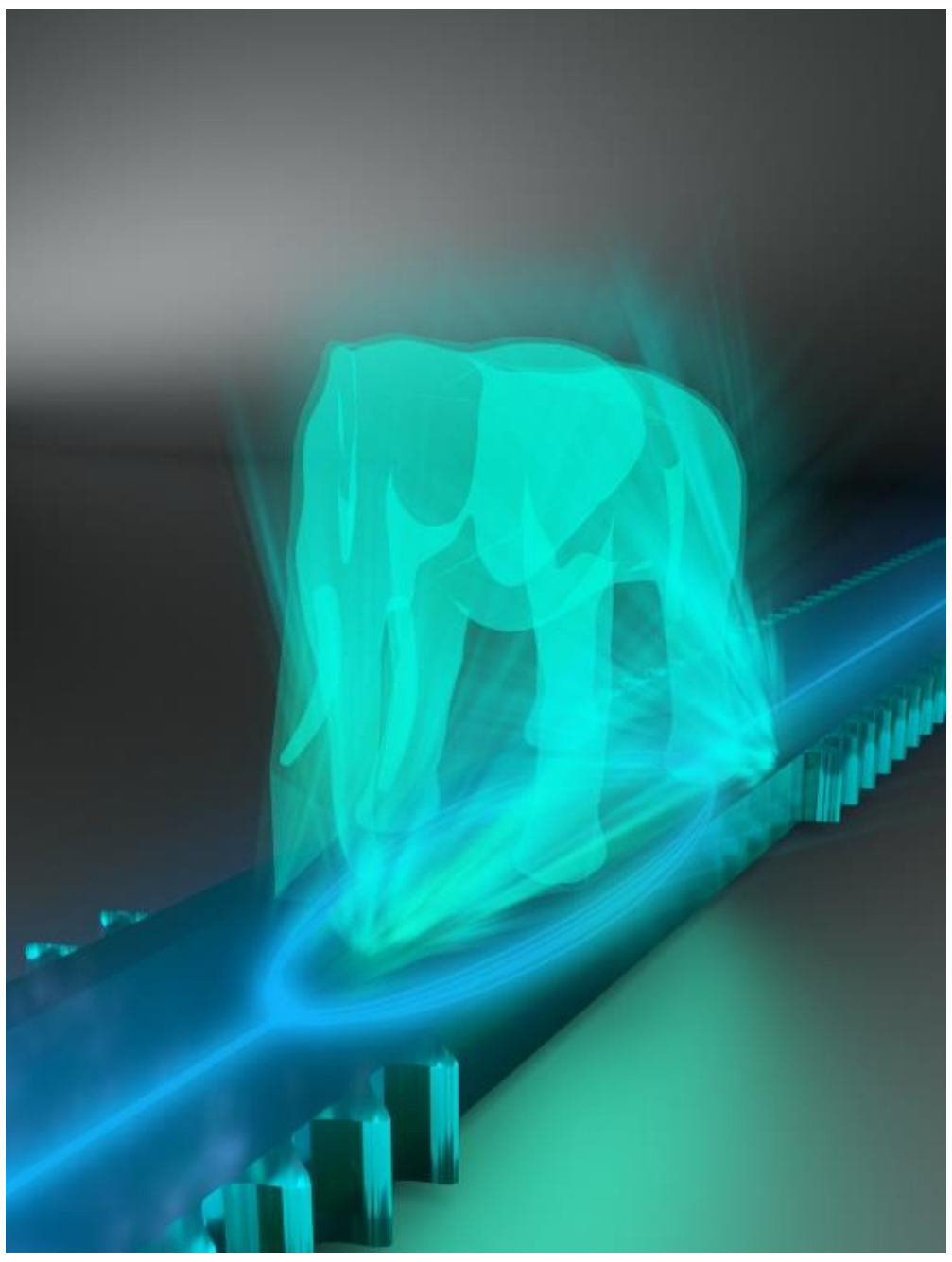

**Figure 2.** An artist impression of an arbitrary near-field landscape shaped by the interference of multiple transverse modes above the arena of a cascaded mode converter. In our work, we show that a one-dimensional, elephant-shaped near-field profile can be obtained using the interference of cascaded mode conversion [5].

Another limitation present in near-field design concerns scalability in terms of the number of usable orthogonal waveguide modes, which ultimately dictates the smallest resolution of the profiles that can be obtained. While the theory of cascaded mode conversion is derived to an arbitrary number of modes, increasing the number of modes in a practical system would require the use of large waveguides and precise control of mode coupling [6]. Considering that the number of spatial frequencies in the intensity profile grows nonlinearly with the number of propagating modes [5], we demonstrated that four independent modes already offer enough design freedom for realizing useful near-field landscapes. A related challenge lies in the conversion between modes of different parity, e.g., from odd to even modes, to exploit mode conversion between all supported modes of the waveguide. Finally, we note that while the first demonstration of cascaded mode conversion showed field structuring along either the longitudinal or the transverse direction, one could aim to control them simultaneously. Summarizing the above considerations, we see an ambitious but promising goal of achieving two-dimensional, dynamically tunable near-field landscapes.

**Advances in Science and Technology to Meet Challenges**

There are several possible approaches to realize dynamical landscapes based on tunable mode converters. The tuning of these mode converters could be achieved electrically, optically, thermally, or even mechanically. The specific technique's choice depends on several constraints, such as the tuning rate, dynamics range, and overall device size. Another tuning parameter that would allow for dynamic landscapes – without tunable mode converters – is the incident mode frequency. In fact, in previous work, we have shown that by changing the frequency of the input field in the waveguide, various landscapes can be created [5]. The extent to which radically different landscapes can be obtained remains an open question.

To increase the number of modes, we note that one could extend not only the width of the waveguide but also its height. Indeed, increasing both transverse dimensions allows for maintaining a smaller footprint for the same number of modes supported by the waveguide, compared to extending the waveguide in one direction only. In addition, asymmetric converters with out-of-phase gratings at both ends of the arena can match modes with different parity.

In the previous experiment, particular emphasis was given to the near-field intensity. Still, it is essential to note that the entire theory is equally valid to sculpt electromagnetic near-field landscapes with multiple degrees of freedom, introducing near-field polarization or orbital angular momentum landscapes [7]. These will be of particular interest in the context of particle manipulation using near-field optical forces [8]. To make this possible, a large arsenal of different reflective mode converters needs to be designed. In this design, there is an important role for inverse design techniques, which have recently become much more powerful through the incorporation of machine learning techniques [9].

Finally, at the level of platform and materials used, we see many possibilities for extensions. The effective indices of the modes that exist can be engineered using epsilon-near-zero materials or nanostructured waveguides [10]. It is also exciting to include so-called confined modes in the continuum in this platform.

**Concluding Remarks**

Cascaded mode conversion provides a versatile path to convert a single incident mode into a superposition of several modes. The accompanying superposition of evanescent fields can be exploited to construct an electromagnetic near field with unprecedented properties. In the coming years, we expect that this concept’s versatility will be further exploited by an expansion of the materials used, on the one hand, and the degrees of freedom of the field that will be designed, on the other hand. Based on these developments, we foresee that exciting research will emerge in the design and the use of the electromagnetic near field.

## 7. Structured illumination microscopy using a photonics chip

Firehun .T.Dullo and Balpreet.S.Ahluwalia, UiT-The Arctic University of Norway

### Status

Fluorescence microscopy is an important tool in biology, however, is bounded with the diffraction limit barrier inherently present in optical microscopy. Recent innovations have led to the development of super-resolution optical microscopy commonly referred as optical nanoscopy [1-2]. Nanoscopy allows observation of fluorescent samples at resolution below the diffraction limit. Nanoscopy consists of a panel of methods and among the existing techniques, structured illumination microscopy, (SIM) is most suitable for live cell imaging. This is due to its high imaging speed, compatibility with standard photostable fluorophores and relatively low phototoxic. SIM uses interference fringes to illuminate the sample, and the fringes are rotated in three orientations to provide isotropic resolution enhancements. Thanks to well-defined periodic illumination generated by the interference fringes, the method benefits from using only 9 or 15 images (2D/3D respectively) to generate a super-resolved image. In SIM, the laser light is spilt in the free space and is made to interfere at the sample plane by the imaging objective lens. Thus, both the illumination and the collection light paths are coupled (see Figure 1a).

The present challenge with SIM is that it uses free-space optics for laser engineering that makes the set-up both bulky and complex. Furthermore, the light paths are prone to misalignment and phase drift. The maximum achievable resolution in SIM is dependent on the fringe period, which is diffraction limited when operated in linear regime using free-space optics. This limits the resolution improvement of SIM to a factor of only 2X over the diffraction limit, i.e., 100-130 nm. The resolution of SIM can be improved by using plasmonic [3] and/or speckle-SIM [4] that carries higher spatial frequency to their optical counterpart and by harnessing blind-SIM [5] reconstruction approaches. However, these methods use a metal interface and uses large number of images losing its edge on temporal resolution. Saturated-SIM (SSIM) [6] that relies on nonlinear higher harmonics can provides higher resolution (50 nm) but requires special dyes and high laser power increasing phototoxicity to the samples. To fully harness the utility of the linear SIM technique, the spatial resolution, the throughput, the system complexity and footprint of SIM need improvement.

### SIM-on-a-chip

Chip-based TIRF-SIM (cSIM) [7] overcome limitations of conventional SIM, by replacing the cover glass with a photonic chip that both generates the illumination pattern and holds the specimen, Figure 1(b-c). The sample is placed directly on top of a chip (optical waveguide) and is illuminated by the evanescent field present on top of the surface of the waveguide. By using two opposing single-mode waveguides, interference fringes are generated inside the waveguide, creating a standing wave evanescent field that can be harnessed for SIM illumination. Further, an array of single-mode optical waveguides, (Fig. 1c, Fig. 2a) is used for three rotational angles at the overlapping imaging area. By shifting the phase of the standing wave of the respective waveguide arm (Fig. 2b), the necessary data for a TIRF-SIM reconstruction, i.e., 9 images (3 angles/3 phases steps) can be acquired. The fringe spacing $f_s$ obtained inside the waveguide is determined by the effective refractive index ($n_f$) of the guided mode and is given by

$$f_s = \frac{\lambda_{ex}}{2n_f \sin\frac{\theta}{2}} \qquad (1)$$

where $\lambda_{ex}$ is the excitation wavelength, and θ is the angle of interference. Thus, the optical resolution supported by cSIM is given as:

$$\Delta_{xy} = \frac{\lambda ex}{2(N.A. + n_f \sin\frac{\theta}{2})} \qquad (2)$$

where N.A. is the numerical aperture of the imaging objective lens. For cSIM, the resolution enhancement can be scaled by changing the fringe spacing of the interference pattern using different pairs of waveguides interfering at different angles (Fig. 1f). Moreover, by using waveguides made of high refractive index material, such as, silicon nitride, $Si_3N_4$ (n=2; $n_f$=1,7) and titanium dioxide, $TiO_2$ (n=2.6; $n_f$=2.4) (Fig. 1g), the $f_s$ can be made smaller than what can be achieved using a high N.A. objective lens (N.A.=1.49). Eq. 2, gives a maximum theoretical resolution enhancement of 2.4 and 2.6 for $Si_3N_4$ and $TiO_2$ materials respectively, for an objective lens of N.A.=1.49. Importantly, the fringe spacing is independent of the imaging objective lens, thus not changing when a lower magnification and lower N.A. lens is used to increase the field of view. This opens the possibilities to achieve super-resolution imaging over large area. The proof-of-principle of cSIM was presented on 100 nm fluorescent beams (Fig. 2 c-d) and biological samples.

**Advances in Science and Technology to Meet Challenges**

To fully harness the impact of cSIM, several technical aspects must be pushed. Firstly, the imaging area (interference area) of cSIM is fixed and cannot be changed freely. Thus, it is desirable to have large imaging areas or to use an array of pre-defined imaging areas. Tapering of single-mode rib waveguides has been demonstrated but suffers from longer footprint. It is possible to use narrow stripe waveguides (nanowires) and let the light diverge in an open slap region (imaging area). Diverging light will allow imaging over large area, however fringe period (spatial frequency) will change with distance due to mode divergence inside the slab waveguide.

Different on-chip and off-chip phase shifting strategies for cSIM have been proposed [7]. A compact set-up and a stable interferometer can be made by using on-chip phase shifters, such as thermo-optical modulator, MEMs-based phase shifter, etc. Here, it is essential that on-chip phase modulators are power efficient, generate less heat, have fast response time, and takes small footprint. For fluorescence microscopy, it is desirable to use entire visible spectra, 405-660nm. Consequently, the chip-platform should possess low propagation losses and low auto-fluorescence for the entire visible spectrum. It was recently reported that tantalum pentoxide ($Ta_2O_5$) possesses lower auto-fluorescence as compared to the $Si_3N_4$ for shorter wavelengths [8]. Further studies are necessary to develop efficient integrated photonic circuits that are capable to support entire visible spectra.

When using cSIM to image large areas for example with a low N.A./magnification objective lens (20X 0.65 N.A.), it would be necessary to use more angles and phase steps to fill the frequency space for isotopic resolution enhancement. Another approach to image large areas would be to let multiple beams (4-8) interfere randomly generating on-chip speckle alike illumination pattern and harness blind-SIM approach if temporal resolution can be a trade-off. The cSIM technology will benefit from development of on-chip integration of light, further miniaturizing the optical set-up.

**Concluding Remarks**

The photonic chip reduces the footprint of the light illumination path of TIRF-SIM to around 4x 4 $cm^2$, enabling compact footprint and provides higher optical resolution than conventional SIM technique. Until now different on-chip TIRF-nanoscopy methods have been developed, such as on-chip single molecule localization microscopy [9-10], on-chip intensity fluctuations based nanoscopy and cSIM. This opens the possibilities of delivering multiple nanoscopy methods using the same photonic chip.

On-chip illumination strategy enables excitation over large area, however, the light collection is still presently limited by the imaging objective lens. Integration of on-chip illumination with micro-lens arrays for light collection could potentially open the avenues for massive parallelization, where illumination and collection both can be done over large areas. The on-chip nanoscopy will find usage for bio-applications where TIRF imaging over large areas are essential, such as in neurosciences, drug discovery, cell migration, etc.

### Acknowledgements

This work was supported by the European Research Council (Grant No. 336716 to B.S.A.). We would like to thank Øystein Ivar Helle, Marcel Lahrberg, Jean-Claude Tinguely and Olav Gaute Hellesø.

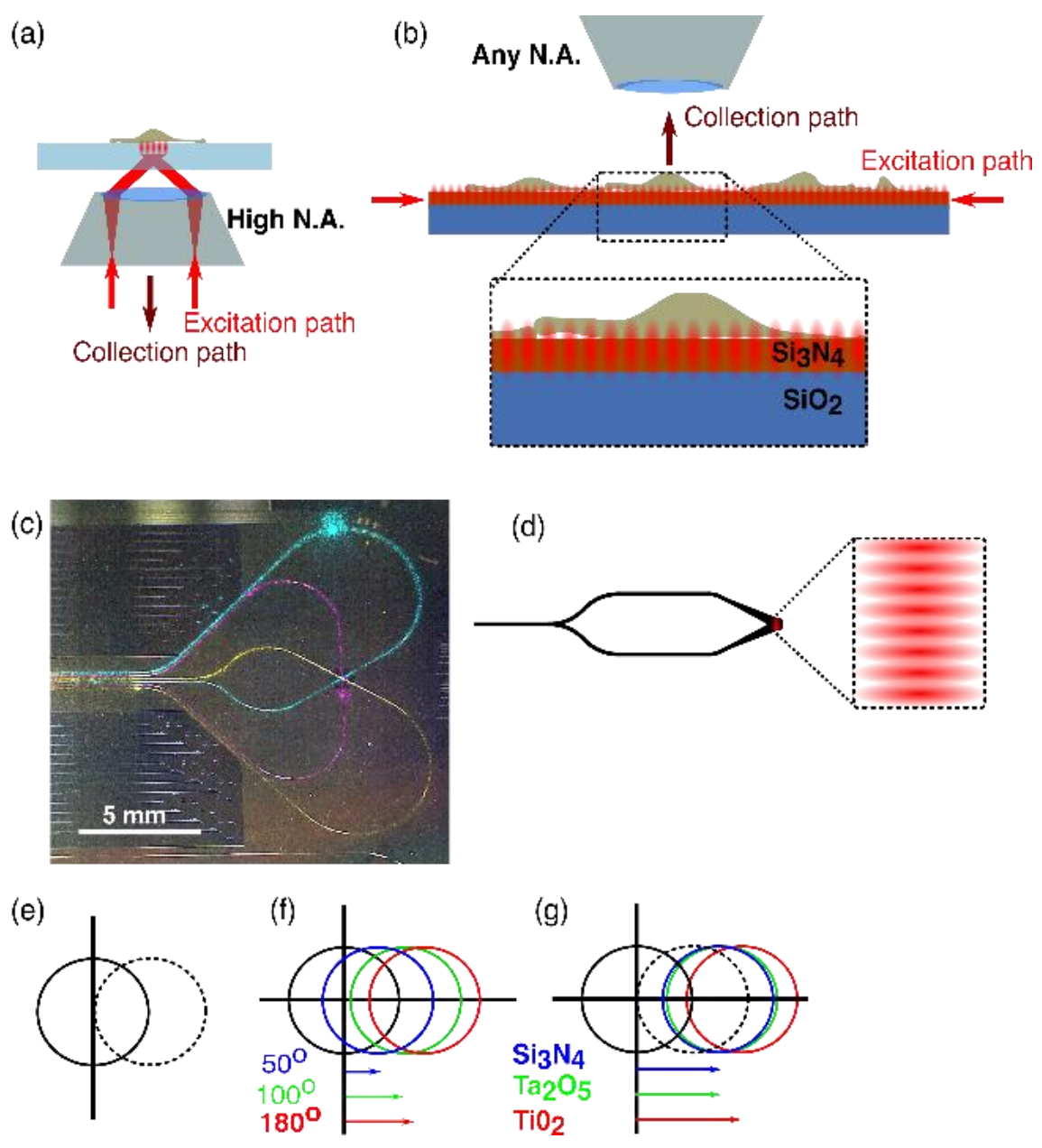

**Figure 1.** (a) Conventional SIM relies on a high-NA objective lens for both excitation and collection. (b) cSIM harnesses interference in a waveguide to excite the specimen via evanescent fields, decoupling the excitation and collection light paths and (c)Three waveguide pairs interfere at different rotational angles (pseudo-colour). (d) A low interference angle creates an interference pattern with a high fringe spacing. (e) The microscope resolution is represented by the optical transfer function (OTF; solid circle). The maximum OTF shift in conventional SIM yields a 2 times resolution enhancement (dotted circle). (f) The OTF shifts with the fringe spacing, and cSIM surpasses 2 times resolution for large interference angles. (g) Using higher-refractive-index materials scales the resolution more, for example $Si_3N_4$ ($n \approx 2.0$), $Ta_2O_5$ ($n \approx 2.05$) and $TiO_2$ ($n \approx 2.6$). Reprinted by permission from Springer Nature: Nature Photonics [7], (2020).

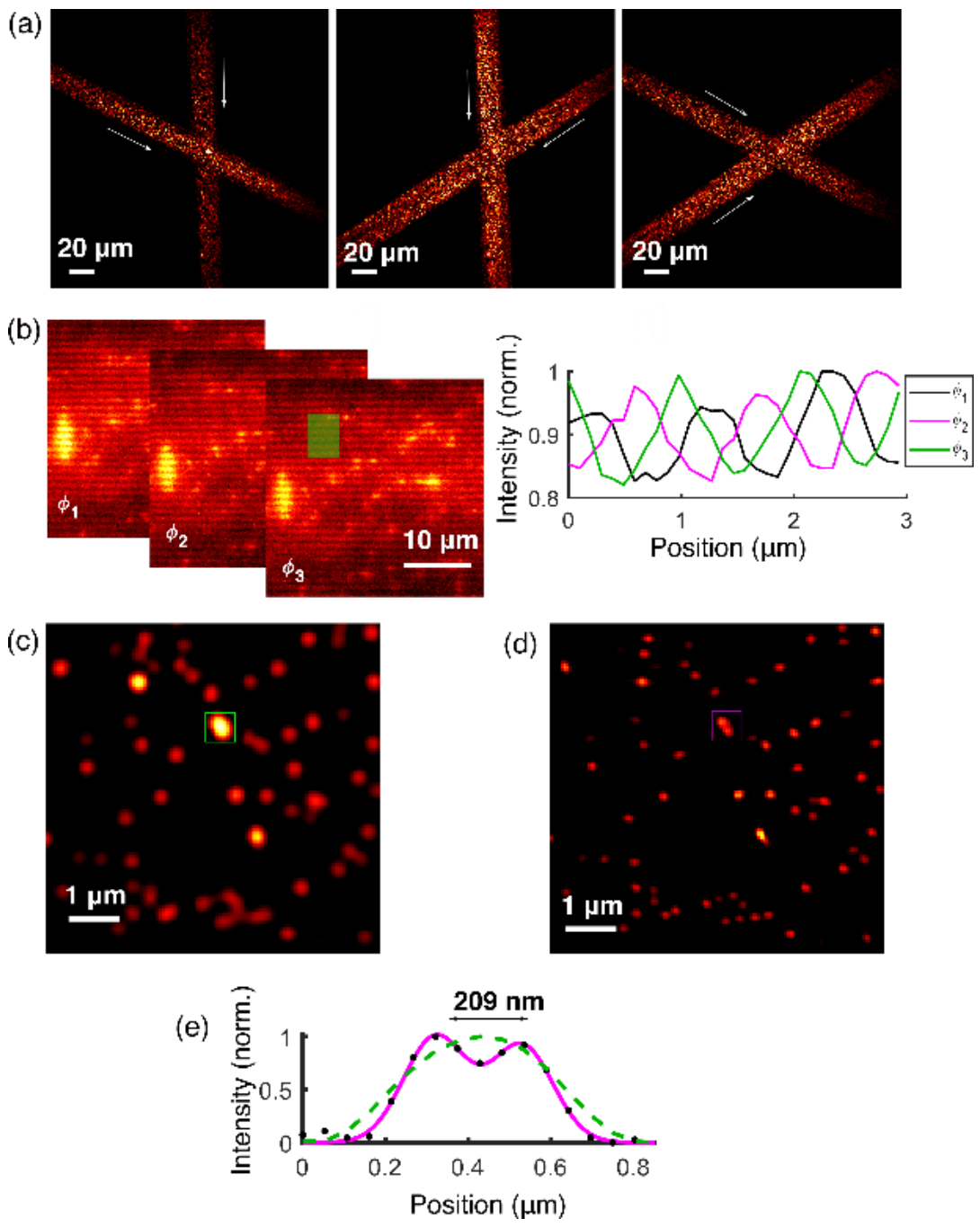

**Figure 2.** (a) The SIM structures with an interference angle of 60° and three angles of rotations for isotropic resolution enhancement. The white arrows indicate direction of light propagation. (b) cSIM patterns at three phase steps ($\phi_1$, $\phi_2$, $\phi_3$), imaged using a fluorescent dye (20° interference was used for increased contrast) and plot of intensity from the position marked with a green box for the three steps. A 100 nm fluorescent bead sample imaged with diffraction-limited resolution (c) and the cSIM reconstruction of the same region (d). The green box in (d) indicating the position of the line profile in (e), where the dashed green curve shows the diffraction-limited result, the black dots the cSIM measurement values and the magenta line a fit. Two beads located 209 nm apart are resolved using cSIM, but not in the diffraction-limited image. The excitation/emission wavelengths were 660 nm/690 nm. 60X, 1.2 N.A. water immersion objective lens was used. Reprinted by permission from Springer Nature: Nature Photonics [7], (2020).

## 8. Vectorial nano-imaging: Ultraresolution in focused light interstices

Marco Piccardo, Istituto Italiano di Tecnologia
Antonio Ambrosio, Istituto Italiano di Tecnologia

### Status

Since the first observations of microbial life by Antonie van Leeuwenhoek, one of the pioneers of the microscope, the main challenge in the field of microscopy has always been to improve the resolution of imaging, to reveal ever finer features. It was established by Ernst Abbe in 1873 on the basis of a theory of light diffraction that resolution cannot be improved arbitrarily, but rather is constrained by an ultimate limit, of the order of the wavelength of the illuminating beam. However, this limit is only valid for far-field techniques, and based on a linear, scalar treatment of light.

Efforts to overcome Abbe's limit have proven successful by concentrating on two main directions. On the one hand, detection in the near-field with small probes systems sensitive to evanescent waves has shown that deeply subwavelength resolution can be achieved, though being limited to imaging mostly near the surface of a sample. On the other hand, far-field techniques have evolved towards sophisticated schemes involving nonlinear and/or vectorial light-matter interactions to increase the resolution below the diffraction limit, even inside a sample. In all these approaches, shaping light using several of its degrees of freedom is key to extracting as much local information as possible from the imaging. Stimulated emission depletion (STED) microscopy is a great example of such multimodal structuration of light for microscopy, as the excitation and depletion processes use beams of different frequency, spatial and phase distribution (typically, a donut beam) to achieve super-resolution thanks to a nonlinear response of the fluorophores labelling the sample.

Vectorial imaging can be another viable approach to break the diffraction limit. Its basic principle is illustrated in Figure 1. After strong focusing of a coherent beam, the focal field intensity, which is inherently scalar, obeys the diffraction limit. However, when considering the full vectorial structure of light one recognizes that the direction of the field, determined by the transverse and longitudinal polarization components [1], can vary on a much smaller scale [2]. For instance, the polarization state changes rapidly at the boundary between the horizontally-polarized donut and the longitudinally-polarized spot (Fig. 1). In presence of a polarization-dependent material response, these photonic spin structures could be used for sub-wavelength excitation and particle manipulation. Vectorial light has been used in recent years to demonstrate several interesting phenomena, such as polarization Möbius strips [3], photonic skyrmions [2], and torque in anisotropic materials due to the longitudinal field [4], but its development is still ongoing.

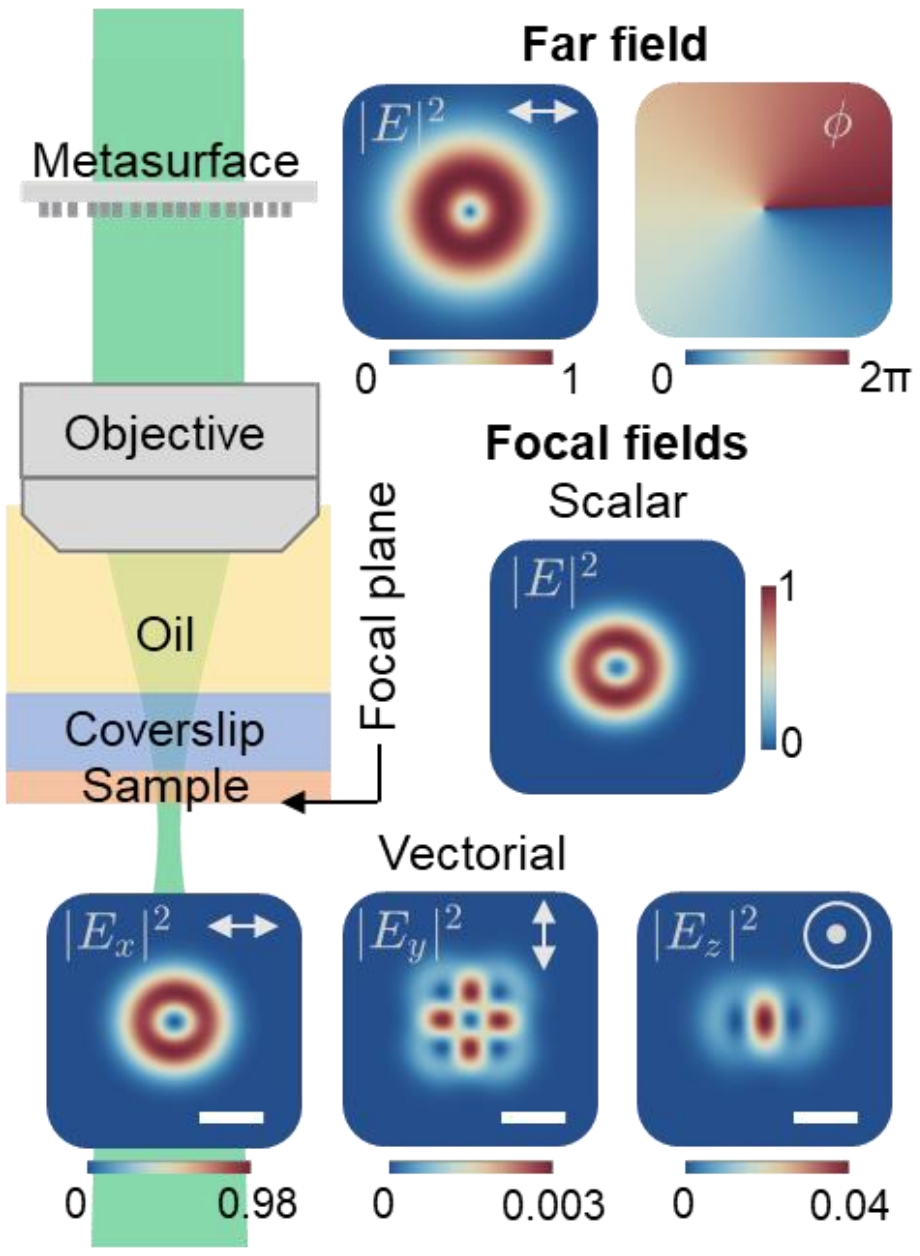

**Figure 1.** A horizontally-polarized vortex beam structured using a metasurface is tightly focused onto a sample through a high numerical aperture objective (NA = 1.4). The total intensity calculated at the focal plane still shows a donut beam but the analysis of its polarization components reveals the existence of other field distributions, which are vertically and longitudinally polarized. The scale bar in the vectorial plots corresponds to the effective wavelength inside the sample.

**Current and Future Challenges**

The characterization of vectorial distributions of light on the nanoscale may follow different approaches. For instance, one can use gold nanoparticles as localized re-emitters to map the transverse component of a strongly focused field, and a near-field scanning optical microscope (NSOM) for the longitudinal component [1]. We identify the biggest difficulties in vectorial nano-imaging as lying in the preparation of the optical beam and in modelling the light-matter interaction. We summarize some of the main challenges in this research in the form of open questions.

***Vectorial light design***

- A strongly focused field carries all the three polarization components, but what is the best combination of polarization and spatial modes for vectorial microscopy? An important problem in this context is to calculate, starting from a desired polarization distribution in the focal plane, the needed far-field pattern. This is equivalent to tackling inverse, rather than direct, design of vectorial focal field distributions, from the sample to the source.
- How can we exploit multicolor illumination in vectorial nano-imaging? This could be used to overlay different polarization distributions in the focal plane for frequency-dependent nonlinear interactions in a sample.
- It was recently shown that the vectorial field can be structured along the optical path for unfocused beams [5]. Could this approach be applied also to the regime of strong focusing, with arbitrary polarization transformations within the focal field region?

***Light shaping device***

- What light-shaping tool is best suited to create a desired distribution of vectorial light? One could consider using either intracavity schemes, such as lasers and resonators shaping light directly at the source [6], or devices operating outside the cavity, such as spatial light modulators and metasurfaces. The first are advantageous in terms of efficiency and output power, which are favourable for microscopy processes relying on nonlinear interactions, while the latter may give more freedom and higher resolution for the realization of arbitrary patterns, as well as ease of alignment.

***Vectorial light-matter interaction***

- How does the sample respond to vectorial light? To answer this question an option is to derive an ab initio physical model of the material, though this can be very complex, as for the case of biological samples. Another approach is to develop a phenomenological model of light-matter interaction using empirical coefficients obtained by fitting experimental measurements [7,8].

## Advances in Science and Technology to Meet Challenges

Advanced structuration of light begins with classical holography, invented by Dennis Gabor in 1948. However, the original concept was based on a scalar notion of the electric field: holographic plates, like other technologies that followed, provide a control only over the amplitude and/or phase of the field. A new tool that emerged in the last decade, disrupting the design of polarization-variant beams are metasurfaces, arrays of birefringent nano-optical elements with subwavelength resolution. Considering that polarization can be used also as a spatial filter to realize amplitude distributions [9], metasurfaces become a very valuable approach for designing vector beams, both intracavity and outside the cavity.

An interesting feature of metasurfaces is that they can effectively operate as multiple optical elements stacked together, just by summing up the corresponding phase profiles. For the purpose of vectorial nano-imaging, in principle one could imagine integrating both the functionality of polarization shaping and strong focalization into the metasurface. However, for the latter a large numerical aperture is needed, which can still be technologically challenging in metasurfaces. Another difficulty, if multicolor illumination is desired, is the chromatic aberration of the metasurfaces. However, when the design needs to be set to only a few frequencies, rather than for a wide spectral band, expedients similar to those employed in RGB metasurfaces could be employed. Finally, regarding the aforementioned problem of inverse design of vectorial focal field distributions, artificial intelligence could come in handy to empower the metasurface design with self-adaptive and learning capabilities.

## Concluding Remarks

Vectorial imaging is a promising path to overcome the scalar limitation of the diffraction limit. Improving our ability to shape light is key to this goal and should head towards four-dimensional (4D) light fields, which are field distributions nano-structured in three-dimensional (3D) space while embedding an additional degree of freedom represented by polarization, useful to address optical functionalization in a 4D material [10]. For this direction of research, we see metasurfaces as a major player on the scene due to their subwavelength control of amplitude, phase, and polarization. Since one of the driving applications in the field is imaging of biological samples, future efforts by optical physicists should go hand in hand with the needs of biologists, e.g. structured light fields should be tailored to the properties of the molecules under study. This direction could open new avenues in super-resolution imaging based on all degrees of freedom of light.

## 9. Multimode shaping with classical and quantum light

Sylvain Gigan, Nicolas Treps, Sorbonne Université, CNRS, ENS-Université PSL, Collège de France

### Status

Optics is by essence multimode, carrying complex information associated to its many degrees of freedom. However, in many applications, spatial information is not simply encoded in the local pixel of the image, but delocalised, would it be as a result of atmospheric turbulences in astronomical applications, or because encoding in higher-order spatial modes is efficient for information transfer, at the classical or quantum level. Hence, a major challenge in the field is to develop lossless complex spatial manipulation able to deconvolute the information, i.e. implementing unitary transformations[1].

A first approach consists in spatial wavefront shaping, that can routinely be done, thanks to spatial light modulators (SLM) or deformable mirrors technology. Through a certain number of actuators or pixels (from a few tens to a few millions) one can locally change the phase of a beam and provide near-arbitrary wavefront, at speed ranging from a few tens of Hz to several tens of kHz. This has found numerous applications, for instance in astronomy, in microscopy for structured illumination, in optical trapping for trapping and guiding multiple particles at once, in quantum optics and communications to exploit higher-order spatial modes to encode information.

However, a SLM performs a single plane transformation, and even if it can consist of many pixels it cannot access the full breadth of light unitaries. In adaptive optics, correcting the atmospheric turbulence (that is intrinsically volumetric) with a single deformable mirror results in a correction over a small field of view only [2]. In optical communications, demixing multiple spatial modes that have been coupled by the propagation through a complex medium such as the atmosphere or a long fiber also requires  a transformation that goes beyond what a single plane SLM can provide. This requires volumetric phase transformation.

### Current and Future Challenges

Propagation of a field, be in free space, or through a structured medium, provides a convenient means to couple different spatial positions together. Free space propagation to a finite distance is equivalent to a spatial convolution, while free space propagation to the far field or to the focal plane of a lens is equivalent to a 2D Fourier Transform.  Using this principle, it has been known for a long time that it is possible to use an SLM to generate arbitrary patterns in amplitude and phase at a distance, even in 3D, albeit at a cost in efficiency and accuracy. This is the basis of digital holography.

Here, we want to discuss several options to exploit existing controlled phase plane technology to generate complex modal transformations, adapted  to modern applications. Combined to controllable single or multi-plane phase transformations, through SLM for instance, one can show that all unitary transformations are, at least conceptually, accessibles.

This is particularly important for quantum optics and quantum information, where complex mode transformation needs to be performed with as little loss as possible, and for application of optical computing, where ad'hoc matrix multiplications need to be implemented optically [3].

### Advances in Science and Technology to Meet Challenges

While unitary transformation can be readily done on-chip using waveguide technologies (albeit on a limited number of modes, few tens at most), such multimode shaping in free space optics remains difficult. Two successful approaches have been implemented and even made it to commercial products. A first one aims at exactly implementing a predefined spatial mode transformation, or basis change. Alternating free space propagation with taylored phase plates allows this arbitrary mode converter. The idea has been applied in fiber telecommunications, to increase the bandwidth capacity of multimode fibers [4], but is also very promising for quantum metrology in order to dramatically improve multi-parameter estimation in astronomical or biological imaging for instance [5, 6].

A second one consists in using a single SLM together with a complex medium, for instance a multiple scattering medium or a multimode fiber. While the SLM corresponds to a (local) vector multiplication on the field, the complex medium can be shown to behave as a dense random matrix multiplication [7]. In combination with each other, it is possible to emulate arbitrary small scale linear optical circuits that have been used for instance to emulate a full table-top linear quantum optics experiment (see Fig. 1 [8] where a coherent absorption experiment is emulated).

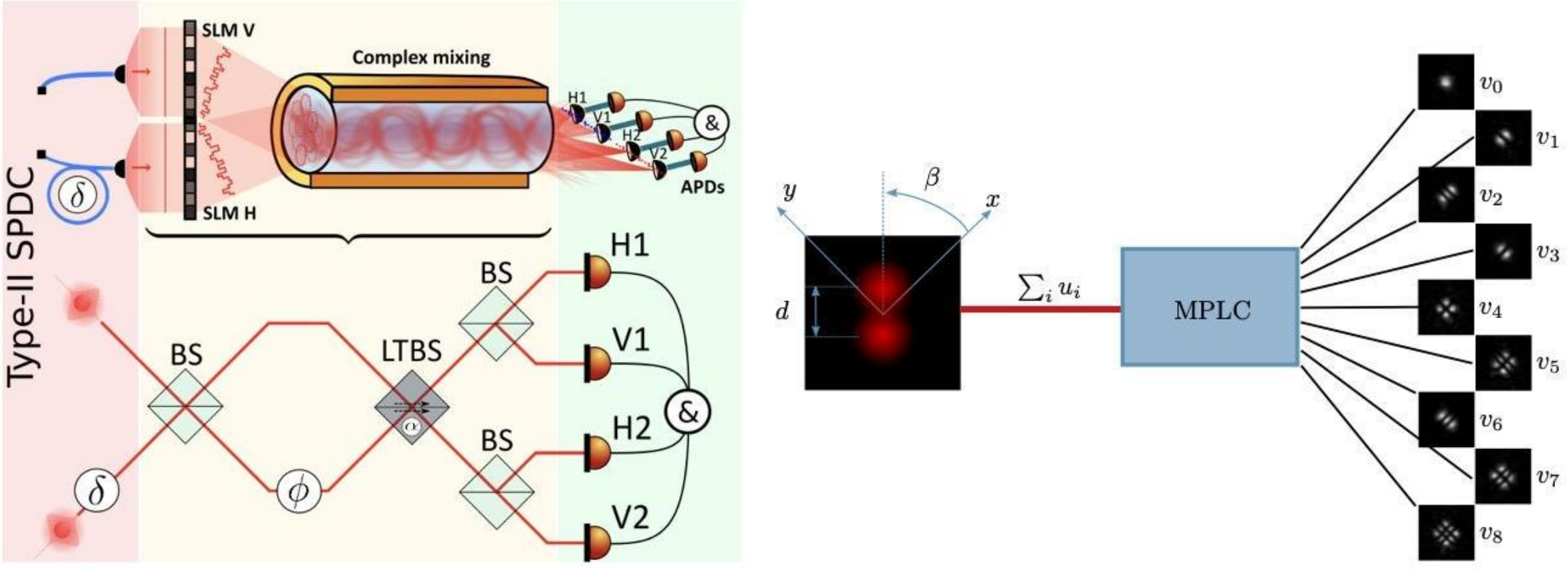

Fig 1: Left. Programming a linear quantum circuit using a complex medium. top: schematic of the experiment. Bottom: example of an emulated linear circuit. (adapted from [8]). Right: Estimating the separation between two incoherent sources, using a multiplane light converter [4] to project the incoming light onto an adapted mode basis (from [5]).

## Concluding Remarks

Going beyond single plane shaping is of paramount importance. Several methods have been introduced, combining planar SLM and propagation. A key parameter is that these approaches are very versatile and near unitary. Beyond spatial shaping, other degrees of freedom can be addressed with these methods, such as spectral ones [9], while tailored mode control in the spectral domain requires non-linear effects [10]. All these open a wide range of applications from quantum information processing to more efficient telecommunications, machine learning, Imaging in complex media, etc.

## Acknowledgements

*S.G. acknowledges support from H2020 European Research Council (SMARTIES-724473).*

## 10. Spatial shaping of flying qudits

Markus Hiekkamäki, Robert Fickler, Tampere University, Finland

### Status

Shaping the spatial structure of photons has also become an important research task in quantum information science. While many quantum information schemes rely on discrete two-level quantum states, i.e. qubits, it is known that encoding quantum information in high-dimensional states, so-called qudits, not only enables novel fundamental research studies but is also beneficial for technological applications. For example, qudits provide an enlarged information capacity, increased noise-tolerance, and simplifications to quantum information processing schemes. To encode qudit information onto a photon, the spectral and temporal (See Sec. 11. and [1], respectively) as well as spatial degrees of freedom (DOF) can be utilized. To use these DOFs for storing discrete quantum information, the DOFs themselves need to be discretized into orthogonal modes. For the spatial DOF, one popular discretization are transverse-spatial modes of light. These transverse-spatial modes allow the largest information capacity within the optical system's limitation and are propagation invariant, thus enabling transmission of quantum information over long distances, hence being dubbed "flying qudits" [2].

Over the last two decades, novel techniques to generate, measure, and manipulate spatially encoded qudits have enabled the demonstration of various benefits, including the ones mentioned above, and pushed the field of high-dimensional quantum information forward. Amongst many other developments, the advent of commercially available spatial light modulators (SLM), in combination with advanced holographic techniques, has enabled a vast improvement in the generation, i.e., the encoding of the high-dimensional quantum state, as well as the detection of the modes, i.e., the read-out of the qudit states [3] (see Fig.1(a)). The development of modulation devices that combine the polarization degree of freedom with the transverse beam profile, so-called q-plates [4] (See Fig. 2(c) and Sec. 5), have further enlarged the usage and fields of application of spatially encoded qudits.

Besides these single component solutions, novel arrangements for implementing complex unitary transformations for qudits have been devised using bulk optics [5] and multi-plane light conversion techniques (MPLC) (See also Sec.9). The latter of these was shown to be an especially powerful scheme for the advanced modulation and read-out of the quantum states, and demonstrated that qudits encoded in spatial modes nicely complement other high-dimensional DOF of photons.

### Current and Future Challenges

Due to the importance of modulation techniques, not only for information processing tasks but also encoding and read-out schemes, improving them might be the most important challenge on the road to larger scale applications of transverse-spatial modes.

For example, the generation of spatial modes is often performed using a well-defined starting state, e.g., a Gaussian transverse profile with a certain polarization, which is modulated by a flexible device, e.g., via a computer-generated hologram, that performs the required complex amplitude modulation. In reverse, similar schemes are used to read-out the encoded qudit in a versatile manner (see Fig.1(a)). However, the performance of these single plane approaches comes at the cost of losses, which is a limiting factor especially for multi-partite, multi-dimensional studies.

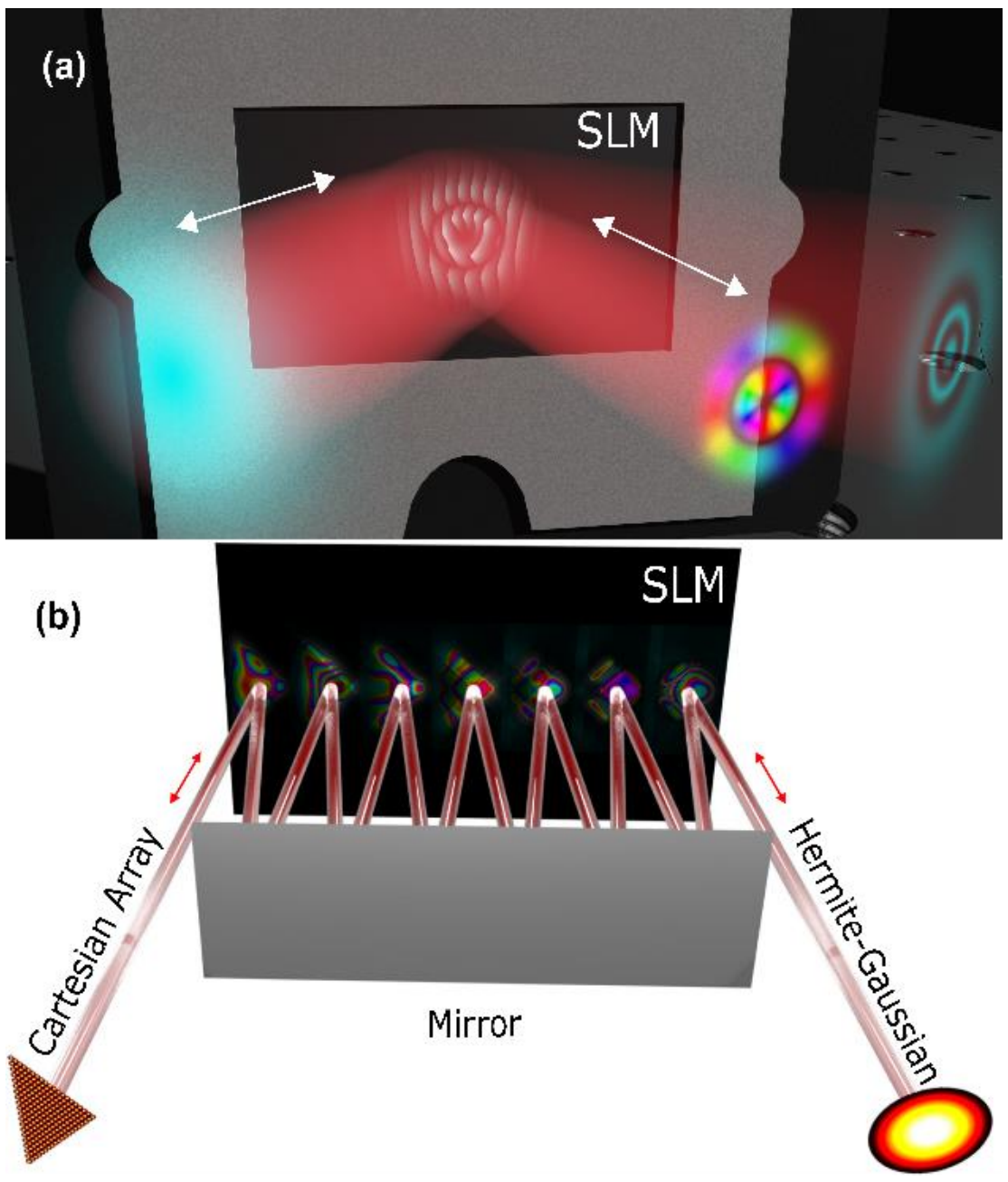

**Fig. 1** Different generation and measurement techniques. a) shows a visualization of a so-called mode carving technique where the wanted mode is carved from a large Gaussian beam with a lossy amplitude and phase modulation technique [3]. In b), MPLC has been used to create a mode sorter directing above 210 different Hermite-Gaussian spatial modes into separate gaussian spots in an array, image adapted from [7]. Both concepts can be used in generation and measurement of transverse-spatial modes.

Another approach to generating, measuring, and also manipulating spatially-encoded qudits, is using high-efficiency bulk-optics setups (e.g. see Fig.2(b)) [5]. They usually rely on the spatial symmetries or other mode-specific properties, such that complex interferometric arrangements are required, which results in challenging setups when scaling to high dimensionalities.

A significantly different but versatile modulation technique is MPLC [6]. It uses multiple consecutive phase modulations to perform any unitary modulation on a given set of input and output modes. The modulation of the phase front of the photons, together with some free-space propagation after each plane, allows a controllable shaping of the complex amplitude (see Fig.2(a)). As this approach operates along a single beamline, no interferometric stability is required. Additionally, only a limited number of phase modulations are enough to realize any unitary transformation for a given dimensionality [6]. Using programmable SLMs to perform these modulations enables a general, flexible transformation for which only the working principles of the used device limit the spatial resolution, modulation efficiency, or switching speed. In the quantum domain, the scheme was used to perform enhanced measurements [8], implement high-dimensional quantum gates [9], and demonstrate high-dimensional two-photon interferences [10]. Multiplexing and de-multiplexing of more than 200 modes was also demonstrated using classical light with only seven phase-modulation planes (shown in Fig.1(b)) [7].

In general, for all the outlined approaches, it will be important to increase the efficiency of the schemes while keeping their flexibility and, moreover, develop means for scaling them up to large state spaces.

## Advances in Science and Technology to Meet Challenges

The bottleneck of nearly all advanced techniques for spatial shaping of flying qudits is the limited performance of the modulation devices. Hence, a strong research focus will be required to improve them in terms of efficiency, resolution, and modulation speed. Although different approaches already achieve best values in one or two of these specifications, a single device operating in a near perfect manner in all aspects would allow the scaling of quantum technological applications to large photon numbers and dimensionalities, while enabling fast switching between operations.

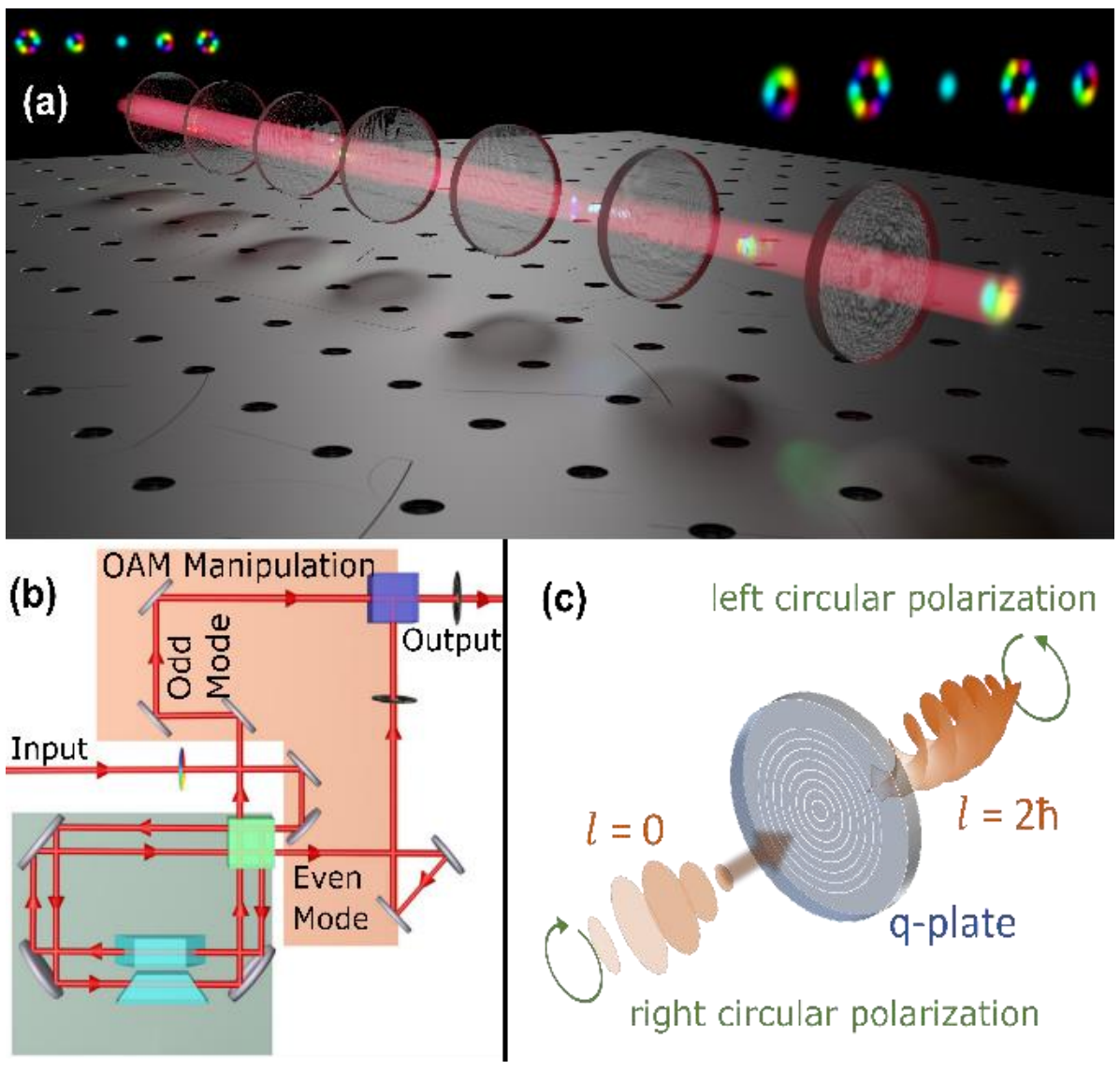

**Fig. 2** Different modulation techniques and technologies. In (a), an MPLC system implementing a unitary operation on a five-dimensional spatial mode qudit, through seven lossless phase modulations, is visualized. In (b), an implementation of a cyclic X-gate using bulk optical elements, for spatial mode qudits, is shown. The system utilizes interferometric schemes to perform the transformations, image adapted from [5]. (c) is a visualization of a q-plate that couples the polarization and orbital angular momentum degrees of freedom of a photon to transform its spatial structure [4].

Another research direction for bringing such schemes closer to real-world applications, is to develop technologies for implementing the operations in an integrated manner. Here, multiple routes seem promising, e.g., adapting bulk-optics components into waveguide-compatible solutions, interfacing bulk-optics schemes with already established integrated circuits, or developing novel integrated ways to perform spatial mode manipulations in a small footprint. Furthermore, to fully harness the benefits of high-dimensional quantum states, the integrated solutions should also be flexibly adjustable and of high efficiency.

Another important path to take, is developing ways to include other DOFs such that the information density of a single photon can be maximized within the limitations of the system. Here, a recent study based on a complex MPLC system invoking all degrees of freedom of a classical laser beam, i.e., the frequency, polarization and space, can be seen as one possible solution that could be adapted to the quantum domain [11].

Finally, along with the important progress of the required hardware, in one of the outlined schemes, i.e. MPLC, there is another challenge to meet. Current realizations rely on wavefront matching ideas, which is an efficient but non-analytic optimization algorithm to determine the required phase modulations. However, it will be interesting to study analytic ways of finding the correct phase modulations. This would not only improve the scheme by finding the most efficient realization reliably, but it might also lead to a deeper understanding of the performed modulation and, therefore, could inspire entirely new modulation schemes.

## Concluding Remarks

A lot of progress in using spatially encoded qudits has been made over the last decades. Starting with first proof-of-principle demonstrations of experimentally-realized high-dimensional quantum states and protocols, the field of structured photons has matured to a level where many implementations outperform conventional qubit-schemes in the laboratory. While these demonstrations can be seen as great successes and hold promises for future developments, they also ask for a continuation of the efforts in order to keep pushing the field of structured photons forward. Together with utilizing other DOFs to encode qudit states [1,4] (See also Sec. 5 and 11), structured photons can be seen as an important player in showcasing and harnessing the benefits of complex photonic quantum states. Building on this progress, high-dimensional quantum information science is on the verge of taking the next steps towards demonstrating its superiority in various applications for next generation quantum technologies and conceivable high-dimensional quantum networks.

**Acknowledgements**

MH and RF acknowledge the support of the Academy of Finland through the Competitive Funding to Strengthen University Research Profiles (decision 301820) and the Photonics Research and Innovation Flagship (PREIN - decision 320165). MH also acknowledges support from the Magnus Ehrnrooth foundation through its graduate student scholarship. RF also acknowledges support from Academy of Finland through the Academy Research Fellowship (decision 332399).

## 11. Integrated quantum frequency combs

Michael Kues, Leibniz University Hannover, Germany
David Moss, Swinburne University of Technology, Australia
Roberto Morandotti, Institut National de la Recherche Scientifique, Canada

### Status

The realization of large-scale (i.e. multi-particle and multi-mode) quantum systems and their precise control in the context of a practical and high-performing technology represent a key challenge to ultimately move quantum applications from proof-of-concept demonstrations to secure communications, metrology and computation applications. In the photonics context, standard approaches to encode quantum information, i.e. bi-dimensional qubits or high dimensional qudits, rely on the use of spatial and polarization modes or the orbital angular momentum of a light beam, respectively. Those modes, however, suffer from fundamental implementation challenges, which pose limitations to achievable complexity increase in realistic systems.

A different and promising approach is the exploitation of frequency mode encoding, enabled by the realization of ‘quantum frequency combs’ (QFCs). Their multimodal characteristics — having many phase-stable frequency modes residing in a single spatial mode, immune to environmental perturbations — allow these frequency structures to store a large amount of discrete variable quantum information.  This is made possible by the joint biphoton complex spectral amplitude and phase, as well as the ability of using temporal and frequency correlations. Such states can be directly processed and manipulated with standard state-of-the-art and/or custom designed fiber-based frequency filter and modulation approaches. Non-classical frequency combs and their processing quickly become a rapidly growing field of research. In particular, recent progress can potentially lead to a mature technology for realizing compact and distinct controllable complex quantum systems.

Entangled quantum frequency combs are commonly created from spontaneous parametric processes (spontaneous parametric down conversion (SPDC) or spontaneous four wave-mixing (SFWM)), which mediates the annihilation of one or two photons from an excitation field and simultaneously generates two daughter photons [1]. Exploiting this process, either directly within a resonator (free-space or integrated) [2] or in a waveguide (and subsequently applying periodic spectral filtering), creates discrete frequency modes (see Figure 1).

Advances in integrated optics recently led to the realization of microring resonators with free spectral ranges accessible via common frequency filters, also featuring mass-producibility, stability, and low-cost. In turn, those devices enabled the realization of frequency multiplexed heralded photon states on-chip [3]. Together with manipulation elements in the frequency-domain, such as electro-optic modulators for frequency superpositions and programmable optical filters for phase amplitude control, these quantum frequency comb systems are capable of creating and transforming multi-photon (currently 4 photons) [4], high-dimensional (currently up to 4 dimensions) [5] as well as hyper-entangled [6] states. In this context, quantum transformations such as the Hadamard gate [7] and CNOT gate [8] have been recently implemented.

In addition to a large scalability, achieved by extending the number of frequency modes, the frequency degree of freedom is naturally stable and inherently compatible with today’s telecommunication technology, hence particularly suited for the processing as well as the transfer of large complex quantum states, e.g. in the context of the future quantum internet. Moreover, high-dimensional quantum state encoding in the frequency domain is robust towards noise, enables beneficial error correction schemes and can thus enhance the performance of quantum systems.

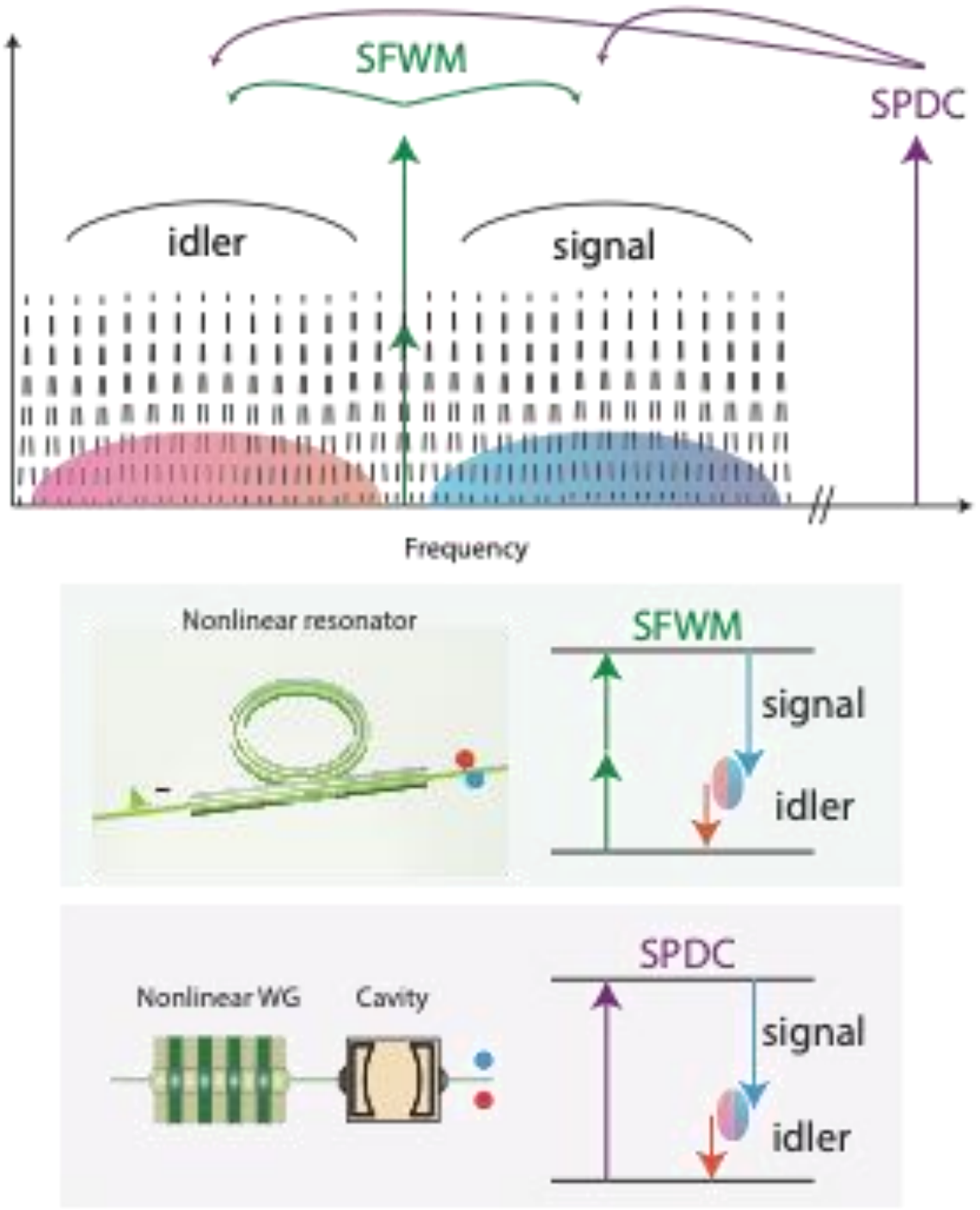

**Figure 1.** Quantum frequency comb generation in the discrete variable picture. Two time-frequency correlated photons are created through SPDC (spontaneous parametric down conversion) or SFWM (spontaneous four-wave mixing) from filtered nonlinear waveguides or nonlinear resonators.

## Current and Future Challenges

In order to reach a practical and high-performing technology for frequency domain quantum processing, key challenges of technological and conceptual nature arise. These involve a common development of comb source and frequency manipulation technology.

*1. QFC scalability in photon number and dimensionality*

While todays integrated quantum comb systems are limited to relatively few frequency modes (ca. 40) a larger number of frequency components can be scalably accessed by increasing the phase matching bandwidth of the spontaneous nonlinear parametric processes (tens of nm) while decreasing the free spectral range of the used cavities (<25 GHz). For a large phase matching bandwidth, cavity dispersion needs to be controlled as it leads to non-equidistantly separated frequency modes, which can cause problems to broadband frequency mixing elements, limiting their spectral operational range. Moreover, scalability in the photon number is required to increase the computational Hilbert space, which necessitates the use of non-deterministic multi-spontaneous nonlinear processes featuring high efficiencies.

*2. Highly performing frequency mixing as well as spectral phase and amplitude control components*

In frequency domain processing, the bandwidth required of the electro-optic modulators is crucial, where commercial phase modulators have reached frequencies of more than 50 GHz. Although using higher-order sidebands can extend this bandwidth, such schemes suffer from low efficiency and versatility, thus severely limiting the circuit depth (the number of operations which can be concatenated sequentially) as well as qubit (or qudit) connectivity (the maximum separation of frequency bins that can be efficiently mixed by a single modulator). Another approach to realize frequency mixing, which has not been explored yet in the QFC context, can be the use of nonlinear optical processes reaching operational bandwidths of more than 1 THz. Moreover, todays commercially available spectral phase and amplitude manipulation elements have a limited spectral resolution of approx. 15 GHz, which needs to be minimized and adapted to smaller FSR sources for improved performances. In general, the optimization of all manipulation components in terms of losses is critical for a performing processing platform in real-world scenarios.

*3. Interface to other quantum systems*

Photonic systems are important for control, transport and communication in other quantum platforms. To exploit the QFC benefits, it is central to develop interfaces between frequency-encoded complex quantum states and other quantum systems such as superconducting devices. This may be enabled though integrated optical device schemes. In this context, frequency to time encoding, or path encoding transformers, are rapidly becoming necessary.

**Advances in Science and Technology to Meet Challenges**

To realize a performing QFC platform, new device concepts and schemes are important. This can be achieved by following, adapting as well as pushing advances in integrated photonics technology and fundamental concepts in nonlinear optics.

*1. Highly efficient integrated QFC sources*

While integrated QFCs have been mainly realized from systems based on silicon and silicon nitride, other material platforms with higher nonlinearity and thus better photon generation rates need to be considered. Very promising platforms for highly efficient QFC sources are for example AlGaAs-OI or lithium niobate waveguide and resonator structures. Fabrication processes will need to be optimized to allow for low losses and small free-spectral ranges with large Q-factors.

To reach fully integrated QFC devices, the excitation laser needs to be implemented on the same photonic system. In this context, also on-chip high-rejection bandpass filters with more than 100dB rejection are required. Here, advances in integrated technology that allow the realization of hybrid approaches and heterogeneous implementation of active and low loss passive structures are increasingly pursued.

*2. Integrated low-loss amplitude and phase control of frequency modes*

In integrated optics, arrayed waveguide gratings allow the spectral splitting of an input field. Advances in their low-loss fabrication and resolution optimization as well as their connection to fast phase and amplitude modulators in the Fourier-plane are envisioned to make them applicable to frequency domain quantum processing applications. Here not only photonic device development is important but also the complex electronic control of phase and amplitude modulators urgently needs further advances.

*3. Broadband frequency mixing elements*

For frequency mixing via electro-optic sideband generation, integrated phase modulators with a low $V_{pi}$ as well as low insertion losses are required. Novel structures with strong chi2 coefficients and new modulator approaches tailored to quantum optical requirements need to be considered, with the aim of a fully integrated approach.

Frequency mixing could be also achieved though nonlinear parametric processes using a structured excitation field. However, advances in tailoring the phase matching conditions for such processes are still required. Also concepts to suppress spurious side-effect such as four-wave mixing residuals need to be first considered.

*4. Fundamental interface concepts for frequency states*

In addition to technological improvements, fundamental advances in coupling frequency-encoded photon states to quantum memories using conversion schemes and multi-level atomic systems are necessary. In this context, coupling via transducer techniques to e.g. superconducting qubit systems is essential to use the frequency domain approach towards building the future quantum internet.

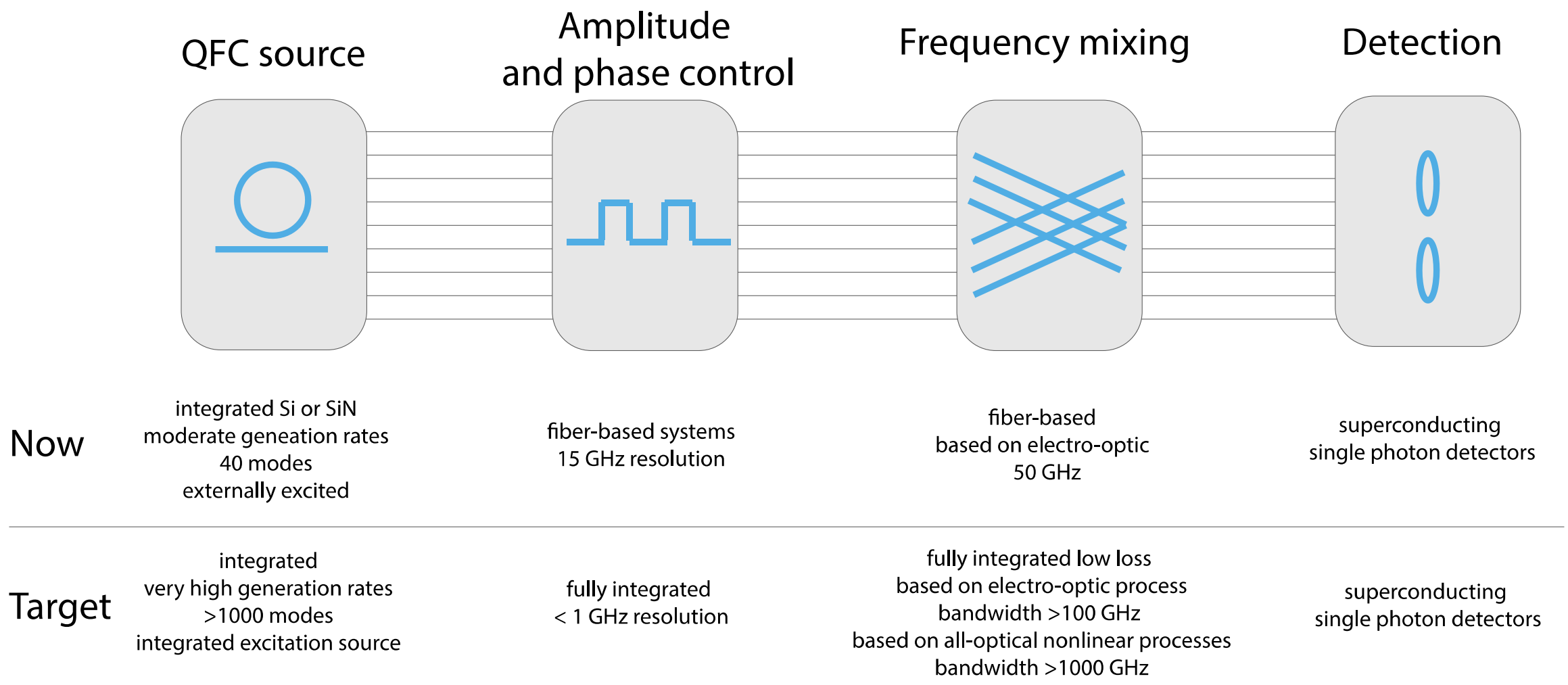

Figure 2. Current state of the art of frequency domain processing and future targets for meaningful applications.

## Concluding Remarks

Quantum frequency combs have achieved significant break-throughs in the generation and control of the highly complex photon states that will be critical for large-scale quantum information processing. Quantum photonic frequency-based platforms will naturally benefit from ongoing advances in classical telecommunications integrated circuits and components. As well, they will profit from access to custom designs with the potential for reduced loss through integrated photonics foundry services. It is clear that, whichever roles are ultimately assumed by photonic-based quantum information processing, whether it be for interlinks, communications, simulations or computing, those systems will tremendously benefit from the ability to generate and operate on large-scale multimode quantum states. In this context, frequency domain processing has indeed demonstrated an attractive and powerful approach towards achieving this goal.

## Acknowledgements

*M.K. acknowledges Funding from the German Ministry of Education and Research; D.M. and R.M. from the National Science and Engineering Research Council and the Canada Research Chair Program.*

## 12. Photonic integrated frequency combs

Johann Riemensberger, Tobias J. Kippenberg – Swiss Federal Institute of Technology (EPFL), Lausanne

### Status

The development of optical frequency combs [1], and notably self-referencing, has revolutionized precision measurements over the past decade, and enabled counting of the cycles of light. Frequency combs, for which John Hall and Theodor W. Hänsch shared the Physics Nobel Prize in 2005, has enabled dramatic advances in timekeeping, metrology and spectroscopy. More so, optical frequency combs are key enabling technologies for atomic molecular and optical Physics experiments, attosecond science, spectroscopy, astronomy and can synthesize the lowest noise microwave to date. They are versatile light sources with major application potential. Yet usage is limited by the low mode spacing, and the complex technical nature of such systems. Indeed, optical combs despite their vast scientific and technological application potential are mostly confined to national laboratories and far from employed widespread – even two decades after their invention.

A new revolution is under way, which could make frequency combs ubiquitous and unlock new application domains, with the discovery that frequency combs can be generated on chip using optical microresonators. First discovered in 2007 it has been shown that parametric frequency conversion in microresonators [2] enable to generate optical frequency combs from a continuous wave laser. Microresonator frequency combs have made it possible to synthesize optical combs in compact devices that can be microfabricated on silicon chips, with unprecedented high repetition rates in the technologically relevant gigahertz regime, and with optical bandwidths that are far beyond what can be achieved with conventional mode locked lasers.

The full potential of such microresonator frequency combs or 'microcombs' have been unlocked by the discovery of dissipative Kerr solitons (DKS) [2], which enabled to synthesize fully coherent, broadband frequency combs with repetition rates from 1 THz to as low as 2 GHz. Such soliton microcombs have moreover unveiled a very large landscape of novel nonlinear phenomena, ranging from soliton crystals, the observation of the Pasta-Ulam-Fermi recurrence relation, Raman solitons, soliton molecules, to emergent nonlinear dynamical phemomena. Equally microcombs enable entirely new regimes in which multiple soliton pulse streams that are mutually coherent are generated within one and the same cavity.

DKS result from the double balance between the dispersive pulse spreading and nonlinear compression due to the optical Kerr effect, and the parametric gain and optical loss in the microresonator to support ultrashort optical pulse generation. Microcombs have synthesized extremely short optical pulses, approaching the single cycle limit. Such microcombs have today been generated in an extremely large number of materials, that now encompass (see [3] for an overview) many integrated photonic platforms, including silicon nitride, lithium niobate, GaP, AlGaAs, SiC or silicon. The remarkable advances in developing ultra-low loss tighthly confining waveguides, in particular $Si_3N_4$, have enabled microcombs to be pumped with on chip III-V planar laser diodes. Crystalline whispering gallery mode resonators, which are fabricated by polishing highly transparent fluoride and oxide bulk crystals on a precision lathe are complementary as they offer optical quality factors in excess of $10^9$ for low power DKS generation [2] and excellent frequency stability for low noise microwave generation [4] and other precision metrology applications.

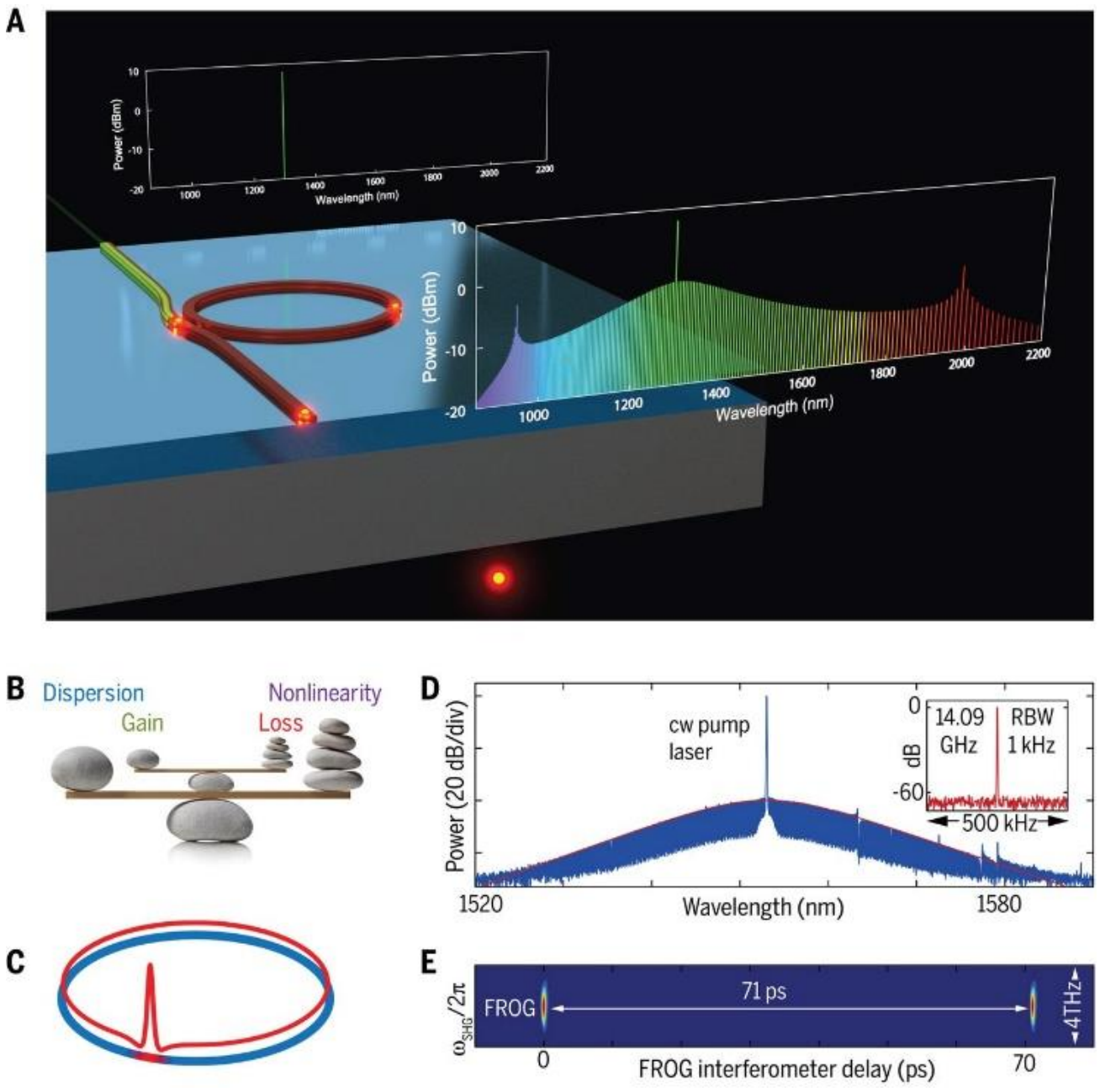

**Figure 1.** (A) Graphic image of dissipative soliton formation in a CW laser-driven crystalline WGM microresonator, enabling conversion of a CW laser to a train of DKS pulses. (B) Principle of DKSs, representing a double balance of dispersion and nonlinearity as well as (parametric) gain and cavity loss. (C) Field envelope of a temporal soliton. (D) Experimental optical spectrum of a DKS in a crystalline $MgF_2$ resonator. (Inset) Detected microwave beatnote of the soliton pulse train. (E) Frequency-resolved optical gating spectrum showing localized optical pulses, separated by the soliton-microcomb line spacing. (Images are adapted from [3]).

DKS are a special form of bright soliton formation in the microresonator that requires anomalous chromatic dispersion, either directly provided by the resonator material or obtained through dispersion engineering of the waveguide cross-section. To date, soliton microcombs have been applied to numerous fields, such as metrology and optical synthesis [5] and dual-comb spectroscopy [6]. Owing to their GHz to THz repetition rates, soliton microcombs have unlocked also entirely new applications of frequency combs that cover numerous applications in wavelength-division multiplexing for massively parallel optical communications [7] and coherent laser ranging [8], neuromorphic computing, astrophysical spectrometer calibration, or microwave photonics, because individual comb lines can be separated with low-cost diffractive optics .

**Current and Future Challenges**

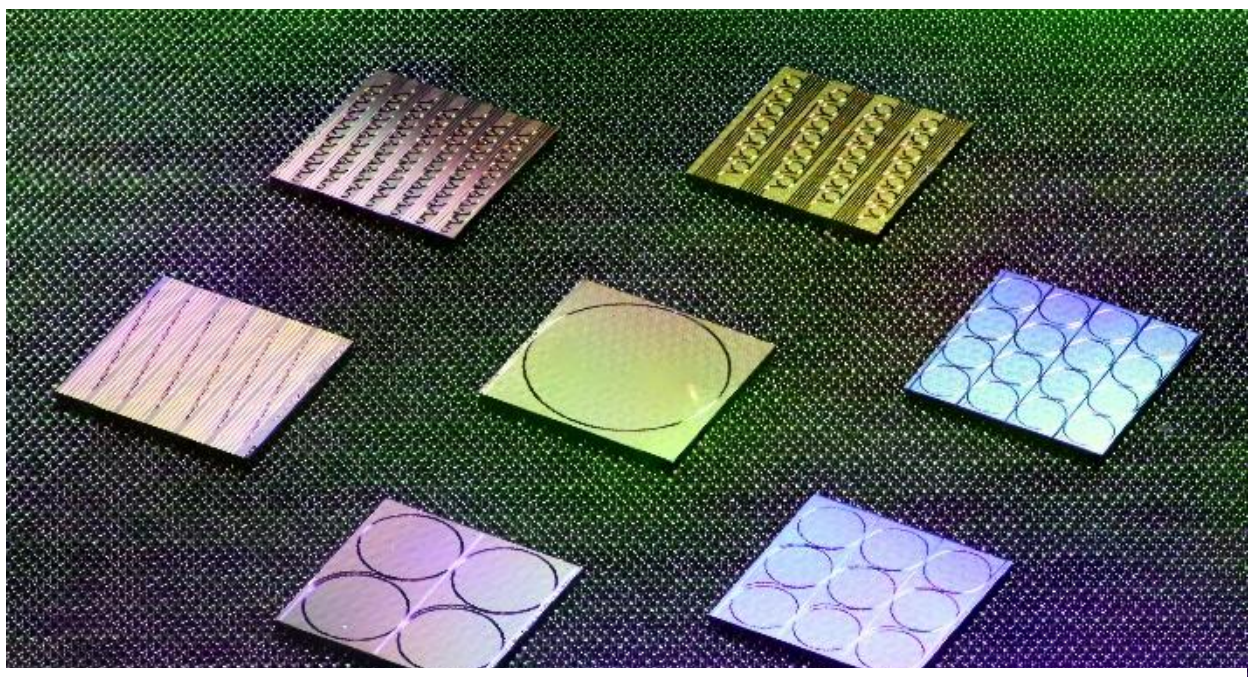

Figure 2: Photo of various silicon chips fabricated with the photonic Damascene reflow process for dissipative Kerr soliton generation (Photo credit: Jijun He, Junqiu Liu (EPFL))

The area of soliton microcombs still has vastly unexplored potential, that could enable to conquer new spectral domains, provide new ways to link the domain of microwave signals (used ubiquitously in electronic signals) to optical signals (as used in fiber communications), make the technology robust enough that it can be deployed at a large scale, and operate with power levels that are relevant for use in satellites or mobile devices. Our ability to engineer low loss nanophotonic circuits, from new materials, is presently still an ongoing frontier. Equally, optical losses in fibers – dB/km- are more than 4 orders of magnitude better than those in integrated photonics, and little is known where the losses stem from, and how they can be mitigated – which will require innovative approaches beyond traditional ones employed to date. Likewise, integrated nonlinear photonics can be leveraged with new materials, such as AlGaAs or GaP, which are promising but have not been widely explored. Innovative inverse design approaches, may herald unprecedentedly flat and complex dispersion profiles, enabling to synthesize light on a chip with virtually any color, any desired spectral envelope, in a fully coherent manner, at low powers – both pulsed and continuous wave. By bringing the precision of optical frequency combs together with integrated photonics, a second revolution in frequency metrology could follow – that could lead to widespread use of frequency combs in laboratories for frequency measurements, allow chip-scale optical atomic clocks to emerge, and lead to a new generation of sensors that measure mid infrared molecular spectra, and devices that may fuel data-centers of the future, offering vastly more bandwidth by offering hundreds of channels at once. Combined with advances in silicon photonics, that allow on chip generation and modulation of light, complex transceivers may come within reach that address the growing demand of energy in data centers that already today is consuming 3% of the world energy. In order to reach widespread technological adoption, the improvement of nonlinear conversion efficiency is a major challenge, especially in the case of broadband DKS generation, where the nonlinear conversion is limited by the small temporal overlap of the CW pump laser and the soliton in the cavity to a few per-cent.

Finally, nonlinear photonics and microcomb generation in integrated devices, may also lead to unforeseen discoveries, as nonlinear interaction between the delocalized optical modes in complex systems of coupled resonators may induce emergent phenomena beyond the description of the single resonator Lugatio-Lefever equation. Because solitons are ubiquitous in nature, optical systems can serve as well controlled testing grounds for the fundamental science of nonlinear systems and their dynamics. The exploration of the soliton dynamics in small coupled resonator systems such as dimers and plaquettes and the nonlinear photonic lattice are the obvious next steps and promise the discovery of rich new physics at the intersections with topology and non-hermiticity. Theory is leading the way here, because the requirement not only of exceptionally low loss, but also excellent uniformity and faithful design reproduction places additional stringent precondition on the fabrication processes.

**Advances in Science and Technology to Meet Challenges**
The key advance, which we foresee will continue to drive forwards the field of microresonator frequency comb generation are the advances in nanophotonics fabrication techniques that facilitate lower optical absorption and scattering losses and allow a higher degree of design fidelity and a larger design space. Firstly, the parametric oscillation threshold power scales with the optical loss rate as $1/\kappa^2$. Breakthrough advances in optical fabrication technology, mostly on the $Si_3N_4$ material system have allowed to reduce the optical power threshold for parametric oscillation and DKS generation into the sub-mW regime and allow direct pumping with low power laser diodes (see [9], for example). The soliton generation setup can also be vastly simplified by using self-injection locking the laser via the backscattering signal of the optical cavity [4], which also avoids the notoriously difficult integration of an optical isolator on the optical waveguide chip. The ultra-high optical quality factor of the microresonator provides narrowband feedback to the laser, reducing its linewidth by up to four-orders of magnitude and hence facilities the construction of a low noise laser and comb source, directly on chip. Active actuation and control can be directly integrated via monolithic integration of piezoelectric thin films[10]. Heterogeneous integration techniques between III-V gain materials and passive waveguides are already commercial in the silicon-on-insulator platform and are in particular appealing to further reduce the cost of microresonator frequency comb generators. The efficiency of microresonator comb generation can be increased either by dark soliton generation in the normal dispersion regime, by recycling the pump light either in a passive resonator or in the gain section of the pump laser/ amplifier. One particular attractive scheme here is the so-called laser-cavity soliton, where the pump laser is replaced by an amplifier in a closed loop configuration. Yet, it remains to be proved that such sources can reach the extremely low optical linewidths afforded by self-injection locked DKS and dark soliton.

**Concluding Remarks**
By bringing the precision of optical frequency combs together with integrated photonics, it may well be possible to establish frequency combs widely, across all spectral regions with integrated photonic technologies, and make them common in virtually all applications which requires multiple frequencies of coherent laser light. It could allow chipscale optical atomic clocks to emerge, lead to new sensors for LiDAR, lead to a new generation of sensors that measure mid infrared molecular spectra, and devices that may fuel data-centers of the future, offering vastly more bandwidth by offering hundreds of channels at once. Finally, nonlinear photonics and microcomb generation in integrated devices, may also lead to unforeseen discoveries, as nonlinear interaction in complex systems may lead to emergent phenomena.

**Acknowledgements**
*J.R. acknowledges support from the EUs H2020 research and innovation program under the Marie Sklodowska-Curie IF grant agreement number 846737 (CoSiLiS).*

## 13. Quantum cascade laser optical frequency combs for spectroscopy

Jérôme Faist, Giacomo Scalari, ETH Zürich, Switzerland

### Status

The initial push for the development of optical frequency combs and their greatest initial impact was in the context of time metrology. Since then, it was realized that such an optical spectrum had also extremely interesting applications in spectroscopy. Because their fundamental vibration modes occur in this frequency range, the mid-infrared (Mid-IR) portion of the spectrum (loosely defined as corresponding to wavelengths between 3 and 12 μm) enables the most sensitive and selective spectroscopy of light molecules. Down conversion of near-infrared frequency combs to the Mid-IR is feasible, but the direct generation of combs using quantum cascade lasers literally unlocked that frequency range for comb applications[1]. The most original and important way to use frequency combs is the so-called dual comb technique[2] where a sample and a local-oscillator combs are mixed onto a single detector (see Fig.1). In this way, very broadband spectroscopy can be performed with no moving parts, while the heterodyne nature of the detection enables a very high sensitivity to be achieved. The relatively large repetition rate (around 10 GHz) and the easy tuning of the offset frequency was found to be very favourable both for the signal-over-noise and acquisition speed of quantum cascade laser based dual-comb spectrometers[3, 4], allowing the achievement of high signal over noise spectra even for free running combs. The high acquisition speed of such spectrometers enabled to follow a photo-induced reaction in real time[5]. A very important feature of QCL-based frequency combs is their frequency modulated emission with only a small amplitude modulation, a behaviour which is very much a result of the subpicosecond recovery time of the gain medium. Actually, it was shown experimentally that the instantaneous frequency increases linearly with time during a period of the comb modulation[6] while the intensity is weakly modulated around a constant value.

Optical frequency comb generation has also been realized in the terahertz region with similar properties, but THz devices still require cryogenic cooling for their operation[7, 8]. One very attractive feature of THz QCLs is that they are able to generate octave-spanning spectra, opening the potential for self-referenced combs[8]. Also, the extensive use of metallic cavities for THz devices makes their interface with RF components easier and more efficient.

### Current and Future Challenges

Today's quantum cascade laser comb devices are capable of producing hundreds of milliwatts of optical output power, which make them very appropriate for experiments with very low optical throughputs. Their offset and their repetition rate can be tuned using either injected current and temperature. In addition, the repetition rate can also be injected using an external electrical oscillator. The challenge faced by QCL combs remains the fact that it is difficult to consistently fabricate devices exhibiting a stable comb over their full dynamical range with high yield. The other main challenges for this relatively new optical source are mainly connected with optical bandwidth (presently limited to 50 and 100 $cm^{-1}$) and stability.

The comb generation in quantum cascade laser is attributed to the interplay between four wave mixing originating from population pulsations, enabled by the very fast recovery time of the gain, and spatial hole burning. Despite recent progress, the understanding and modelling of the device is still very limited, especially compared to more established comb sources such as fibre lasers and microresonator-based combs. One fascinating aspect of QCL combs is the fact that the electron transport inside the active region of quantum cascade lasers is occurring at a time scale which is much faster than the roundtrip frequency. At the practical level, it enables the detection of the beatnote by merely extracting the high frequency component generated by the device itself on the injected current. In turn, injection of a radio frequency component allows an additional control “knob” on the device that not only modifies the round trip frequency but also the comb state and its optical bandwidth. At the conceptual and modelling level, the addition of ultrafast transport in the Maxwell-Bloch formalism is rather unique to this kind of device and has only be tackled very recently and in a limited manner.

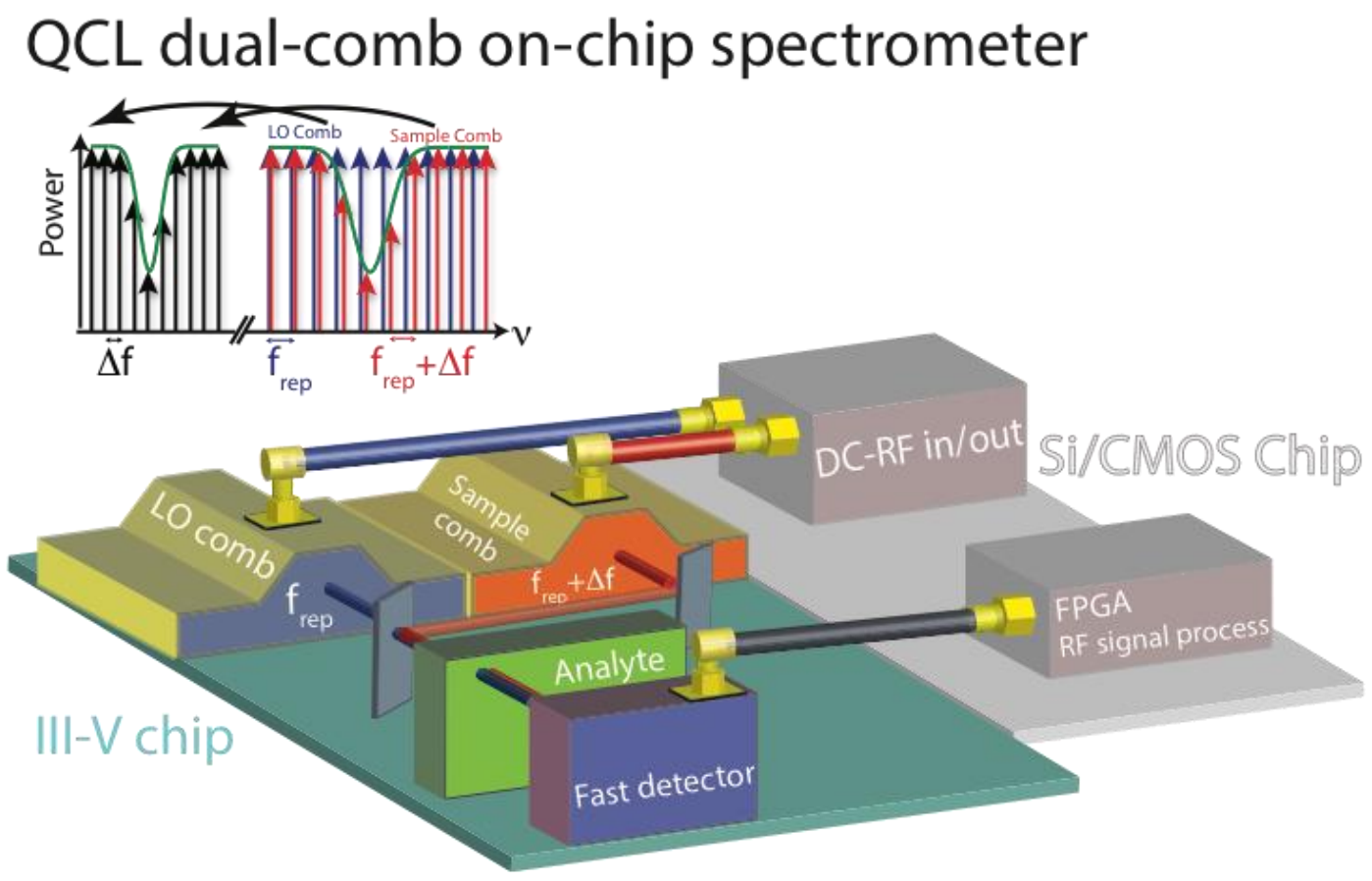

**Figure 1** Schematic of a QCL-based on-chip dual comb spectrometer. CMOS-based electronics controls the III-V chips carrying lasers and detector. The optical signals are produced by the two QCL combs located on the same chip and controlled via RF. Beams are combined through low loss optical elements and sent through the analyte. A fast detector downcoverts the mixed optical signals that are processed on the Silicon-based chip. electronics as well as the on-board signal processing.

**Advances in Science and Technology to Meet Challenges**

Quantum cascade laser frequency combs have a great potential for further developments in both their physics and applications. One aspect is the richness of the different active regions and geometries that can be utilized. So far, most of the active regions are single or multistacks derived from gain chips originally optimized for their use in external cavity tuneable laser and not for their non-linear properties. The recent realization of comb devices utilizing a ring cavity demonstrated that the phenomenology that has been observed in microcavity Kerr combs could also be found with a QCL active region[9, 10]. Very interestingly, these devices exhibited a much more stable beatnote than devices using Fabry-Perot cavities albeit with up to now a significantly narrower spectrum.

Progress in these devices will arise from the development of active regions and geometries that will enable broader operation spectra combined with excellent stability. Much of that development will also require further progress in the understanding and modelling of these devices. Indeed, while it is possible – at least conceptually–to create finite-element models that will take into account all the microscopic interactions and non-linear behaviour of the active region, combine it with the optical waveguiding and the fast electron transport in the cavity this model will very likely be numerically too heavy to be useful in practical systems. For this reason, a better understanding of the basic physics is needed to guide the design of the next generation of cavities and active regions.

On the technological side, exploiting the optical integration techniques originally developed for the near-infrared will enable the fabrication of complete dual comb receivers on a single chip, as schematized in Fig.1. This will require the development of very low-loss passive elements and waveguide. If successful, such integrated device could use a mixture of gain-generated and optical non-linear broadening to achieve extremely large optical bandwidths. In the Terahertz, the recent progress in the QCL operation on Peltier elements open the possibility of THz QCL combs operating without cryogenic cooling.

**Concluding Remarks**

The active control of mid-infrared and THz non-linearities using intersubband devices on an InP or GaAs platform will enable a whole spectrometer to be integrated onto a chip. By the easy control of the repetition rate and the offset frequency using temperature tuning and RF injection, these devices can be tuned continuously with very high accuracy and narrow linewidth, easily resolving absorption features of gases even at low pressures.

On the other hand, the relatively high repetition rate of these combs enables also the fabrication of dual comb spectrometers with extremely fast spectral acquisition times, presently of about 1us but potentially much faster. This, in turns, enables the tracking of complicated reactions like protein configuration changes or combustion in a single shot and in real time.

As a result, the research on new systems based on QCL combs and their application is still wide open and hold promises for a very strong growth in the coming future.

**Acknowledgements**

*The Authors acknowledge funding from European Research Council Consolidator Grant (724344) (CHIC).*

## 14. Laser frequency combs for broadband spectroscopy

Nathalie Picqué & Theodor W. Hänsch, Max-Planck Institute of Quantum Optics, Garching, Germany

### Status

The generation of arbitrary waveforms can be a powerful tool in spectroscopy, as illustrated by Fourier Transform Nuclear Magnetic Resonance spectroscopy, which exploits complex sequences of radio-frequency pulses. Optical spectroscopy still encounters significant limits on what waveforms can be generated and observed: because the involved frequencies are much higher, optical devices often generate and measure intensity waveforms, without control of the phase of the electric field.

In the late 1990's, the advent of laser frequency combs made it possible to measure and control the phase of an optical electric field with respect to the corresponding intensity-envelope waveform. Frequency combs are spectra of evenly spaced narrow phase-coherent lines. Often such combs are generated by mode-locked lasers. The pulse train emitted by a mode-locked laser can have a periodic envelope, for one pulse is emitted every cavity round trip. The electric field of the carrier wave of the pulses is not in phase with the pulse envelope, owing to the difference between group and phase velocities. In a frequency comb generator, it is shifted by a constant amount from pulse to pulse, that researchers have learnt to control (Fig.1a). Consequently, the spectrum of the electric field is composed of equidistant sharp lines of frequencies $\nu_n$ that obey the simple relation $\nu_n = n f_r + f_0$ (Fig.1b). Laser frequency combs have initially been invented as frequency rulers in precision spectroscopy and they have revolutionized precise measurements of frequency and time [1].

Very quickly, applications beyond the original purpose have been explored. Spectroscopy over broad spectral bandwidths where the comb directly interrogates the sample is one of the fields that strongly benefits from the new opportunities offered by frequency combs [2]. Frequency combs improve the performance of state-of-the-art spectrometers such as Fourier transform interferometers [3] or high-resolution dispersive spectrographs [4]. Interestingly, they have also enabled a new class of instruments, the dual-comb interferometers, which hold much promise for new approaches to linear and nonlinear spectroscopy and to coherent control. In a typical dual-comb spectrometer, a frequency comb interrogates an absorbing sample and is heterodyned against a second comb of slightly different line spacing at a single fast photodetector. The measured time–domain interference leads to a spectrum through the harmonic analysis provided by Fourier transformation.

### Current and Future Challenges

Dual-comb spectrometers rely on the precise control of the frequency combs. The beat notes between pairs of comb lines, one from each comb, enables one-to-one mapping of the discrete optical lines into the radio-frequency domain (Fig.1c). The comb structure of the spectrum is key to a meaningful spectral measurement. Furthermore, for a high resolution limited by the comb line spacing and for a high signal-to-noise ratio increasing with the square-root of the measurement time, the coherence between the two interfering combs must be maintained, a challenging requirement.

Dual-comb interferometers preserve all the characteristics of interferential spectrometers and especially their multiplexed recording: all the spectral elements are simultaneously measured on a single photodetector, so that, in principle, the approach can be applied in any spectral region with excellent overall consistency of the spectral measurements. Moreover they add distinguishing features: absence of moving parts and use of laser beams of high brightness provide short measurement times; self-referenced frequency combs make it possible to directly calibrate the frequency scale within the accuracy of an atomic clock, the interrogation of the sample by sharp laser lines allows to neglect the contribution of the instrumental line shape; and uniquely, the absence of geometric limitations theoretically leads to unlimited spans and resolutions [2].
Today, linear absorption dual-comb spectroscopy can successfully measure high-quality broad-spectral-bandwidth spectra in the near-infrared and in the mid-infrared regions, from 0.8 µm to 5 µm, at resolutions limited by the Doppler-broadened width of small molecules in the gas phase at room temperature. Reaching longer or shorter wavelengths is still hindered by substantial instrumental challenges. Moreover, in dual-comb spectroscopy, the resolution in a single measurement is fundamentally limited by the comb line spacing. For comb generators well suited for dual-comb spectroscopy often have a repetition frequency of a few hundreds of MHz, gas-phase spectroscopy of Doppler-broadened transitions has been explored most of the time. Finally, as frequency comb sources often involve intense ultra-short pulses, nonlinear transient phenomena can be generated in the observed medium, showing an exciting potential for nonlinear [5, 6] and multidimensional [7, 8] spectroscopy and imaging, which remains largely unexploited.

**Advances in Science and Technology to Meet Challenges**
The past decade has witnessed exciting progress in frequency-comb laser technology, in dual-comb interferometry and in demonstrations of a variety of sampling techniques and applications.
Still, the lack of appropriate lasers hinders the easy deployment of dual-comb spectroscopy over the entire electro-magnetic spectrum. In the mid-infrared range, self-referenced frequency combs are still complex to generate and one has to rely on nonlinear frequency generation. Despite its scientific interest for molecular science and for precision spectroscopy, the ultraviolet region has been poorly explored because of technical challenges: at wavelengths shorter than 190 nm, low-efficiency harmonic generation has to be performed in a rare gas rather than in a nonlinear crystal. Reaching sufficient power at repetition frequencies higher than 10 MHz is challenging.
Moreover, dual-comb and multi-comb spectroscopies potentially address a very wide range of topics and applications, which involve a large variety of parameters in the comb synthesizers. Extensions towards higher resolution, as required for spectroscopy of cold molecules [9] or for Doppler-free broadband spectroscopy [5] (Fig.2), would benefit from combs of low repetition frequency (1 MHz) whereas spectroscopy in the condensed phase, of relevance to chemistry and biology, requires a broad spectral span, a large repetition frequency (100 GHz) and a flat-top spectrum. Additionally, nonlinear spectroscopy, which promises new approaches to coherent control and precision spectroscopy for tests of the fundamental physical laws, needs peak powers sufficiently high to generate nonlinear phenomena and/or sensitive detection at the photon level [10].
In dual-comb interferometry, the requirements in relative phase and timing jitters scale inversely proportional to the wavelength, whereas, in the meantime, the power spectral density of the phase fluctuations in a harmonic frequency comb multiplies with the square of the harmonic, requiring clever strategies to approach the short-wavelength region. The situation becomes even more challenging if more than two comb generators are implemented [7]. Furthermore, the data processing may be overwhelming in multidimensional spectroscopy, in time-resolved spectroscopy or in imaging and calls for new signal-processing tools for analysing the observed chemical or physical processes.

On a longer term, advances in integrated optics promise on-chip dual-comb spectrometers for portable or even handheld comb-based spectral analysis, with potential for integration into lab-on-a-chip devices.

**Concluding Remarks**

Through the control of the phase of the optical carrier wave relative to the pulse envelope, frequency combs provide a powerful means of sampling optical spectra, with a discrete structure of sharp laser lines of accurate frequencies. Over the past decade, the field of frequency comb spectroscopy has blossomed thanks to tremendous progress in laser science and in measurement and sampling techniques: wideband precise spectra of the linear response of a sample have become measurable across the infrared and visible region. While part of the efforts is now shifting towards the new physics and chemistry that can be explored with the novel tools, even more exciting advances can be envisioned as the instrumentation reaches new frontiers in precision and control of nonlinear phenomena.

**Acknowledgements**

Funding by the Max-Planck Society and the Carl-Friedrich von Siemens Foundation is acknowledged.

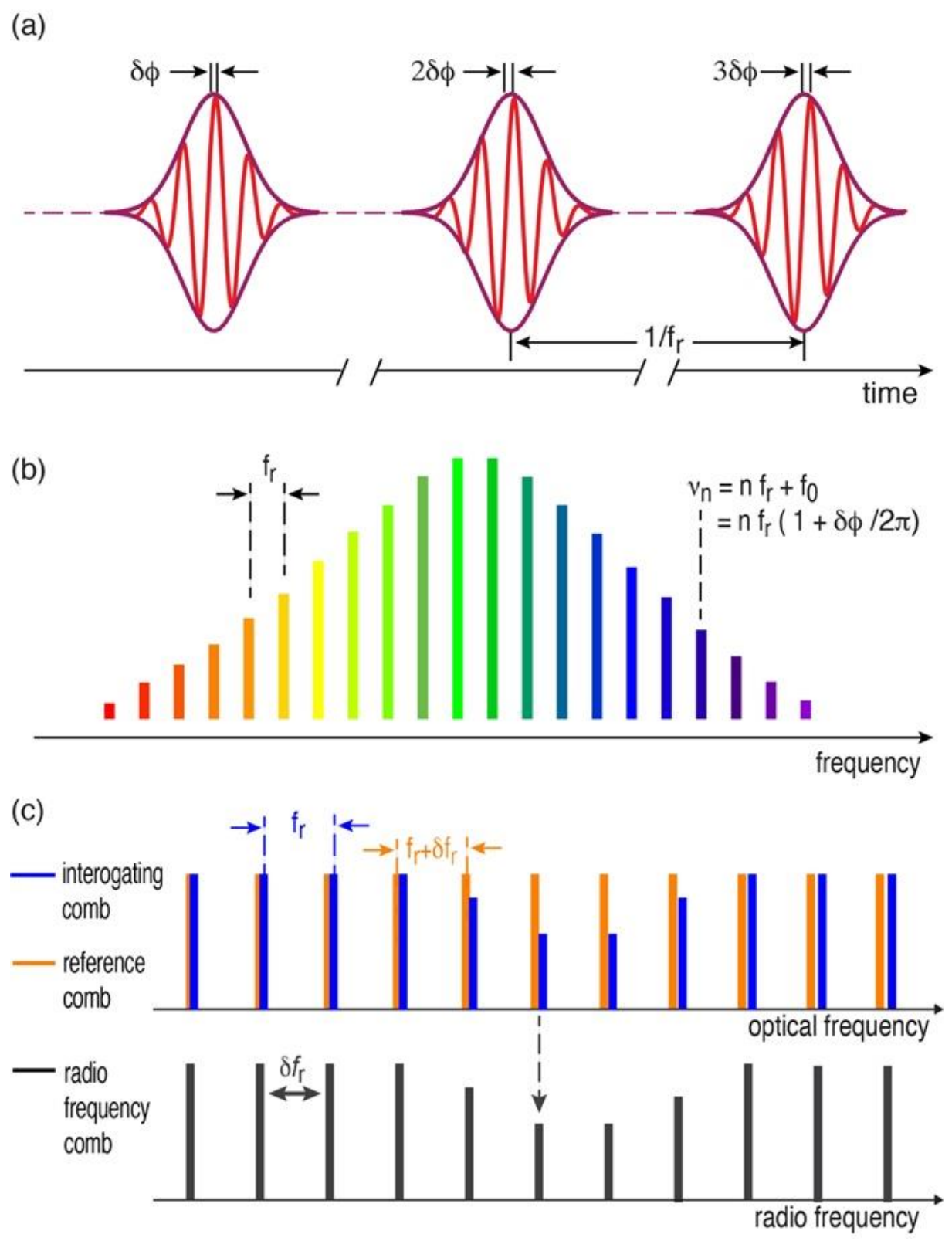

**Figure 1.** Frequency comb: (a) In the time domain, a regular train of pulses is emitted with a pulse-to-pulse phase-shift of the carrier to the envelope of the pulses. (b) In the frequency domain, the spectrum is a frequency comb, with discrete equidistant lines (c) Frequency-domain principle of dual-comb spectroscopy. Beat notes between pairs of optical comb lines, one from each comb, generate a radio-frequency comb that can be precisely measured using digital electronics.

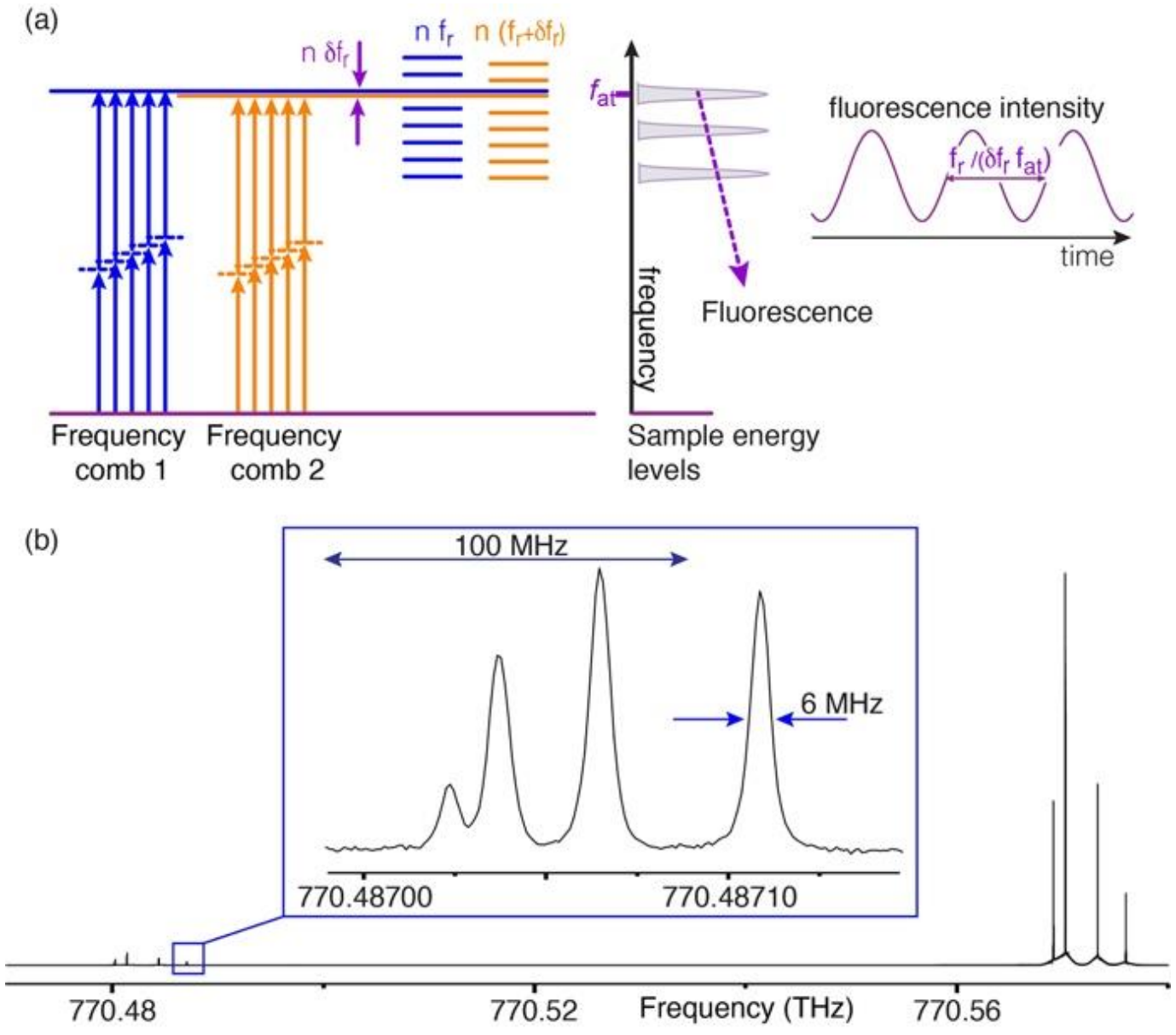

**Figure 2.** Non-linear dual-comb spectroscopy with the illustration of two-photon excitation. (a) Frequency domain principle: the atomic or molecular transitions are excited by two frequency combs of slightly different repetition frequencies. The fluorescence of the levels is intensity modulated at a rate that unambiguously identifies the optical frequency of the transition. (b) Experimental two-photon-excitation dual-comb spectrum at a resolution of 1 MHz (through spectral interleaving) of the sub-Doppler hyperfine the $5S_{1/2}$-$5D_{3/2}$ and the $5S_{1/2}$-$5D_{5/2}$ transitions in $^{85}$Rb and $^{87}$Rb [5].

## 15. Temporal shaping of ultrafast lasers

Giulio Cerullo and Cristian Manzoni, Politecnico di Milano, Italy

### Status

The invention of the laser and the discovery of mode-locking, coupled with amplification and nonlinear frequency conversion techniques, have enabled the generation of ultrashort coherent light pulses, with duration down to a few femtoseconds in the visible and wide frequency tunability from infrared to ultraviolet (UV). Such pulses, which have intrinsically broad bandwidths, are described by an electric field temporal profile E(t) whose Fourier transform can be written as $\widetilde{E}(\omega) = \widetilde{A}(\omega)e^{i\varphi(\omega)}$where $\widetilde{A}(\omega)$ is the spectral amplitude and $\varphi(\omega)$ is the spectral phase. For many applications the spectral phase is adjusted in such a way that all frequency components of the pulse arrive simultaneously, obtaining the shortest possible pulses, with so-called "transform-limited" duration. In other cases, however, one wishes to control the spectral amplitude and spectral phase in order to custom tailor the temporal intensity profile $I(t) \propto |E(t)|^2$of the pulse. Femtosecond pulse shaping finds numerous applications in fundamental science and technology, which range from coherent control of chemical reactions [1] to nonlinear spectroscopy and microscopy and high-speed fiber optics communications [2].

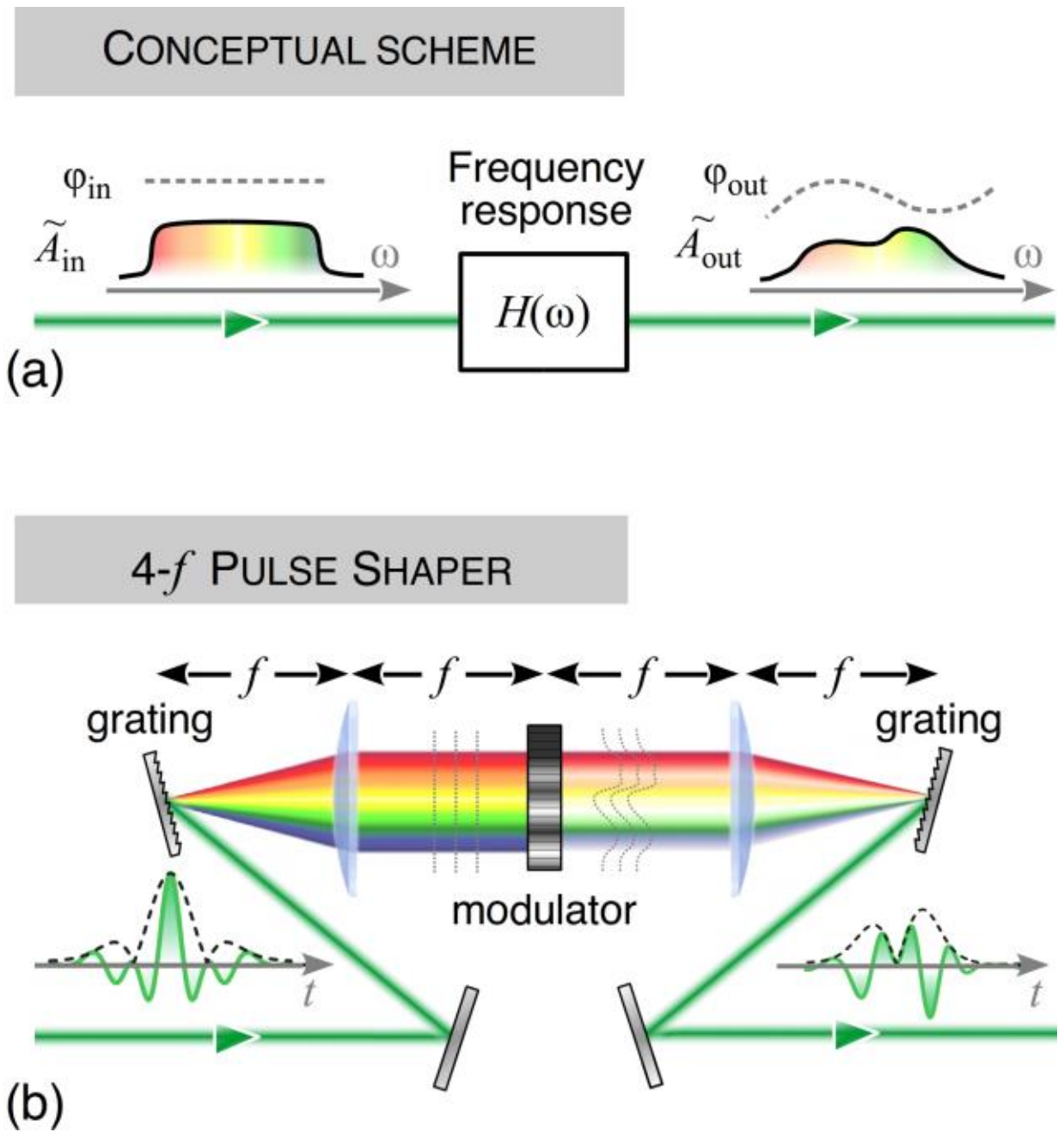

**Figure 1.** (a) Conceptual scheme of pulse shaping; (b) scheme of a frequency-domain pulse shaper based on the 4-*f* configuration. The modulator in the Fourier plane acts on the amplitude and the phase of each spectral component.

Typically, a pulse shaper works in the frequency domain, according to the conceptual scheme shown in Fig. 1(a), and manipulates the input pulse spectrum $\tilde{E}_{in}(\omega)$ with a linear, time-invariant spectral filter, resulting in $\tilde{E}_{out}(\omega) = H_f(\omega)\tilde{E}_{in}(\omega)$, where $H_f(\omega) = A_f(\omega)e^{i\varphi_f(\omega)}$ is the transfer function of the filter [3]. In the most general case, the filter controls both the spectral amplitude (through $A_f(\omega)$) and the spectral phase (through $\varphi_f(\omega)$) of the output pulse. Figure 1(b) shows a typical experimental setup for a femtosecond pulse shaper, based on a so-called 4-*f* configuration, also known as zero dispersion pulse compressor [3]. A first optical element (a grating or a prism) angularly disperses the different frequency components of the pulse, which are then focused by a first lens (or mirror to avoid chromatic aberrations) which performs a spatial Fourier transform of the beam. The individual frequencies of the pulse, which correspond to plane waves propagating at different angles, are then available in the focal plane of the lens –the Fourier plane- for amplitude and phase modulation. Finally, a second lens/mirror performs an inverse spatial Fourier transform and a second dispersive element cancels the angular dispersion of the beam, producing the temporally shaped output intensity profile. Among the modulators in the Fourier plane, the most commonly used are liquid crystal spatial light modulators (LC-SLMs) [3] and acousto-optic modulators (AOMs) [4]. A LC-SLM consists of an array of pixels, each of which is a nematic liquid crystal cell whose birefringence can be varied by an applied voltage, allowing control of phase and amplitude of the transmitted light, if the cell is inserted between polarizers. In an AOM an acoustic wave, travelling perpendicularly to the incident light beam, induces a refractive index grating which diffracts the beam, controlling its amplitude and phase. AOMs have two main advantages with respect to LC-SLMs: (i) they allow a spatially continuous, rather than pixelated, modulation; (ii) they have much higher refresh rates, limited by the transit time of the acoustic wave in the modulator, up to tens of kHz. On the other hand, due to the time-varying nature of their modulation, AOMs do not work with high repetition rate (MHz) ultrafast lasers. The efficiency of acousto-optical interactions decreases quadratically with increasing wavelength; for this reason most AOM operate in the visible and near-infrared frequency regions. In the infrared, the most commonly used AOM materials are germanium, silicon, and gallium arsenide, which have low acoustic absorption and high acousto-optic figure of merit, enabling operation up to 15 µm [5]. The operating wavelength region of LC-SLMs ranges from 260 to 1100 nm mainly limited by the transparency range of the indium tin oxide layer used as transparent electrode. Operation up to 5 µm was obtained by developing alternative materials, such as Ni and Cu ultra-thin metallic films, CuxO thin films, and chemical vapour deposition grown mono-layer graphene.
A different pulse shaping approach is the acousto-optic programmable dispersive filter (AOPDF) [6], in which the light wave and the acoustic wave propagate collinearly in a birefringent photoelastic crystal. Under phase matching conditions, the acoustic wave causes an additional stress-induced birefringence that couples light into the orthogonal polarization state. By controlling the acoustic wave in such a way that a given optical frequency is coupled at a specific longitudinal coordinate, one can generate almost arbitrary optical waveforms. Similar to AOMs, materials for AODPFs are available from the UV to the mid-infrared frequency range.

**Current and Future Challenges**

Standard pulse-shaping techniques control the temporal intensity profile of the ultrashort pulse, but do not generate optical waveforms with a reproducible electric field profile $E(t)$. To this end, one crucial ingredient is the control of the carrier-envelope phase (CEP) $\varphi$, i.e. the phase of the carrier wave with respect to the pulse envelope. CEP control extends to the optical domain the capability, customary at radio frequencies, to generate waveforms with reproducible and user-controlled electric (and magnetic) field profile, thus ushering the era of optical waveform electronics. CEP control becomes especially relevant for light pulses with duration approaching (or exceeding) the limit of a single cycle of oscillation of the carrier wave, since for such pulses the amplitude of the envelope varies significantly between one oscillation cycle and the next. The generation of precisely controlled light waveforms with durations of the order of one optical cycle [7] opens a completely new regime of light-matter interaction, especially for what concerns highly nonlinear optical effects, such as high harmonic generation and the production of isolated attosecond pulses.

CEP control is intimately linked to the generation of pulse trains whose spectrum is a series of equally spaced lines at fixed frequency positions, a so-called "frequency comb". Conventional pulse shapers have rather coarse frequency resolution, of the order of tens to hundreds of gigahertz, corresponding to maximum durations of the shaped pulses of the order of picoseconds to tens of picoseconds. Pushing the frequency resolution down to the GHz regime or below would allow to control in amplitude and phase the individual "teeth" of a frequency comb, producing shaped pulses with durations of hundreds of picoseconds or longer. This line-by-line shaping capability would enable to generate an optical waveform which "fills time", i.e. has a duration matching the period between consecutive pulses. If the shaping capability is dynamic, i.e. if the transfer function of the shaper can be varied from one pulse to the next, one talks of optical arbitrary waveform generation (OAWG) [8], which would extend to the optical domain those waveform shaping capabilities which are customary at radio frequencies. OAWG finds numerous technological applications, such as optical communications with synthesized terabit-per-second signals, high accuracy light detection and ranging with precisely frequency swept sources or synthetic aperture radars at optical frequencies.

**Advances in Science and Technology to Meet Challenges**

Both the previously mentioned challenges in pulse shaping rely on the measurement and control of the CEP, which allows to generate reproducible electric waveforms, rather than intensity waveforms which differ by the relative phase. The generation of CEP-controlled sub-cycle optical waveforms can be achieved by coherent pulse synthesis [9], i.e. the phase-coherent combination of broadband pulses with different frequencies produced by separate sources. The individual spectral channels can be obtained from mode-locked laser oscillators, by spectral broadening in a $\chi^{(3)}$ medium or by optical parametric amplification in a $\chi^{(2)}$ medium. Pulse synthesis requires control of the individual CEPs of the pulses and of their relative phases, both of which can then be adjusted allowing shaping of the optical waveform on the sub-cycle level (see Fig. 2).

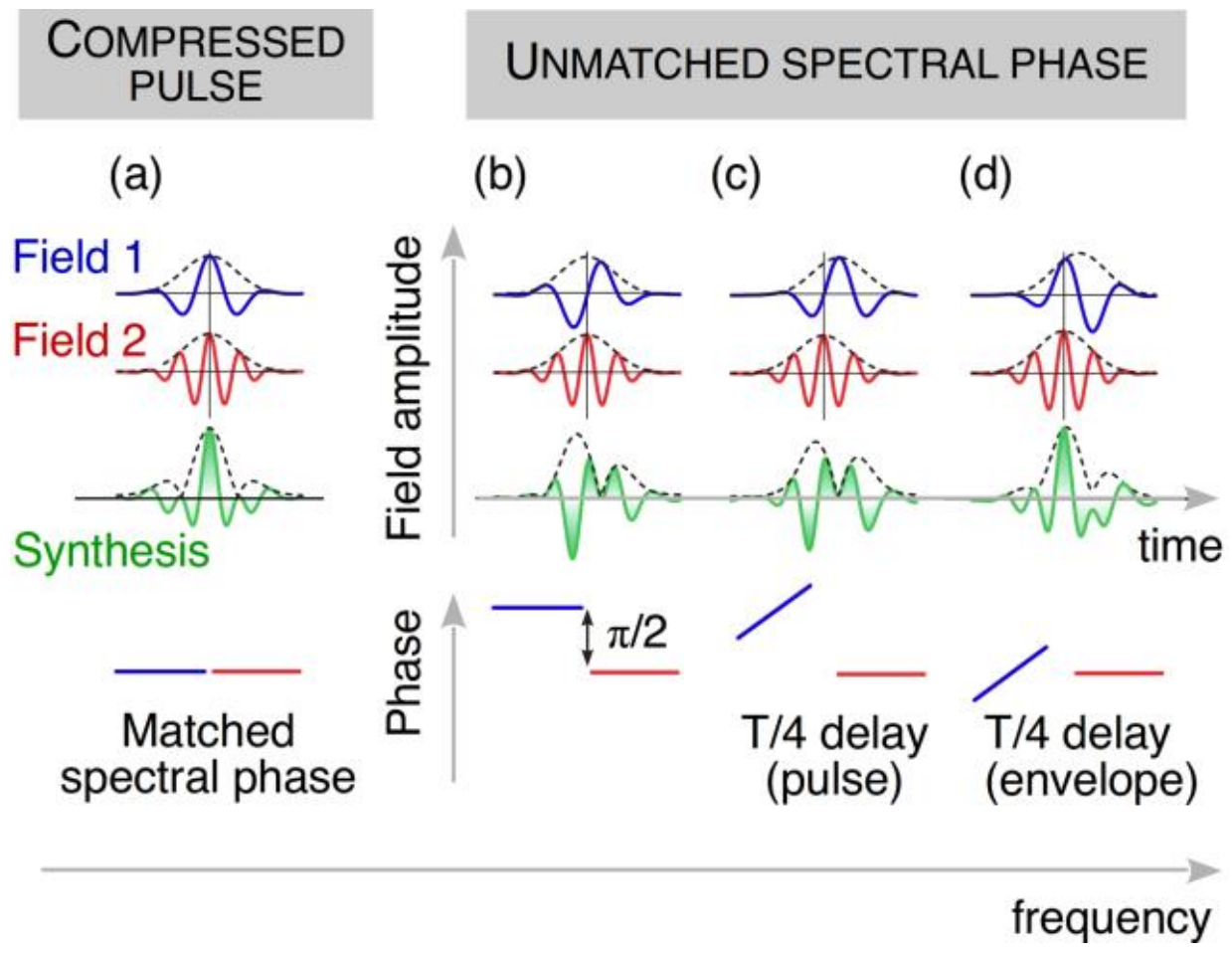

**Figure 2.** Role of the CEP for the generation of controlled electric field transients. Dashed line: pulse envelope; solid line: electric field. The first line shows the electric field transient of pulses with the same envelope, but different spectral phases and CEPs. The spectral phase plays a crucial role also for the synthesis of pulses from the combination of different light transients (a) Compressed pulse; both fields are cosine pulses, and the spectral phases are perfectly matched; (b) the CEP of field1 is shifted by a quarter cycle; (c) field1 is shifted by a quarter cycle; (d) only the envelope of field1 is shifted by a quarter cycle.

OAWG requires to dramatically increase the spectral resolution of pulse shapers. Grating-based approaches have spectral resolutions limited to ~5-10 GHz, for which a rather bulky optical setup is required. Higher resolutions in a more compact setup can be achieved using a virtual imaged phased array (VIPA), which is a tilted Fabry-Perot etalon with a side entrance. Coupled with a diffraction grating, a VIPA allows to disperse the individual lines of the comb in a plane. Alternatively, high resolution pulse shaping can be achieved using arrayed waveguide gratings with thermo-optic or electro-optic phase control. An intense research is currently devoted to increasing the refresh rate of line-by-line pulse shapers, so as to achieve true OAWG.

One final challenge is to transfer with high fidelity the shaped pulse from one frequency range to another using nonlinear optical effects. Spectral phase transfer has been successfully applied to generate shaped pulses in the UV by sum-frequency generation and in the infrared by difference frequency generation. For second-harmonic generation, a frequency domain approach has been proposed in which the nonlinear effect takes place in the Fourier plane of a pulse shaper, thus allowing phase transfer to the second harmonic and the ability to perform a temporal convolution. It would be extremely interesting to extend this capability to free-electron lasers (FELs), which generate coherent XUV and hard-X-ray radiation. For FELs seeded by UV pulses this capability has already been demonstrated, leading to the recent generation of a phase-locked XUV pulse pair whose relative delay is controllable with attosecond precision [10].

## Concluding Remarks

Shaping of ultrashort light pulses is a mature technology, which is used for different purposes: either to accurately control the spectral phase introduced by a sequence of optical components, to retrieve the transform-limit pulse duration, or to tailor the spectro-temporal properties of the pulse for applications to nonlinear light-matter interaction or spectroscopy. The frontiers of pulse shaping involve pushing this capability to the extremes: on the one hand, to control the electric field profile of the pulse on a sub-cycle timescale; on the other hand, to control each line of a frequency comb, thus allowing the shaped temporal waveform to extend over a timescale comparable to the pulse repetition period. A further important challenge is to extend pulse shaping capabilities to the short wavelengths generated by FELs.

# 16. Future of optical pattern formation

L.A. Lugiato, Università dell'Insubria, Italy
M. Brambilla, Università e Politecnico di Bari, Italy
L. Columbo, Politecnico di Torino, Italy
A. Gatti, Università dell'Insubria, IFN-CNR, Italy,
F. Prati, Università dell'Insubria, Italy

### Status

Shaping coherent light in time and space or, symmetrically, devising materials and devices to achieve it, is a prominent challenge in the future sketched within this Roadmap. Soliton formation in 2D and 3D in advanced coherent sources is one of the most promising avenues to achieve such goals.

The topic of spontaneous optical pattern formation has a very long history since it started in the sixties [1]. An idea of its historical evolution can be obtained from [2]. Recently the usefulness of the results in this topic emerges especially from the fields of frequency comb in Kerr microresonators (reviews can be found, e.g., in [3]) and of quantum cascade lasers [4].

Pattern formation occurs in nonlinear media included in optical cavities [1-5], in systems with a single feedback mirror [6] (see also Sec. 27.4 of [2](a)) and in cavityless configurations [7] (see also Fig. 27.5 of [2]a). In this paper we discuss only systems of the first class.

Optical patterns in time, space, or space-time, display an incredible variety of phenomena. Some of them arise from the nonlinearity under conditions of singlemode operation, while most of them arise from the multiform multimodal aspects of such systems. Some investigations pointed out fundamental relationships and analogies among this vast phenomenology. Examples are:

1) The general relationship between multimode and singlemode instabilities that exists in the case of ring cavities, both with and without population inversion (see [2](a), Chapter 23).
2) The general relationship between phenomena occurring in the transverse planes orthogonal to the direction of light propagation originated by diffraction and phenomena occurring in the longitudinal direction originated by dispersion, with the consequence that the equation formulated in 1987 in [5] (LLE) for diffraction applies equally well to the dispersive case (see [2](a), Chapter 28).
3) The "order parameter" approach, which has allowed to relate relevant optical systems, governed by sets of partial differential equations, to general equations in physics such as, for example, the complex Ginzburg-Landau equation or the Swift-Hohenberg equation.
4) The analogy between propagation of longitudinal structures, e.g. cavity solitons; in ring cavities as those of Ref. 3 driven by an external stationary field and the propagation of transverse structures (e.g. solitons) when the external driving field (holding beam) is doughnut-shaped. In the latter case the structures circulate in the ring corresponding to the crater of the doughnut mode (see [2](a), Fig. 30.6).

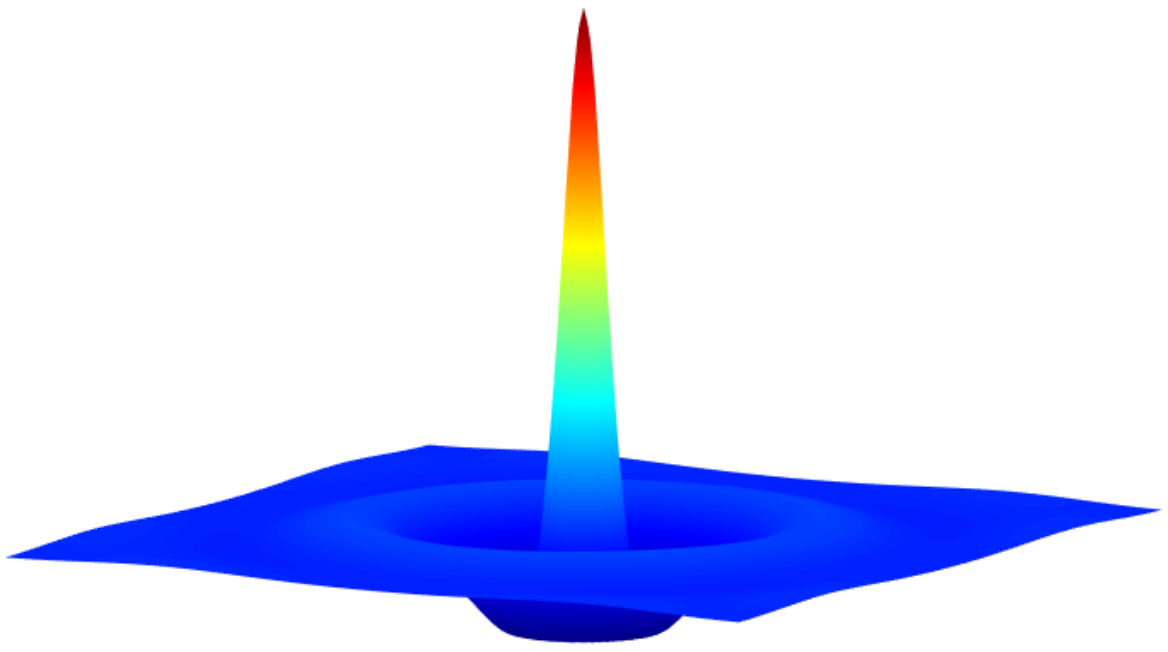

Figure 1. A typical 2D Kerr cavity soliton obtained from the LLE, showing a bright peak on a darker homogeneous background with a weak diffraction ring.

## Current and Future Challenges

Optical pattern formation was born as a topic of fundamental character, object of theoretical/numerical investigations accompanied by experimental verifications in the laboratory.

In the sixties there was the start of a surge of interest in dynamical phenomena in the very general context of systems governed by a nonlinear evolution. Disciplines such as Nonlinear Dynamics, Prigogine's Theory of dissipative structures and Haken's Synergetics were formulated and widely developed. Especially chaotic phenomena were the object of a very lively interest. In such a general framework, optical systems offered a remarkably interesting area of investigation. A special feature of optical systems is that they are very fast and display a large frequency bandwidth, which opens the possibility that theoretical research in the field of optical pattern formation produces a practical impact. An example of this is offered by the LLE [5], which describes a Kerr medium contained in an optical resonator of high-Q driven by an external coherent field. This model is perfectly realized by Kerr microresonators [3], with the additional advantage represented by the miniaturization of such devices.

For the present and future activities in the area of optical pattern formation, an obvious possibility is to activate novel theoretical/numerical research that will in turn activate new experimental verifications toward timely applications in, e.g., the fields of all-optical and high-capacity information transmission and processing, high resolution imaging, high precision spectroscopy.

On the other hand, a further essential challenge is to develop activities that, hopefully in a few years, will lead to commercial devices. These pages on optical pattern formation will not address the latter challenge.

## Advances in Science and Technology to Meet Challenges

A first obvious possibility is to identify novel fundamental relationships and analogies in the scenario of the multiform multimodal aspects of optical systems., thus developing the wisdom mentioned in the "Status" section.

After saying that, let us mention a few possible directions for investigations.

The research on optical pattern formation has mainly concerned two kinds of contexts: In the first, one considers configurations that involve only time and the longitudinal spatial variable (e.g. in [3,4]), so that only longitudinal structures arise; in the second, only time and the transverse variables are involved (see e.g. [8] in which the cavity length is so small that only one longitudinal cavity mode can be excited, see also Fig. 1), so that only transverse structures arise. It is natural to consider a full 3D+time context in which 3D spatial structures can arise (see Fig. 2). Of course, this last context presents much higher numerical difficulties and there is not so far clear experimental evidence, but it is pregnant of novelties.

Along that vein, interesting new scenarios can emerge in active systems when well-known longitudinal instabilities are analyzed in a full 3D context. We refer to instabilities due to physical effects such as, for example, resonance with the Rabi frequency [1].

So far, in the studies of pattern formation in optical cavities the attention has been mostly devoted for simplicity to homogenously broadened systems and ring cavities. We believe that an important room for future research is offered by the case of inhomogeneously broadened systems (spectral hole burning) and systems contained in a Fabry-Perot cavity (spatial hole burning). The recently formulated travelling wave formalism [9] offers the opportunity of addressing the Fabry-Perot case in a relatively simple way and of comparing systematically the scenarios for the two kinds of cavity.

Other possibilities are offered by the study of pattern formation in systems like VCSELs where there is an additional degree of freedom related to the polarization of light and systems governed by different nonlinearities.

A system that deserves to be further explored with respect to optical pattern formation is the laser with saturable absorber. While the formation of pulses due to passive mode locking is a well-known phenomenon, only recently the transition to independently addressable localized structure (even in 3D [10]) when the system exhibits bistability between a lasing and a nonlasing state has been investigated. These localized structures have a much higher contrast with respect to those produced in driven systems. Moreover, the connection between these structures and the pulses that arise from the resonance with the Rabi frequency, which is a coherent phenomenon, has still to be clarified.

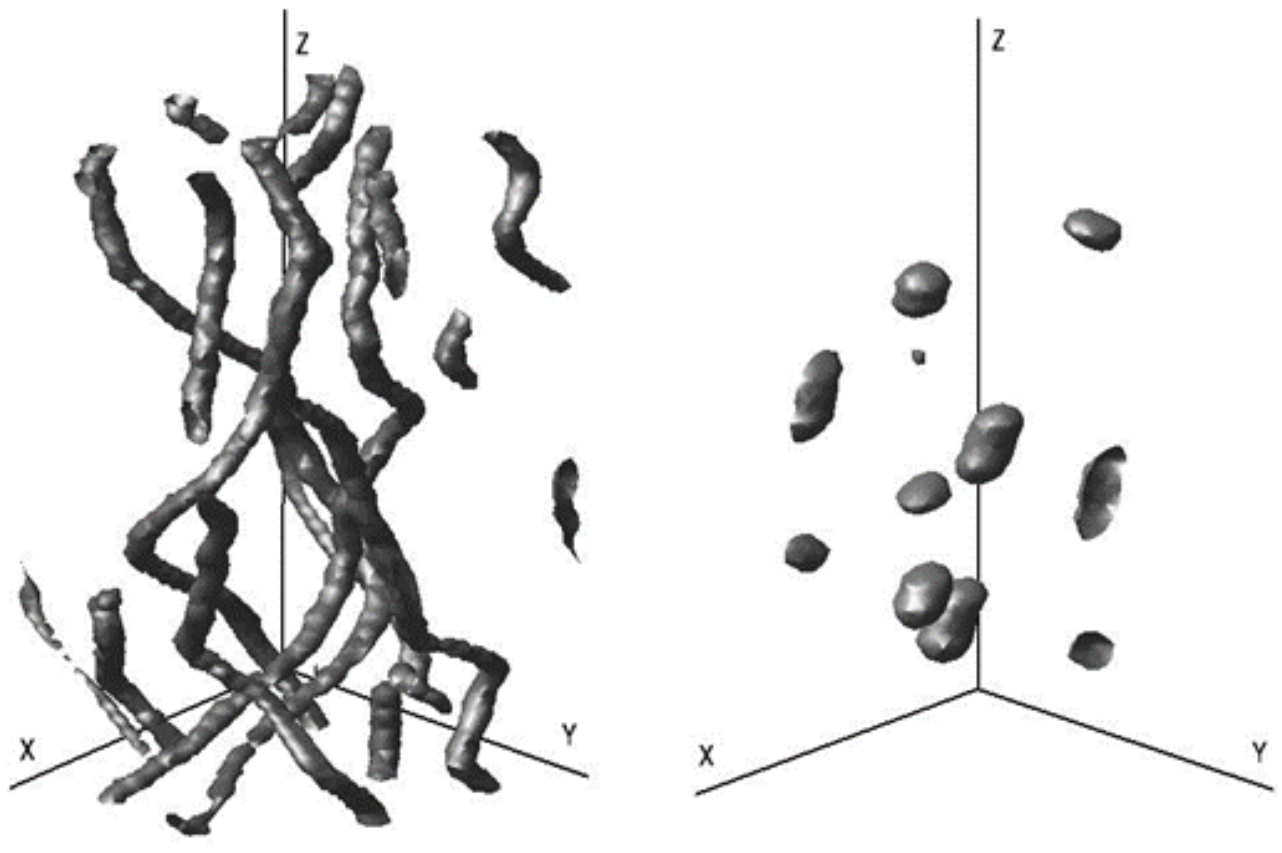

Figure 2. 3D Self-confined structures in a 2-level, saturable absorber resonator driven by an external, transversally homogeneous, CW field. The shaded surfaces show isointensity contours of the intracavity field, the structures propagate along the optical axis, Z, and are confined in the transverse plane (X,Y). Left: for high values of the external intensity, irregularly modulated filaments are realized at regime; right: decreasing the input field intensity, the filaments shrink and eventually stabilize in fully self-confined 3D solitons (Cavity Light Bullets – CLB) which perform regularly periodic roundtrips (under periodic boundary conditions). Multiple CLB could be individually turned on and off by external field pulses, both in the transverse plane and in pairs along Z. [Reprinted from M. Brambilla, T. Maggipinto, G. Patera, L. Columbo, "Cavity light bullets: three-dimensional localized structures in a nonlinear optical resonator", Phys. Rev. Lett. 93, 203901 (2004)]

## Concluding Remarks

Optical patterns include an intrinsic connection between time/frequency, space and space-time. For example, the longitudinal patterns that arise in cavities are simultaneously modulated in space and time, because they travel along the cavity with the light velocity.
Because of the six-decades history of the topic of optical pattern formation, it is not straightforward to predict its future evolution. However, in the previous section we have outlined a few possibilities. In the past, the main trend in the theoretical/numerical investigations has been to go from models that provide a complete description of the dynamics in space and time to simpler models that include time and the longitudinal variable or time and the transverse variables. For the future it is natural to expect an opposite trend towards models that include a complete spatial 3D description.

## 17. Towards Space-Time Photonics

Abbas Shiri and Ayman F. Abouraddy, University of Central Florida

### Status

Many of the features of photonic devices, including some of the most ubiquitous components such as resonators and waveguides, are usually thought to be intrinsically dependent on their geometry and constitutive materials. As such, the behaviour of an optical field interacting with such devices is dictated by the boundary conditions imposed upon the field. For instance, the resonant wavelengths and linewidths of a planar cavity are expected to be set by the mirror reflectivities, cavity length, and refractive index. Henceforth, satisfying a longitudinal phase-matching condition allows for incident light to resonate with the cavity; see Fig. 1(a). As another example, consider the planar waveguide shown in Fig. 2(a). The field is confined along one transverse dimension, but diffracts along the other unbounded dimension.

We have recently introduced several strategies for challenging these long-held intuitions that may be collected under the moniker 'space-time (ST) photonics', whereby the response of a photonic device is tailored post-fabrication in useful ways by sculpting the spatio-temporal structure of the incident optical field, rather than modifying the device itself. In fact, introducing a prescribed relationship between the spatial frequencies (or propagation angles) and the temporal frequencies (or wavelengths) can help overcome the constraints imposed by the boundary conditions. We refer to such pulsed beam configurations as ST wave packets [1] (or simply ST fields when incoherent [2]).

In one scenario, introducing carefully designed angular dispersion into a pulsed field (or a broadband incoherent field) allows the realization of *omni-resonance* [3]: the pulse traverses the cavity without spectral filtering even if the pulse bandwidth is larger than the cavity resonant linewidth after the entire bandwidth resonates with it; see Fig. 1(b). Omni-resonance is achieved by restricting the spatio-temporal spectrum of the field to a one-dimensional trajectory on the cavity light-cone corresponding to the dispersion relationship of a single resonant mode; see Fig. 1(c) [4,5].

A similar strategy enables a new class of planar waveguide modes we refer to as 'hybrid guided ST modes', as shown in Fig. 2, where the field is confined along the unbounded dimension through ST coupling [6]. Crucially, the spatio-temporal structure introduced into the field along the unbounded dimension enables overturning the impact of the boundary conditions along the other dimension. For example, the modal size, index, and dispersion can all be engineered independently of the thickness and refractive index of the planar waveguide; i.e., the impact of the boundary conditions is overturned [6]. These two examples, omni-resonance in a planar cavity and hybrid guided ST modes in planar waveguides, are only two examples of this emerging area of ST photonics.

### Current and Future Challenges

The concept of omni-resonance addresses a long-standing challenge in optical physics: how to deliver resonant enhancement over a continuous broad bandwidth, rather than to discrete, spectrally narrow resonances. For example, judicious design of a cavity structure containing a thin weakly absorbing layer can enable coherent perfect absorption (CPA); that is, full absorption of incident light independently of the layer thickness or intrinsic absorption, albeit only on resonance [7]. By combining omni-resonance with CPA via pre-conditioning of the incident optical field, broadband CPA is achieved. We recently realized such a configuration by embedding an amorphous silicon PIN junction in a CPA cavity and observed a doubling in the harvested electrical power from near-infrared light in the range where the absorption of silicon drops rapidly, thereby confirming the potential impact of ST photonics on solar-energy harvesting [8].

This proof-of-principle demonstration helps set the stage for the next challenges that must be addressed to capitalize on this capability. The first challenge is to extend the omni-resonant bandwidth to encompass the visible in addition to the near-infrared. This can open up applications in transparent solar cells, high-speed thin-film photodetectors, and imaging in photon-starved environments. To date, we have demonstrated omni-resonance by increasing the resonant bandwidth associated with a single cavity resonance. Further broadening of the omni-resonant bandwidths can make use of the concept of 'spectral-recycling' [9], whereby the full spectrum is divided into narrower segments, each of which is then assigned to a distinct cavity resonance, thereby alleviating some of the difficulties involved in broadband omni-resonance.

With regards to hybrid guided modes, several challenges need to be met to fully benefit from the low losses and controllable characteristics of these novel mode structures. First, the controlled excitation of one mode or a superposition of modes in few-mode planar waveguides has yet to be demonstrated. Second, sustaining such modes in dispersive medium modifying their spatio-temporal design. Thirdly, new avenues can be explored by demonstrating nonlinear interactions that harness the controllable group indices and potentially long interaction lengths of multi-wavelength and multi-mode field configurations in planar waveguides.

**Advances in Science and Technology to Meet Challenges**

The chief advance required to meet these challenges is in the area of compact, precise, yet convenient and versatile synthesis of ST wave packets [1] and ST fields [2]. To date, we have relied on free-space-based spectral synthesis using bulk optical components, which requires either a specially designed phase plate of a spatial light modulator to manipulate the spectrally resolved wave front. Replacing this approach with a single-surface photonic structure, such as a metasurface, can dramatically simplify the overhead in applications that benefit from omni-resonance, especially in the area of solar energy. Alternatively, this entire endeavour can benefit from developing lasers, whether solid-state or semiconductor-based, that emit ST wave packets directly rather than conventional pulsed beams in which the spatial and temporal degrees of freedom are separable. Indeed, combining an on-chip source of ST wave packets with planar waveguides can help produce compact delay lines [10] needed for constructing optical buffers, which are critical components for future all-optical networks.

**Concluding Remarks**

We are currently witnessing the emergence of a nascent field of research in which the precise spatio-temporal structure of the optical field is tailored to modify its mode of interaction with photonic devices, including such basic and fundamental components such as cavities and waveguides. New modes of behaviour consequently become available: omni-resonant interaction with planar cavities, and hybrid guided modes in planar waveguides with modal indices that are independent of the boundary conditions. This new behaviour is *not* a result of nano-scale structuring of the devices or strong field confinement; instead, it is a consequence of restricting the spatio-temporal spectrum to a reduced dimensionality domains with respect to those of conventional pulsed beams. Rapid progress in this emerging field is expected to yield breakthroughs in nonlinear and quantum optics, and applications ranging from improved solar-energy harvesting to optical imaging.

## Acknowledgements

The work was funded by the U.S. Office of Naval Research (ONR) under contract N00014-17-1-2458 and ONR MURI contract N00014-20-1-2789.

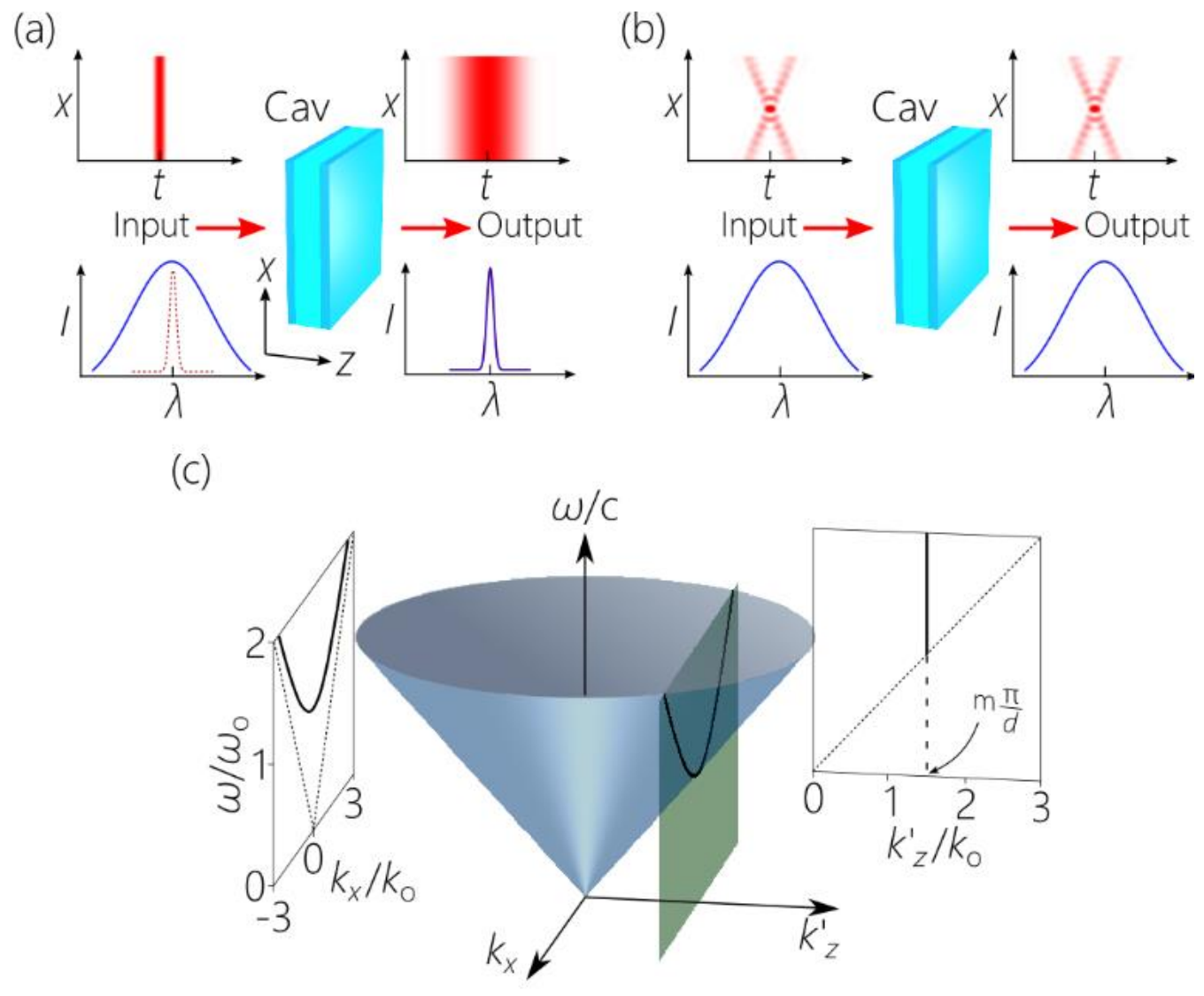

**Figure 1**. Omni-resonant ST wave packets. (a) A conventional short wave packet (in which the spatial and temporal degrees of freedom are separable) is spectrally filtered by a narrow-linewidth cavity, and thus the pulse broadens in time. (b) An omni-resonant ST wave packet having the same pulse width and bandwidth as in (a) traverses the same cavity without spectral filtering or temporal broadening. (c) The spectral support domain of the omni-resonant ST wave packet on the light-cone is restricted to the dispersion relationship associated with a constant axial component of the wave vector inside the cavity.

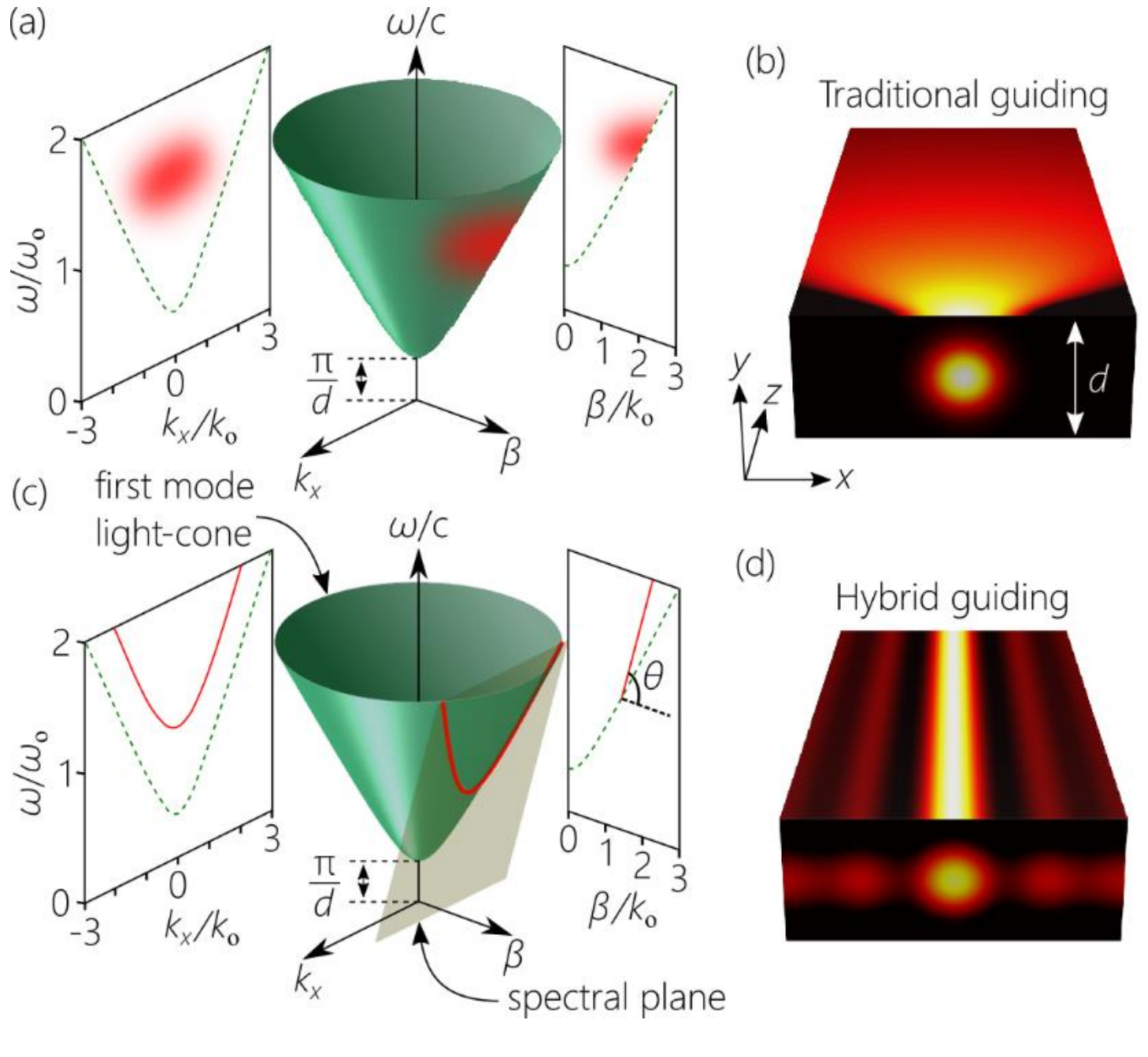

**Figure 2.** Hybrid guided ST modes. (a) A conventional wave packet in a planar waveguide is represented by a patch on the surface of the light-cone. This light-cone corresponds to the dispersion relationship of a single mode. (b) The field is confined along one transverse dimension by the planar waveguide via a conventional guidance mechanism. However, the field diffracts along the other unbounded transverse dimension. (c) The spectral support domain of a hybrid guided ST mode is restricted to the red curve on the surface of the light-cone from (a). (d) The hybrid guided ST mode is now guided along both transverse dimensions.

## 18. Space-Time Metasurfaces

Andrea Alù, City University of New York

### Status

Metasurfaces have been offering exciting tools to control and tailor the electromagnetic wavefront. Their operation is typically based on patterning the surface aperture at the subwavelength scale with suitably designed meta-atoms, which can imprint the desired spatial response relying on local and nonlocal resonant phenomena. Despite the significant progress in this research field and the various opportunities for applications, it is important to realize that metasurfaces based only on spatial engineering are fundamentally limited in their operation by linearity, time-invariance and passivity. Power conservation poses limitations on the total energy scattered by the metasurface towards the available diffraction channels, while linearity and time invariance are typically associated with frequency conservation and reciprocity constraints. When a suitably tailored temporal modulation along the surface is added to the picture, these constraints can be relaxed, and exciting opportunities emerge in the context of wave manipulation and engineering [1-5]. These space-time metasurfaces are not bound to conserve energy or frequency, i.e., they can extract or dump power from/to the modulation network across the surface, and they can modify the frequency content of the signal with large flexibility. The schematic geometry in Fig. 1a shows a metasurface modulated by a periodic traveling wave, which supports the emergence of a discrete number of space-time diffraction orders, through which the incoming wave can be steered with large flexibility and efficiency towards a wide range of directions and frequencies [5]. These phenomena are not constrained by Lorentz reciprocity [1-3], implying the possibility of implementing free-space circulation of waves, as shown in Fig. 1b, and isolation over a surface. These functionalities can be realized not only for free-space radiation, but also for surface-wave guided propagation [6]. The techniques used for time modulation enable also real-time reconfigurability, opening a new paradigm for metasurfaces in which the energy, time, frequency, spatial and momentum content of the incoming signals can be widely controlled and manipulated in real time, with opportunities for wireless communications, energy harvesting, integrated technology, nanophotonics and quantum optics, spanning a broad frequency range from radiofrequencies to visible light.

### Current and Future Challenges

Space-time metasurfaces currently face several challenges before they can become mainstream technology. First, their design and optimization are inherently complex due to their time-varying nature, which makes challenging their theoretical modelling and full-wave simulations. Time-domain techniques are inefficient, given that the involved phenomena are typically resonant, whereas frequency-domain approaches need to consider the emergence of several new harmonics and mixing among them. The control of all generated harmonics presents additional design and optimization challenges. In addition, the inherent combination of multi-physics domains emerging from the involved time modulation schemes requires the use of complex and often inefficient co-simulation techniques.

As a function of the operating frequency, temporal modulation schemes require various approaches. At radio-frequencies, varactors, switches or transistors can be used to impart the required temporal variations, with a relatively straightforward path towards integration. Fig. 2 shows the measured response of a radio-frequency space-time metasurface [3] implemented through an array of varactors loading a transmission-line, enabling largely asymmetric transmission and reception patterns (Fig. 2c):

incoming and outgoing signals can have very different transmission strengths, beyond the constraints dictated by reciprocity for time-invariant surfaces. This operation (Fig. 2b) is ideally suited to neutralize jamming and unwanted noise and echoes. Similar concepts translated to higher frequencies can enhance the efficiency of energy harvesting devices, overcoming the symmetry between absorption and thermal emission (Fig. 2a) [3]: different from conventional absorbers, an efficient space-time absorbing metasurface does not need to re-emit a portion of the absorbed energy towards the source as it becomes hot.

At these higher frequencies electro-optical or all-optical modulation techniques, through the use of nonlinearities, appear to be best suited. Large quality factor resonances and low loss are required to lower the wave speed as it interacts with the surface, in order to maximize light-matter interactions and to make most effective use of the typically weak and slow temporal modulations. In turn, this requirement poses constraints on the material properties and fabrication techniques for space-time metasurfaces, and on their overall available bandwidths of operation. Strong nonlinearities come at the price of material loss, implying important trade-offs in the design of optimal structures. Graphene, transition metal dichalcogenides, multiple quantum wells and other innovative photonic materials offer interesting prospects in this context, providing large nonlinearities, limited loss, and opportunities for efficient temporal modulation schemes.

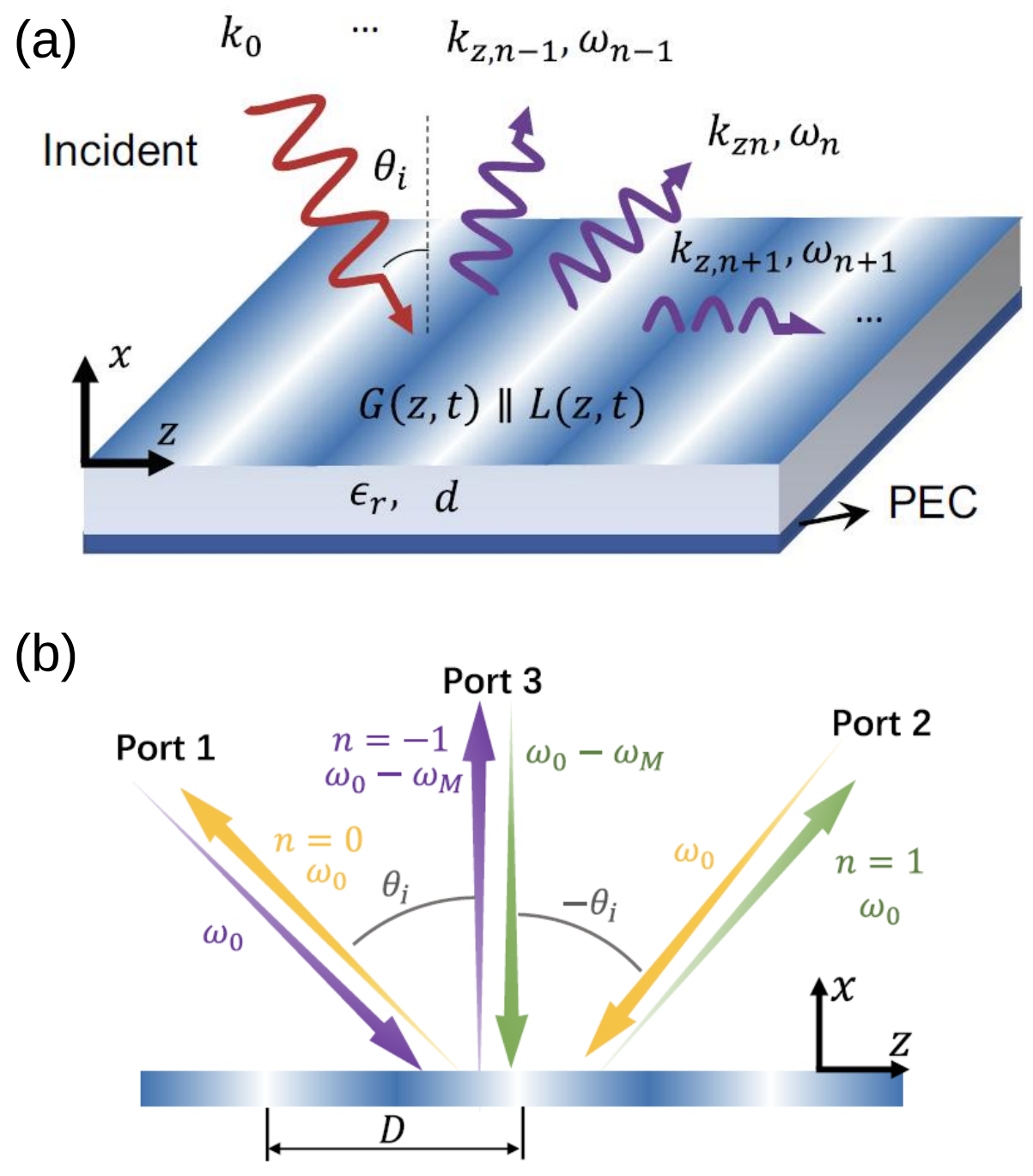

**Figure 1.** (From 5]) (a) A space-time metasurface formed by a spatio-temporally-varying surface impedance backed by a ground plane. (b) A free-space circulator implemented through the suitable engineering of the space-time diffraction orders of the metasurface.

## Advances in Science and Technology to Meet Challenges

Recent discoveries in the context of metasurfaces have been offering new tools to address some of these challenges. Light-matter interactions can be drastically enhanced over broad bandwidths exploiting hyperbolic phenomena, which are however challenging to establish in passive, linear, time-invariant devices. In [6] hyperbolic surface waves with the added benefit of large nonreciprocity were demonstrated over a space-time metasurface in which the spatio-temporal modulation scheme mimicked a very fast translational motion of the surface. This example showcases the powerful opportunities offered by temporal modulations to go beyond the limitations of time-invariant systems, which can be leveraged for space-time metasurfaces.

Suitable forms of temporal modulations have been found to overcome the conventional trade-off between bandwidth and light-matter interaction enhancements, largely overcoming the delay-bandwidth limit of linear, time-invariant structures [7], and they can also provide parametric gain that can be used to overcome passivity limitations and broaden the bandwidth of operation beyond the conventional limits [8], offering interesting paths towards broadband, compact, efficient space-time metasurfaces. These possibilities, which may be implemented within the same modulation schemes discussed above, need to be fully explored in the context of space-time metasurfaces. All-optical pumping through nonlinearities can be particularly exciting in the near-infrared and visible range, especially when non-monochromatic modulation schemes are considered. In [9] a photonic topological insulator was envisioned within a space-time metasurface exploiting the natural nonlinearity in silicon and an optical pump characterized by the beating between two tones with different frequency and opposite polarization handedness. This specific all-optical modulation scheme imparts the required slowly rotating modulation mimicking the effect of a dc magnetic bias and enabling nonreciprocal topological responses. An extreme form of non-monochromatic modulation is abrupt temporal switching, a technique recently explored for metamaterial and metasurface technology. A temporally switched metasurface can enable, for instance, broadband absorption [10] beyond the limits of passive thin absorbers, and more generally efficient frequency transformations and energy amplification. All these possibilities demonstrate the powerful opportunities offered by space-time metasurfaces, and the potential of this technology for a wide range of applications.

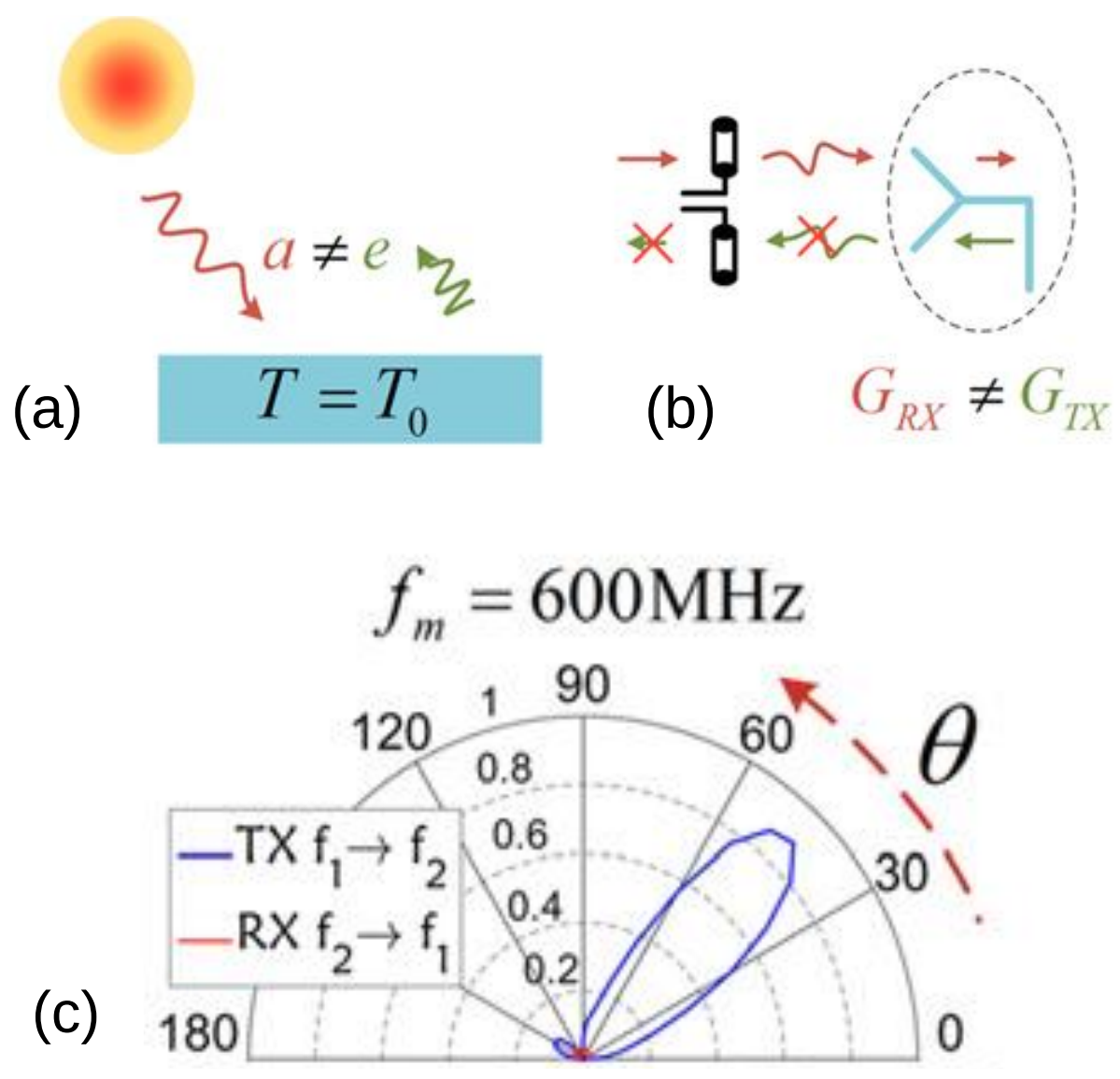

**Figure 2.** (From [33]) Space-time metasurface for (a) breaking the symmetry between absorption and thermal emission, (b) breaking reciprocity in antenna systems. (c) Large asymmetry between radiation and reception patterns in a space-time leaky-wave metasurface.

## Concluding Remarks

The field of metasurfaces has matured into a powerful platform to manipulate waves and enable extreme phenomena. Yet, only spatial transformations severely limit the overall reach of this technology. Adding to this platform suitably tailored temporal modulations offer new opportunities for extreme wave manipulation, overcoming the limitations and restrictions of passive, linear, time-invariant metasurfaces, including reciprocity constraints, power conservation and bandwidth limitations. By exploring new material and integration techniques, we envision a bright future for space-time metasurfaces, with applications spanning radio-wave and mm-wave technology for wireless communications, including new opportunities in the emerging field of reconfigurable intelligent surfaces, and a plethora of nanophotonic, classical and quantum photonic reconfigurable technologies for communications and computing, energy harvesting, sensing and imaging, and efficient sources.

**Acknowledgements**
Our work on this field of research has been supported by the Air Force Office of Scientific Research, the Office of Naval Research, a Vannevar Bush Faculty Fellowship, the National Science Foundation and the Simons Foundation.

## 19. Space-Time Metamaterials

Emanuele Galiffi, J.B. Pendry, Imperial College London, London, UK
Paloma A. Huidobro, IST-University of Lisbon, Portugal

**Status**

Metamaterials and metasurfaces enable extraordinary wave effects, such as negative refraction. Despite the unprecedented advances in wave control they have enabled, most metamaterials are static in nature: their properties are fixed upon fabrication. Novel materials and techniques now allow for large and fast tunabilities of their optical response even at moderately high frequencies: this is being exploited to realise active metastructures, with time-modulation (TM) offering a new tuning knob for shaping light-matter interactions. TM also bears fundamental consequences: Energy conservation is no longer guaranteed, as external energy inputs must drive any change in the electrodynamic parameters of a medium. However, other fundamental symmetries, such as electromagnetic reciprocity, cannot be violated by TM alone.

In a temporally periodic medium, the common frequency band gaps of spatial crystals [Fig. 1(a)] are replaced by wavevector-gaps [Fig. 1(b)], originating from the mixing of positive and negative frequencies. However, the symmetry of both band structures under reflection about either axis signifies their electromagnetic reciprocity. Breaking reciprocity constitutes a major challenge in electromagnetics, commonly met in microwave technology using ferromagnetic components, unsuited to miniaturisation. Recently, the quest for compact devices for full-duplex telecommunications has steered unprecedented interest towards combining spatial and temporal modulations into travelling-wave modulations [Fig. 1(c)], capable of inducing a directional bias in the response of a medium, and hence of breaking reciprocity [Fig. 1(d-f)]. This opportunity sparked the rise of space-time metamaterials (STMs), leading to the realisation of nonreciprocal devices in silicon photonics, microwaves, and acoustics [1].

However, non-reciprocity is only one direction opened by STMs, and these systems offer fundamentally new pathways for the advanced manipulation of wave-matter interactions, combining spatial and temporal degrees of freedom for multimode light structuring [2]. Metamaterials are usually defined as artificial structures with arbitrary effective parameters, thus generalising the concept of *matter*. STMs introduce a new gear by generalising the concept of *motion*. From travelling-wave amplifiers in RF to free-electron lasers spanning the electromagnetic spectrum, moving matter is uniquely capable of interacting strongly with light. However, while relativistic matter is unsuited to most applications, STMs enable *synthetic motion*.

**Current and Future Challenges**

As recently shown [3], synthetic motion is capable of producing radiation, accelerating any incident field to realise strong optical pulse trains, an effect termed luminal amplification. This could be achieved via a travelling-wave modulation synthetically moving at speeds similar to that of waves in the unmodulated medium, realising a forward-band degeneracy with resulting strong forward-forward scattering [Fig. 1(e)]. The experimental realisation of luminal amplification is an open challenge.

Another characteristic feature of moving matter is its ability to drag light. Synthetic motion extends the concept of light drag, breaking reciprocity in the quasi-static limit [4]. Curiously, this requires the modulation of both electric and magnetic material parameters, and the direction of the resulting drag may be tuned through the relative phase of electric and magnetic modulations [Fig. 2(a)]. The possibility of modulating multiple parameters is in fact unique to synthetic motion, as physically moving matter carries both electric and magnetic dipoles with identical speed and phase. Similar drag effects associated with the combination of multiple travelling-wave modulations have also been recently proposed for thermal waves [5], and measured in charge-diffusive systems [6], demonstrating the potential of this concept for systems beyond photonics [Fig. 2(b)]. However, experiments with multiple electromagnetic modulations have only been proposed theoretically [4] and their realisation constitutes a key challenge.

Further degrees of freedom may still be unlocked by empowering STMs with anisotropic modulations. STMs may also provide a new route to engineer geometric phases, such as Floquet topological insulators [1]. Moreover, the concept of synthetic dimensions in the frequency domain, enabled via TM [7], may be further extended to STMs, realising nonreciprocal synthetic dimensions consisting of combinations of spatial and temporal degrees of freedom. Finally, being energetically open, STMs may enable the implementation of exceptional points and rings, constituting a new exciting playground for non-Hermitian Physics.

**Advances in Science and Technology to Meet Challenges**

Similarly to TM media, achieving high modulation speeds constitutes a major challenge for the realisation of STMs. However, STMs present one let out clause: in many instances, the desired effect relies on the *velocity* $\Omega/g$ of the modulation, which needs to be significant compared to the speed of light $\omega/k$, while the modulation *frequency* $\Omega$ may still be much lower than $\omega$. The key figure of merit then is the lifetime of the relevant mode.

Moreover, modulation strengths are crucial: At near-infrared frequencies, epsilon-near-zero materials and metasurfaces currently offer the best platforms thanks to their high sensitivity to external actuation, combined with relatively low losses, with relative index modulations of order unity recently reported in indium-tin-oxide, see Fig. 2(c) [8]. Nevertheless, the nonlinear processes involved in such large nonlinearities have picosecond-timescales, and new efforts and breakthroughs will be needed to push comparable modulation efficiencies towards the femtosecond barrier. Further exploration of charge and exciton dynamics in other two-dimensional materials will be a promising route [9].

At microwave frequencies, varactors and nonlinear inductors offer possible pathways towards fast and efficient TM in transmission lines. These can also be combined with the high tunability of 2D materials, exploited via gate-biasing or photo-excitation, with modulation rates already surpassing hundreds of GHz. In addition, STMs bear an endemic additional challenge: The need for phasing the temporal and spatial components of the modulation into a travelling-wave. Efficient schemes to achieve this experimentally will be essential for these new concepts to progress towards practical applications, such as metasurfaces with isolating, circulating and non-reciprocal phase-shifting capabilities [10].

Finally, theoretical developments will be needed. Microscopically, a clear fundamental understanding of TM in epsilon-near-zero materials is still in its infancy, a challenge stretching from photonics to material science. Macroscopically, although STMs are commonly modelled with Floquet-Bloch expansions and the first analytical theories have been introduced, most models assume isotropic, dispersionless media: Understanding dispersion and causality in time-dependent systems is key to assessing their ultimate potential. Finally, new fundamental advances will be needed to allow STMs to enter the realm of quantum light-matter interactions.

**Concluding Remarks**

The study of STMs is a very promising area, fuelled by the prospects of realising a plethora of effects such as frequency conversion, time reversal, negative refraction, nonreciprocal light-matter interactions, relativistic effects based on synthetic motion, and topological physics. Despite the challenges to progress from current microwave realisations of STMs for non-reciprocity to the exploitation of their full potential, there has been great progress in achieving temporal modulations of the permittivity or conductivity, which established solid foundations for further investigation of additional degrees of freedom. Meanwhile, the rise of epsilon-near-zero materials has been catalysing the recent progress leading to the first experimental realisations of TM media in the near-infrared. STMs bring their own challenges, and the capability of realising potentially complex spatio-temporal modulation patterns will be key to the advancement of the field, although the high number of new degrees of freedom available in STMs may open the gates to truly new and unexpected opportunities.

**Acknowledgements**

E.G. acknowledges funding by the Engineering and Physical Science Research Council, grant number EP/T51780X/1. P.A.H. acknowledges funding from Fundação para a Ciencia e a Tecnologia through the CEEC Individual program with reference CEECIND/03866/2017 and Instituto de Telecomunicações under project UIDB/50008/2020. J.B.P. acknowledges funding from the Gordon and Betty Moore Foundation.

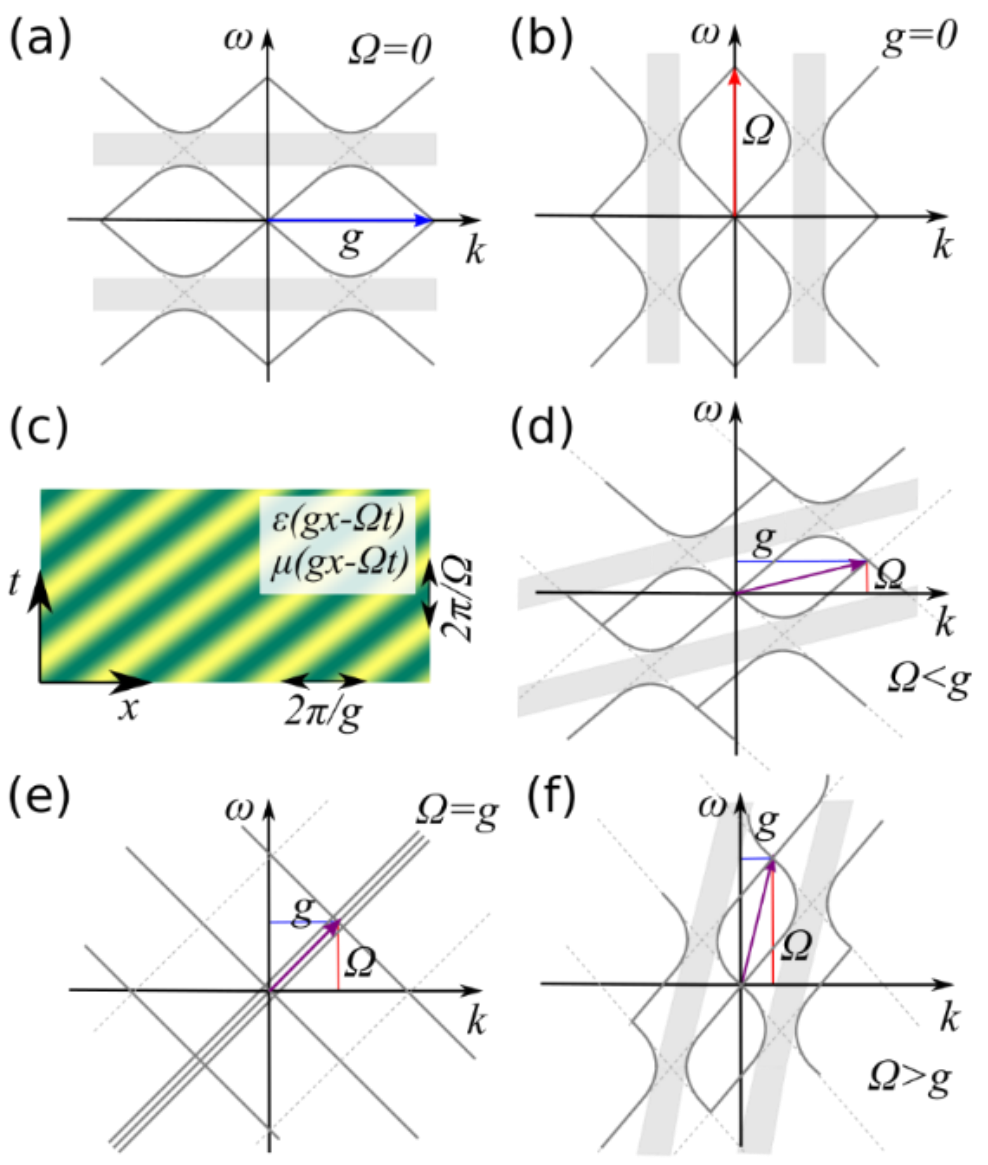

**Figure 1** Schematics of wave dispersion in space, time and spacetime—periodic metamaterials, with $g$ and $\Omega$ the spatial (blue arrow) and temporal (red arrow) modulation frequencies. (a) Spatially periodic media like photonic crystals produce conventional frequency band gaps. (b) Periodic frequency modulations give rise to vertical wavevector gaps hosting parametrically amplified states. (c) A modulation in space-time produces (d-e) non-reciprocal (tilted) band gaps: (d) frequency gaps open for subluminal modulation, and (e) wavevector gaps open for superluminal disturbances (f). Finally, luminal modulations, which synthetically travel close to or at the speed of waves in the medium, result in a degeneracy of all forward waves.

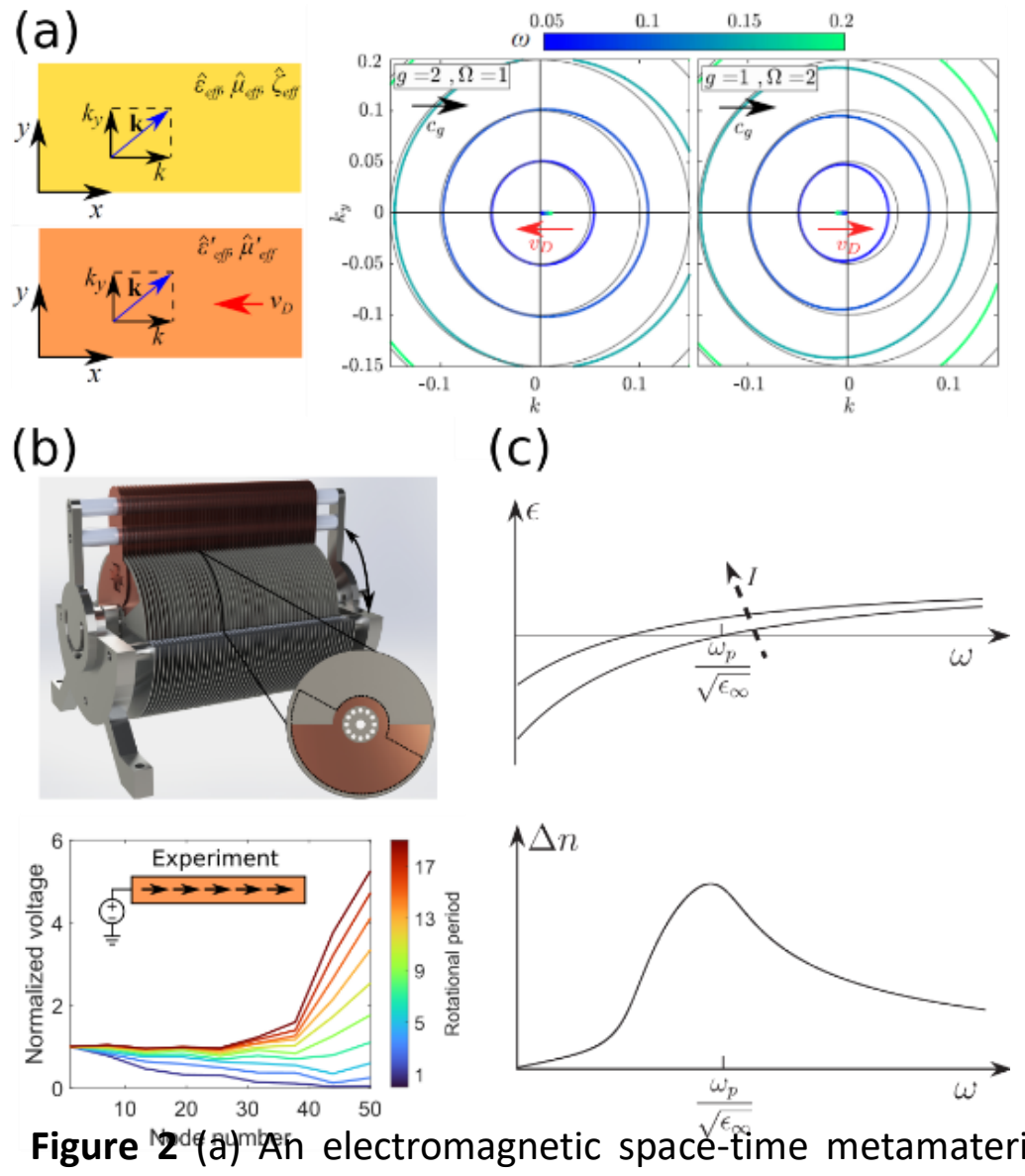

**Figure 2** (a) An electromagnetic space-time metamaterial behaves effectively as a bianisotropic medium, and modulations of the permittivity and permeability result in a synthetic Fresnel drag of light revealed by displaced frequency contours. The direction of the synthetic light drag can be tuned by changing between subluminal and superluminal modulations, or via the modulation phase. Reproduced from Ref. [4] with permission. (b) Experimental demonstration of a spatiotemporal diffusive metamaterial. Reproduced from Ref. [6] with permission. (c) Tunability of the refractive index in indium-tin-oxide. Reproduced from Ref. [8] with permission.